\documentclass[prd,nofootinbib,floats,aps,superscriptaddress,preprint]{revtex4}
\usepackage[hypertex]{hyperref}
\usepackage{graphicx} 
\usepackage{amsbsy}   
\usepackage{epsfig}

\newcommand{\sphi}{S_{\phi K}}
\newcommand{\seta}{S_{\eta' K}}
\newcommand{\spsi}{S_{\psi K}}
\newcommand{\spsph}{S_{\psi \phi}}
\newcommand{\ADK}{A_{DK}}
\newcommand{\srr}{S_{\rho\rho}}

\newcommand{\epsK}{\varepsilon_{K}}

\newcommand{\beq}{\begin{eqnarray}}
\newcommand{\eeq}{\end{eqnarray}}

\begin{document}

\pagestyle{plain}
\vspace*{-1.5cm}
\preprint{\footnotesize LBNL-58627}
\preprint{\footnotesize UCB-PTH-05/26}
\preprint{\footnotesize MIT-CTP 3680}

  \vspace*{1.05cm}

\title{Next to Minimal Flavor Violation}

\author{Kaustubh Agashe} 

\affiliation{\small Department of Physics and Astronomy, Johns Hopkins University,
Baltimore, MD 21218-2686}

\affiliation{\small School of Natural Sciences,
Institute for Advanced Study, Princeton, NJ 08540}

\affiliation{\small Physics Department, 
Syracuse University, Syracuse, NY 13244}

\author{Michele Papucci} \affiliation{\small Theoretical Physics Group,
  Ernest Orlando Lawrence Berkeley National Laboratory, University of
  California, Berkeley, CA 94720} \affiliation{Department of Physics,
  University of California, Berkeley, CA 94720}

\author{Gilad Perez} \affiliation{\small Theoretical Physics Group, Ernest
  Orlando Lawrence Berkeley National Laboratory, University of
  California, Berkeley, CA 94720}

\author{Dan Pirjol} \affiliation{\small Center for Theoretical Physics, 
Massachusetts Institute of Technology,
Cambridge, MA 02139}


 \vspace*{-.54cm}

 \begin{abstract}
  \vspace*{-.15cm}
  The flavor structure of a wide class of models, denoted as next to
  minimal flavor violation (NMFV), is considered.
  In the NMFV framework, new physics (NP), which is required for
  stabilization of the electroweak
  symmetry breaking (EWSB) scale, naturally couples (dominantly)
to the third generation
  quarks and is quasi-aligned with the Yukawa matrices.
  Consequently, new sources of flavor and CP violation are present in
  the theory, mediated by a low scale of few TeV. However, in spite
of the low flavor scale,
  the most severe bounds on the scale of NP are evaded since
  these are related to flavor violation in the first two generations.
Instead, one typically
finds that the NP contributions are comparable in size to SM loop
processes. 
We argue that, in spite of
the successful SM unitary
  triangle fit and contrary to the common lore, such a
  sizable contribution to $\Delta F=2$ processes of 
$\sim 40 \%$ (with arbitrary phase)
compared to SM is 
  presently allowed 
since B-factories are only beginning to
  constrain these models.
Thus, 
it is very interesting that
in the NMFV models 
one is not forced to separate
the scale of
NP related to EWSB and the scale of flavor violation.
We show briefly that this simple setup includes a wide class of supersymmetric
  and non-supersymmetric models all of which solve the 
 hierarchy  problem.
  We further discuss tests related to $\Delta
  F=1$ processes,
  in particular the ones related to $b\to s$ transition. The 
$b \rightarrow s$ 
processes are
  computed using two different hadronic models to 
estimate 
the uncertainties involved. In addition,
we derive constraints on the NP from $B \rightarrow K\pi$ data 
using only SU(3) flavor symmetry and minimal dynamical assumptions.
  Finally we
  argue that in many cases correlating  $\Delta F=2$ and
  $\Delta F=1$ processes is a powerful tool to probe
  our framework.

\end{abstract} \pacs{Who cares?} \maketitle


\section{Introduction}
\label{intro}
The most pressing puzzle in modern particle physics
is the origin of electroweak symmetry
breaking (EWSB) and the relative
hierarchy between the EWSB and the Planck scale.
In the last three decades several ideas were proposed
towards the resolution of these mysteries.
They include among others supersymmetry~\cite{SUSY}, 
technicolor~\cite{TC, TCreview},
composite Higgs~\cite{CH}, topcolor~\cite{topcolor}, little Higgs models~\cite{LH} 
in 4d and also Arkani-Hamed-Dimopoulos-Dvali (ADD)~\cite{ADD} and 
Randall-Sundrum I (RS1)~\cite{RS} models\footnote{Based on the AdS/CFT correspondence, RS1 
is conjectured to be dual to 4d composite Higgs models.}
with extra dimensions.
All of these scenarios have 
new physics (NP) at the TeV scale which can be weakly coupled (as in SUSY or Little
Higgs models) or strongly coupled (as in most other solutions).

It is very interesting that
even though flavor physics does not have a direct link with the problems mentioned
above, it plays a crucial role in constraining the 
frameworks proposed to solve them, and might help in the future to distinguish between
the various scenarios. 
The relevance of flavor physics to the resolution of the
above puzzles, if for nothing else, is tightly related to the top quark.
The closeness of the top mass to the EWSB scale $\Lambda_{\rm EW}$
strongly suggest that it has a sizable coupling to the Higgs sector or
to the particles which unitarize the scattering amplitude of the
longitudinal modes of the weak gauge bosons.
In fact, it is the heaviness of the top quark that yields EWSB in many models.
In addition the left handed top is accompanied by
its isospin bottom partner. Thus it seems almost inevitable that
the new degrees of freedom, required for EWSB stabilization, will have
sizable couplings to the SM third generation quarks.
This in turn raises the issue of flavor physics, since
non-universal coupling between the different generations
and the NP sector would induce new sources of flavor and CP violation,
which are tightly constrained.

The fact that, in general, the third generation quarks couple
to a new sector, however, does not necessarily imply additional sources of
flavor and CP violation. If in a model the scale related 
to mediation of flavor
physics is very high ($\gg$ TeV) (see {\it e.g.}~\cite{GMSB, AMSB}
in SUSY) 
then the new spurions which break flavor symmetries
at $\gg$ TeV become irrelevant at low 
energies. Thus, the theory would flow to
a minimal flavor violation (MFV)~\cite{MFV} model in which 
the only relevant source of flavor
and CP violation (i.e., flavor violation in the
NP at TeV) originates from the Yukawa matrices and most of the present
constraints can be evaded~\cite{MFVmore}.
However, the class of such MFV models which naturally account for the hierarchy
problem, the flavor puzzle and present a consistent picture of EWSB
(passing the various electroweak precision tests) is rather limited.
Furthermore, in the MFV scenario
there is no clue about the solution to
the flavor puzzle from the observation of NP
at the TeV scale (expected to resolve the hierarchy problem), 
{\it i.e.}
the EWSB and flavor sectors are decoupled.
Thus we focus below on
the possibility that new sources of flavor violation
are present in the TeV scale physics. We extend
the MFV framework in a rather minimal way, covering many more models  
with TeV scale NP.
Specifically, we assume that NP dominantly couples
only to the third generation quarks (as we argued above, due to heaviness of
the top quark,
NP is very likely to couple {\em at least} to the third generation)
and is {\em quasi}-aligned with the up and
down
Yukawa matrices.
We denote this framework as next to minimal flavor
violation (NMFV).

Within NMFV,
the effective scale mediating flavor violation could be as low as
a few TeV.
Thus, the EWSB and flavor sectors can be more intimately connected
than in MFV, avoiding the latter's unappealing feature of two
vastly different scales.

In order to better understand this point, let us briefly review the
usual argument for a high-scale flavor-violating NP. The most
stringent constraints come from the kaon system.
To study them it is useful to work in the language of effective
theory.
Within the SM the dominant, short distance, contribution
to $\varepsilon_K$ is due to box diagrams with intermediate top
quarks. These induce a four fermion operator,
$\left(\bar sd\right)^2/\Lambda^2_{\rm F}$,
where roughly $\Lambda_{\rm F}\sim 4\pi S_0 M_W/g^2\lambda^5\sim 10^3-10^4\,$TeV
with $M_W$ being the $W$ mass and $S_0$ is the corresponding Inami-Lim~\cite{IL}
function (here $d,s$ are in the mass basis and here and below Lorentz
indices are being suppressed).
It is clear then that, if there is NP which mediates the non-universal
contributions to the first two generations, then such states 
cannot be much lighter than
$\Lambda_{\rm F}$. Such heavy particles cannot
be involved in regularizing the Higgs mass quadratic divergences. This
is a manifestation of the well known tension between the generic lower
bound on the flavor mediation scale and the EWSB scale.

However, suppose
that
NP only couples dominantly to 
the third generation quarks. Or,
equivalently, the NP approximately respects a U(2)$^3$ flavor symmetry
which is a 
subgroup of the U(3)$^3$ SM quark flavor symmetry. 
This implies that in the effective theory one can go to a special
interaction basis\footnote{For instance this can be view as the basis
  in which the horizontal charges in case of alignment models and
anomalous dimensions of fermionic operators 
in the case of composite Higgs models (or equivalently 5D quark-mass matrices
in the RS dual)
  are flavor diagonal or the different
  generations build irreducible representations of a non-abelian flavor
  group.}
in which four fermions operator 
(if we are considering $\Delta F = 2$ processes) induced by the NP
involve only third generation quarks. 
At first sight, the above tension with Kaon system is clearly
avoided, but one needs a more careful
analysis to see effects induced via 3rd generation.

Let us, for example, consider the case in which the dominant non-universal NP
couplings are with the quark doublets, $Q_i$.\footnote{As
 long as the analogue of the operator in (\ref{NMFV}) is quasi-aligned with the down Yukawa matrix, similar arguments would
apply for the case in which the flavor violation is dominantly in the
singlets sector or if it is of a 
mixed type. For a detailed
discussion see  section \ref{RS1}.} Then, in the special interaction basis, the theory
contains one type of new non-universal operators
\beq
{\cal L}^{^{\Delta F=2}}_{\rm NMFV } = 
c_3( \bar Q_3 Q_3)^2/ \Lambda_{\rm NMFV}^2\,,
\label{NMFV}
\eeq
where $c_3 \sim O(1)$ and $\Lambda_{\rm NMFV}$ is the scale of mediation
of flavor violation.
Generically the presence of the above additional term in the theory
implies new sources of flavor and CP violation. The strength of these
is related to the
orientation, in flavor space, of the above term relative to the Yukawa matrices.
For example the
MFV case  corresponds to up (or down) Yukawa matrix 
being {\em exactly} diagonal in this special
basis. 
As we know the up and down Yukawa matrices  are quasi-aligned (from
the left) by themselves
where the misalignment between the first [second] and third
generations are characterized 
by the corresponding CKM mixing
angles, of $\left(V_{\rm CKM}\right)_{13}\sim {\cal}
O(\lambda^3)$
[$\left(V_{\rm CKM}\right)_{23}\sim{\cal} O(\lambda^2)$]
respectively, where $\lambda$ is the Cabibbo angle. 
Thus, in this basis,
the down-Yukawa (up-Yukawa)
is diagonal up to {\em
  exactly} the CKM matrix.

As just described, MFV requires a very restrictive flavor
structure. Here instead we 
assume that the
up-Yukawa matrix is not diagonal in this special basis.
Within our framework, the NP distinguishes between
the third generation quarks, especially the top one, and the other
lighter quarks. Thus it would be natural to assume that in this special
basis, up-Yukawa matrix is still {\em quasi}-diagonal, 
{\it i.e.}, diagonal up to small rotations.
Motivated by the CKM misalignment 
between up and down Yukawa matrices (from the left) and by the phenomenological constraints, 
we take these small rotations to be
CKM-{\em like}. 
In the same basis, 
down-Yukawa matrix is also diagonal up to a CKM-like unitary rotation matrix
which we denote by $\left( D_L \right)$
with $\left( D_L \right)_{13}\sim {\cal}
O(\lambda^3)$
[$\left( D_L \right)_{23}\sim{\cal} O(\lambda^2)$]. 
Clearly, there are new CP violating phases in $\left(D_L \right)$
(since it is not exactly the CKM matrix).
Note that we assume that, to
leading order, the 
interaction (\ref{NMFV}) does not distinguish between the first two generations 
(the couplings are either very small or equivalently approximately
degenerate) so
that the rotation in the 12 plane is unphysical\footnote{See 
later for effects of such small non-degeneracies.}.

Let us estimate now what is the size of the contribution to
various flavor changing neutral current (FCNC) processes.
We shall mainly focus below on constraints coming from
FCNC related to Kaons and B mesons since they
yields the most severe constraints. 
For
the same reason we first
focus on $\Delta F=2$ processes.
In the mass basis
the operators in Eq. (\ref{NMFV}) 
induce flavor violation where the most stringent
bounds are related to the down type sector. In the mass basis, Eq. 
(\ref{NMFV})
will be of the form
\beq
{\cal L}_{\rm NMFV}^{^{\rm mass}}= c_3 {\left( \bar Q_i
Q_j\right)^2\over \Lambda_{\rm NMFV}^2}\, \left(D_{3i}^* D_{3j}\right)^2\,,
\label{NMFVm}
\eeq
where $i,j=1,2,3$ are flavor indices.
This implies that the contributions to $\epsK$ are suppressed by
$\left(\lambda^5/\Lambda_{\rm NMFV}\right)^2$ and the ones related to
the B system (such as $\Delta m_d$, the CP asymmetry in $B\to\psi K_S$
and others)
are suppressed by $\left(\lambda^3/\Lambda_{\rm NMFV}\right)^2\,.$
Comparing this to the SM contributions one find that with
\begin{equation}
\Lambda_{\rm NMFV}\sim 2-3\,{\rm TeV}\,.\label{Lam}
\end{equation}
the NP contribution to all $\Delta F = 2$
are similar in size to the SM short-distance (top quark dominated) ones.
It is rather remarkable that $\Lambda_{\rm NMFV}$ is similar
to a scale either generated by one loop diagram with ${\cal O}(100)\,$GeV
mass intermediate particle as typically induced in supersymmetric
models or little Higgs models (with T parity) or a tree level exchange
of composite particles with ${\cal O}(3)\,$TeV masses.
In both of the above cases this mass scale is the scale required for
EWSB stabilization without being excluded by electroweak precision 
tests\footnote{Just like flavor violation,
contributions to EWPT from loops of particles
with few $100$ GeV masses (as in SUSY) and
and tree-level exchanges of
few TeV particles in models
with strong dynamics/composite Higgs are comparable.}.

Based on the success of SM unitarity triangle (UT) fit,
the lore is that the presence of such 
a NP effects, comparable in size to the SM ones, are ruled out.
However, we show that up to $O(30 \%)$ NP effects (relative to SM) are still allowed
by current data without any significant restriction on the the new CPV
phases.
Therefore, within the NMFV, the usual tension of having the flavor scale
coincide with the one of NP required for stabilizing the EWSB scale does
not exist!

We can also
consider NP effects in
$\Delta F= 1$ processes.
In the language of effective theory these will be induced by
the following Lagrangian (again assuming that flavor violation is in
the left handed sector)\footnote{In that case, due to the presence
  of strong phases, the exact form of the NP operators will modify the
results. This is discussed below in more detail.},
\beq
{\cal L}^{^{\Delta F=1}}_{\rm NMFV } = 
{( \bar Q_3 Q_3)\over \Lambda_{\rm NMFV}^2}\,(c_Q\bar Q_l Q_l+c_u \bar u_l u_l+c_d \bar d_l d_l  )\,,
\label{NMFV1}
\eeq
where $u,d$ stands for up and down quark singlets and $l=1,2$ stands for
flavor index and Lorentz and gauge indices were suppressed (each of
the terms in the above equation stands for all the possible operators allowed by
reshuffling these indices).
$\Delta F=1$ transitions are induced by ${\cal L}^{^{\Delta F=1}}_{\rm NMFV }$
once we move to the mass basis,
\beq
{\cal L}^{^{\Delta F=1,mass}}_{\rm NMFV } = 
{( \bar Q_i Q_j)\over \Lambda_{\rm NMFV}^2}\,(c_Q\bar Q_l Q_l+c_u \bar u_l u_l+c_d \bar d_l d_l  )\,
\left(D_{3i}^* D_{3j}\right) \,,
\label{NMFV1m}
\eeq
where subdominant corrections due to rotation of the U(2) invariant
part were neglected.
Note that given (\ref{Lam}) we expect the NP
contribution to $\Delta F=1$ processes to be roughly of the order of
the SM EW penguins.
Below we will discuss how the ``anomalies''
in the CP asymmetries in  $B \rightarrow ( \eta^{ \prime }, \phi,\pi )
K_s$ can be easily accommodated
in our framework.
Finally, we can consider correlation between
$\Delta F = 2$ and $\Delta F= 1$ processes.

The article is organized as follows: in the next section we describe in some detail
the experimental tests that are considered in what follows for the NMFV
framework.
We also summarize our main results.
The discussion related to the $\Delta F=2$ processes, given in section
\ref{sec:deltaf2}, is general, {\it i.e.}, without
any specific assumptions on the structure of the NP operators. It
applies to a a very broad class (even wider than the NMFV class)
of SM extensions. 

In section \ref{sec:deltaf1} we move to discuss $\Delta F=1$ processes.
In order to get meaningful constraints, further assumptions, beyond
the ones related to the NMFV, will be
required. Below we shall adopt one set of assumptions which cover a sub-class of
NMFV models and then we explain how our analysis can be extended to
include other models as well. We will basically assume, motivated by the experimental current
data, that helicity flipping and right handed operators are subdominant.

In section \ref{Correlations} we describe the possible correlation
between
$\Delta F=2$ and $\Delta F=1$ in our framework.
Such correlations are quite common (but not always present) in NMFV models.

In section \ref{RS1} we give some more formal description of our
framework and list several supersymmetric and
non-supersymmetric models which satisfy the definition of NMFV.
We finally demonstrate how
the Randall-Sundrum (RS1) framework belongs to
the specific subclass 
(with regard to $\Delta F = 1$ processes) 
analyzed below and discuss how it is being currently tested by
this data.
We conclude in section \ref{Con}.

Let us summarize our main messages for this work.
\begin{itemize}
\item[(i)] There is a wide class of models which flows to what we denoted 
as NMFV. In this class the usual 
tension associated with 
the scale required for EWSB stabilization 
being (roughly) the same as 
 the scale in which sources of flavor and CP violation are
 induced is largely ameliorated. Within the NMFV framework (which includes among others SUSY
  alignment~\cite{align}, non-abelian SUSY models~\cite{nonabelian}, 
Little Higgs models~\cite{LH},
  Composite Higgs~\cite{CH}, RS1 models~\cite{RS}, 
  Top-color models~\cite{topcolor, TCreview} and various 
hybrid models~\cite{hybrid}) the
scale of flavor violation $\Lambda_{\rm NMFV}\sim 4\pi M_W/g^2 \sim$
few TeV and the
  flavor violating contributions are quasi-aligned with the Yukawa matrices.
Since NP dominantly couples to the 3rd generation,
too large contributions to $K - \bar{K}$ mixing
from such low flavor scale are avoided. 
\item[(ii)] The mixing
of 1st and 2nd generation with 3rd generation 
still generates NP contributions
to $\Delta F=2$ processes
  with size similar to the SM loop effects. However,
unlike the common lore, the present data can accommodate  
  such NP contributions. This is in spite of the SM
  successful unitarity triangle fit.
\item[(iii)] It seems that, within the NMFV framework, in the 
near future the best constraints on
  the scale of the the new degrees of freedom required for EWSB
  stabilization will come not from electroweak precision tests but
  from flavor physics.
  \item[(iv)] We demonstrate how the data from the $B$ and $K$ system
    help to 
probe models with NMFV. In particular we
    consider: (a) $\Delta F=2$ processes. (b)  $\Delta F=1$
    processes. (c)  Correlation between $\Delta F=2$ and $\Delta F=1$ processes.
\end{itemize}

\section{NMFV, Overview and  experimental tests}
\label{Over}

With the data coming from the BELLE and BaBar experiments, the SM
flavor sector has entered into a new phase of precision tests.
Our main point in this work is to study
to what extent the data really point towards
the SM and how well can we use it in order
to really constrain physics beyond the SM 
in particular the NMFV framework.

In particular the question we have in mind is: what is the maximal size of the NP
contributions so that no conflict with present data is obtained?  This
is provided that, at any stage of our work, we allow for the presence
of arbitrary NP phases. Note that this is the situation expected
within the NMFV as demonstrated in Eq. (\ref{NMFVm}).
Within our framework the spectrum contains only three light
generations so that CKM unitarity is maintained.
Furthermore since flavor violation is
mediated at scale $\Lambda_{\rm NMFV}\sim 3\,{\rm TeV}$, NP effects
cannot compete with SM tree-level effects.
This actually covers a very wide class of models, even broader than
the NMFV (for more details see {\it
e.g.}~\cite{Frame}).

We expect NP contributions
to modify the predictions regarding observables that are related to
$\Delta F=2$ and $\Delta F=1$ processes.  
To be more explicit let us
consider $\Delta F=2$ processes first.  These enter the unitarity
triangle fit and includes
$\epsK,\Delta m_d,S_{\psi K},A_{\rm SL},\Delta m_s \,$.  
On the contrary
tree level observables which enter the fit such as measurements of
$V_{ub}/V_{cb}$ are unaffected by assumption.

It is instructive to consider the status of the new physics
contributions before and after the 2004 results.  We claim that our
understanding of the flavor structure of the quark sector was
dramatically improved during this time as follows.  During the last
year several new exciting measurements, such as the CP asymmetry in
$B\to DK$, $B\to D^*\pi$ etc, have entered into a precision phase.
These new observables, just like $V_{ub}/V_{cb}$, are mediated in the
SM by tree level processes and therefore insensitive to NP
contributions, thus providing a direct measurement (independent of 
$V_{ub}/V_{cb}$) of the CKM
elements.
 
We can parameterize our ignorance of the NP contributions 
to $\Delta F=2$
processes 
by a set of
six parameters $h_{d,s,K},\sigma_{d,s,K}$.  These just stand for the
magnitude and the phase of the NP contributions, normalized by the SM
amplitudes, in the $K-\bar K,\, B-\bar B$ and
$B_s-\bar B_s$ systems respectively~\cite{Para}.\footnote{Note that
  the above parameterization is more transparent than 
  $r_i-\theta_i$ one defined in Eq. (\ref{rdtd}) which is commonly used.
 This is due to the fact that, as
  discussed below, it is directly related to the amount of fine
  tuning implied by the various measurements. Thus $h_i\ll1$ implies
  that some cancellation between, dimensionless, unrelated parameters is required.}  This implies that
the predictions for the above observables is modified as 
follows\footnote{For a related discussion within the SM see~\cite{ENP}.}:  
\beq
\Delta m_d&=&\left|1+h_d e^{2i\sigma_d}\right| \Delta m_d^{\rm
  SM}\,,\qquad \spsi=\sin\left[2\beta+{\rm arg}\left(1+h_d
    e^{2i\sigma_d}\right)\right]\,, \nonumber\\
\Delta m_s&=&\left|1+h_s e^{2i\sigma_s}\right| \Delta m_s^{\rm
  SM}\,,\qquad \spsph=\sin\left[ 2\beta_s + {\rm arg}\left(1+h_s
    e^{2i\sigma_s}\right)\right]\,,\nonumber\\
A_{\rm SL}&=&Im\left[{\Gamma_{12}^d\over M_{12}^d\left(1+h_d e^{2i\sigma_d}\right)}\right]\,,
\label{par} 
\eeq
where in the above we
added $\spsph$ for completeness, $\beta_s\sim \lambda^2$ with
$\lambda\sim0.22$ is the Wolfenstein parameter and $M_{12}^d$ ($
\Gamma_{12}^d$) is the SM dispersive (absorptive) part of the
  $B^0-\bar B^0$ mixing amplitude~\cite{LLNP}.  The short distance corrections
to the $K^0-\bar K^0$ mixing amplitude, $M_{12}^{K}$, are given
by~\cite{epsKc}
\begin{equation}
  M_{12}^{\rm K} \propto \left[ {\lambda_t^*}^2 \eta_2 S_0\left(1+ h_K e^{2i\sigma_K}\right)+\dots \right],
  \label{M12SD}
\end{equation}
where $\lambda_t=V_{ts}^* V_{td}$, $S_0\simeq2.4$,
$\eta_2=0.57\pm0.01$~\cite{epsQCD} and the dots stands for
contributions involving the charm quark.  Given the above modification
to the SM amplitude, the constraint yielded by $\epsK$ is given
by~\cite{epsKc}
\begin{equation}\label{epsK}
  \eta \left\{(1-\rho)\left[1+h_K\left(\cos2\sigma_K+{1\over2}\sin2\sigma_K\left({1-\rho\over\eta}-
          {\eta\over1-\rho}\right)\right)\right] A^2 \eta_2 S_0
    + P_c(\varepsilon) \right\} A^2 \hat B_K = 0.187\,,
\end{equation}
where $P_c(\varepsilon)=0.29\pm0.07$~\cite{charm}, and $\hat
B_K=0.86\pm0.15$~\cite{CKMfitter}.

We stress that this $\Delta F = 2$ analysis is quite model-independent 
in the following sense. In general,
NP induces a set of new $\Delta F = 2$ operators. However, since
to a good approximation strong phases are not involved, 
the relative {\em magnitude} of the matrix elements of the
NP vs. SM operators can be simply absorbed into $h_{K, d, s}$. Then,
$\sigma_{ K, d, s}$ is the relative {\em weak} phase between NP and SM.

Let us now briefly discuss $\Delta F=1$ processes.
Even in the SM the structure of the effective weak Hamiltonian
which governs these processes is much richer than the one related to
$\Delta F=2$ processes and for NP effects,   
the weak phases can,
in general, be different in $\Delta F = 2$ and $\Delta F = 1$ transitions.
Thus in order to simplify the analysis and obtain non-trivial results 
we consider the $\Delta F = 1$ processes 
within a narrower class of NMFV models, which satisfies the following
{\em additional} assumptions.

\begin{itemize}
\item [(i)] NP induce only LH flavor-changing operators. This is 
plausible
in a wide class of models since a large $m_t$ can result in 
anomalous couplings of {\em left}-handed $b$;
Also the CP asymmetries in $b\to s$
transitions seem to prefer LH currents~\cite{Endo:2004dc} (see
however~\cite{LMP}). In addition the presence of chirality flipping
operator is highly constrained by measurements such as $b\to s\gamma$ and
the bounds on the strange electric dipole moments. 
In that sense we can view 
the class of models in which only LH operator are induced as truly NMFV.
\end{itemize}
This assumption has a very important implication. As evident from
Eqs. (\ref{NMFVm},\ref{NMFV1m}) observables
related to $\Delta F=2$ and $\Delta F=1$ transitions have the same weak phase and 
hence
are correlated.
As discussed below even this assumption is not constraining
enough to get meaningful results from present data. 
Thus in our analysis below we add the following assumption:
\begin{itemize}
\item [(ii)] NP in the $\Delta F=1$ processes is aligned with the SM Z-penguin operators,
  {\it i.e.},
only the non-photonic and non-box part of the electroweak
operators is modified. This is motivated by $Z'$ and RS1 models.
We also describe how this assumption could be relaxed
and how our results are still useful in other situations.
\end{itemize}

Following the analysis related to the $\Delta F=2$ processes we
analyze the $\Delta B=\Delta S=1$ ones such
as $B\to \phi K_s,\eta^{ \prime } K_s,K \pi$ and we will look for correlations
with the $\Delta F=2$ ones. Schematically, the amplitudes including the
NP contributions will be
parameterized as before by multiplying the SM contribution by the
factor $(1+h_s^1 e^{i \sigma_s})$.
In order to disentangle the NP short distance parameter a specific
hadronic
model must be used which implies that our results will suffer from
systematic uncertainties. We therefore choose to calculate each
transition using two hadronic models and compare the results as
discussed in more details below.
In general we expect that the
magnitude $h_s^1$ entering here will differ by an $O(1)$ factor from
$h_s$. Moreover, choosing $h_s^1$ to be positive, then the sign of
$h_s$ is physical and we have to scan over it.
Other interesting $\Delta F=1$ processes are the one which mediate
$K\to\pi \nu\bar\nu\,,$ for which NP is 
parameterized by multiplying the SM short distance contributions by
$(1+h_K^1 e^{i \sigma_K})$, while NP in the NMFV framework is
subdominant in the $b\rightarrow d$ system since
all the presently measured quantities are tree-level effects in the SM.   

One way to check whether only subdominant NP contributions are allowed
would be to estimate what are the allowed ranges for the above
parameters $h_{d,s,K},\sigma_{d,s,K}, h_s^1,h_K^1$ which are consistent with
the experimental data.  This is the first purpose of our analysis {\it
  i.e.}  to estimate what are the allowed range for $h_{d,s,K},h_s^1$
(independent of the value of the phases) before and after the summer
of 2005 (the Lepton-photon and EPS 2005 conference).
Even before going to the details of our analysis we want to state our
results.  These are the allowed regions before 2004: 
\begin{eqnarray}
&h_K& \lesssim 6\,\,\,\,{\rm and}\,\,\,\,
-\pi/3\leq 2\sigma_K \leq \pi/2 \,\,\, {\rm or} \,\,\,2\pi/3 \leq 2\sigma_K \leq
7\pi/6 \nonumber \\ 
&h_d& \lesssim 6\,\,\,\,{\rm and}\,\,\,\,
0\leq 2\sigma_d\leq\pi\,,\qquad 
h_s=0-\infty. \nonumber
\end{eqnarray}
The allowed regions after 2005: 
\begin{eqnarray}
&h_K&=0-0.5\,\,\,\,{\rm and}\,\,\,\,
0\leq2\sigma_K\leq 2 \pi\,,\nonumber \\
&h_d&=0-0.4\,\,\,\,{\rm and}\,\,\,\,
\pi\leq 2\sigma_d\leq 2 \pi\,,\qquad h_s=0-\infty, \nonumber
\end{eqnarray}
and $h_K^1,\sigma_K$ are basically unconstrained given the above range
for $h_K$.

Note that $\sigma$ between $(0, \pi)$ is the physical range
for $\Delta F = 2$ processes, 
whereas the corresponding physical range for $\Delta F = 1$ transitions
is $(0, 2 \pi)$.

\section{$\Delta F=2$ processes}\label{sec:deltaf2}

\subsection{Before 2004: coincidence issue?}

Let us now consider the experimental data before the 2004 summer
results.  The set of observables which enter the fit contains five
measurements, $\epsK,\Delta m_d,S_{\psi K},\Delta m_s$ and
$V_{ub}/V_{cb}\,,$ while the number of free parameters is eight: two
SM parameters $\rho,\eta $ and $6$ NP parameters
$h_{K,d,s},\sigma_{K,d.s}$!  We begin with a qualitative
discussion.

The best constraints are found when 
we consider the subset of
three observables in $B_d$ system, ($V_{ ub}$, $\Delta m_d$ and $S_{ \psi K }$) 
which depend on
only 
4 unknown parameters: $\rho$, $\eta$, $h_d$ and
$\sigma_d$. Even in this case
the system is under-constrained, {\it i.e.}, there is $1$ free parameter.
For example, an $O(1)$ or more value of $h_d$ is allowed (other parameters are
then fixed assuming the theory and experimental errors
are small: see later), {\it i.e.} NP not constrained.

\begin{figure}[htc]
\begin{center}
$\begin{array}{c@{\hspace{0.2in}}c}
\includegraphics[width=.5\textwidth]{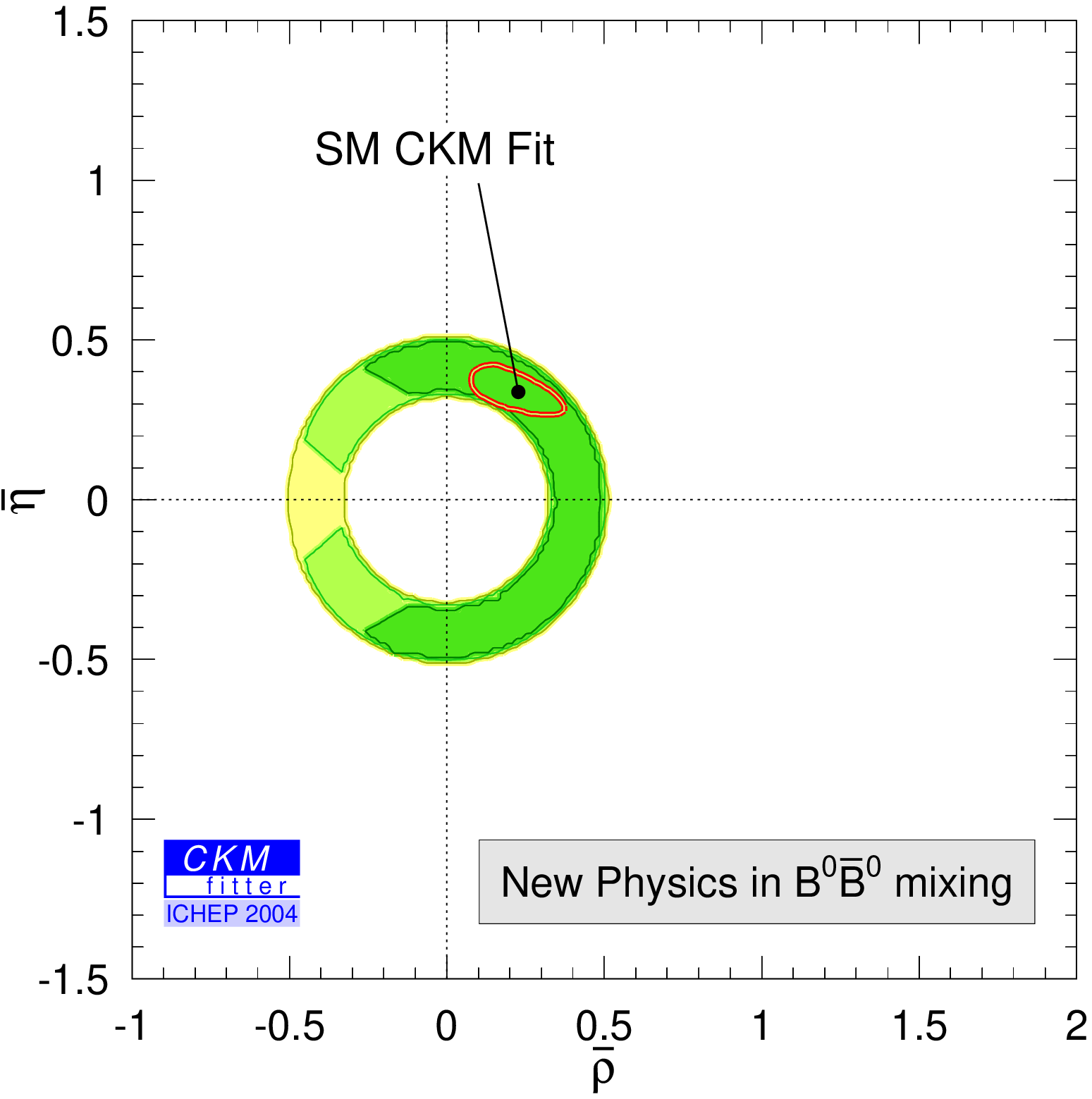} & 
\includegraphics[width=.5\textwidth]{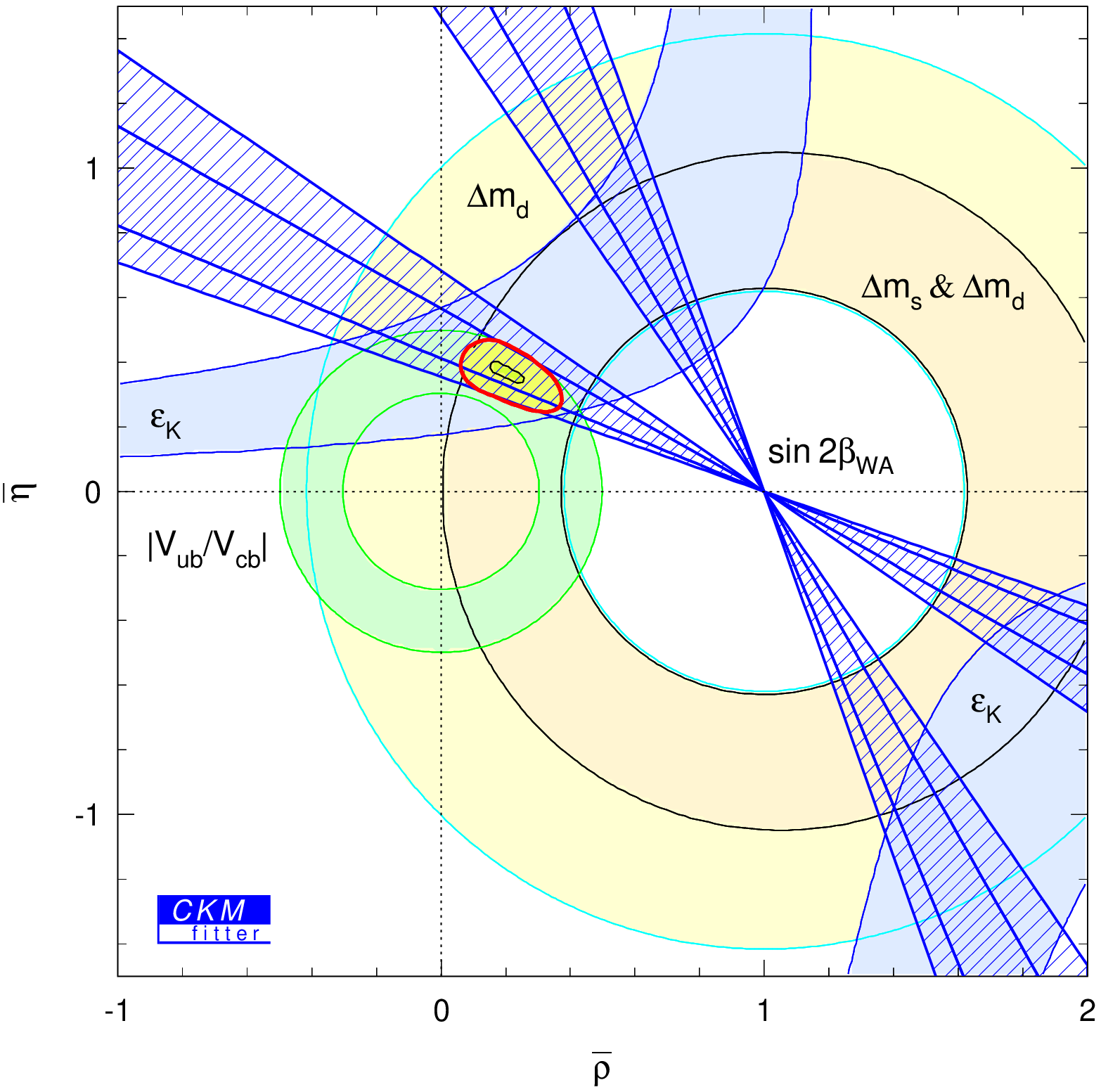} \\ 
\mbox{\bf (a)} & \mbox{\bf (b)}
\end{array}$
\end{center}
\caption{The constraint on the $\rho$ and $\eta$ 
Wolfenstein parameters before 2004. Left: allowing for NP in $\Delta F=2$~\cite{Ligeti04}.
Right: within the SM.}
\label{figrhoetaold}
\end{figure}

The crucial point is that only $V_{ ub }$ is independent of NP so that
only one combination of $\rho$, $\eta$ is fixed.  Thus it is not
surprising that the favored region in the $\rho-\eta$ plane covers the
whole annulus allowed by $V_{ub}/V_{cb}$ as shown in fig.
\mbox{\ref{figrhoetaold}(a)} ~\cite{Ligeti04}.

We can add a $4$th measurement to this analysis, namely, $\epsK$, which
depends on $\rho$, $\eta$ but, in general, this introduces $2$ more
(NP) parameters as well: $h_K$ and $\sigma_K$. So, the system is still
under-constrained and moreover it is clear that $h_K$ is also not
constrained.

Note that the $B_s$ system is (almost) decoupled since the SM
contribution to $B_s$ mixing does not depend\footnote{Recall 
that the only reason that $\Delta m_s$ is usually included in
  the unitarity triangle fit is that the ratio of the $\Delta B =
  \Delta S = 1$ and $\Delta B = 1, \Delta S =0$ hadronic matrix
  elements (bag parameters) is better known (due to flavor $SU(3)$
  symmetry) than the individual matrix elements.} on $\rho$, $\eta$.
The SM contribution depends on $V_{ ts }$ which is known fairly well
based on $V_{ cb }$ and unitarity.  However, both NP parameters ($h_s$
and $\sigma_s$) are not constrained since there is only one data,
namely $\Delta m_s$, presently bounded by a lower limit only.
 
This situation with NP is to be compared with the standard SM fit
shown in fig. \mbox{\ref{figrhoetaold}(b)}~\cite{CKMfitter}.  Even before summer
2004, this fit was already non-trivial since $2$ parameters fit $4$
data~\cite{Nirera} (including $\epsK$).  This implies that while NP
$\gtrsim$ SM is allowed, in such a scenario ({\it i.e.} with
$\rho$, $\eta$ laying somewhere else on $V_{ ub }$ annulus than where
they are in the SM fit), the good SM fit is an accident or a coincidence:
we will refer to this as the ``coincidence issue''. Said another way, although
there is no fine-tuning involved, what is discomforting is that given a 
size for NP (comparable to SM),  
the NP phase and $\left( \rho, \eta \right)$
have to conspire (or have to be orchestrated) 
with this NP size in order to fit the data, whereas the same data can be
fit without NP, i.e., in the SM 
(and with different $\left( \rho, \eta \right)$) (This issue was discussed in the context of RS1 in 
ref.~\cite{APS}).

\begin{table}[b!]
\begin{center}
\begin{tabular}{cc}
\hline
\hline
Parameter & Value \\
\hline
$\bar m_t(m_t)$ & $162.5 \pm 2.8$ GeV \\
$f_B \sqrt{B_d}$ & $223 \pm 50$ MeV \\
\hline
$|\varepsilon_K|$ & $(2.282 \pm 0.017) \times 10^{-3}$ \\
$\lambda$ & 0.22 \\
$|V_{ub}|$ & $(4.22 \pm 0.11 \pm 0.24)\times 10^{-3}$ \\
$|V_{cb}|$ & $(41.58 \pm 0.45 \pm 0.58)\times 10^{-3}$ \\
$\Delta m_d$ & $0.502 \pm 0.006$ ps$^{-1}$ \\
$S_{\psi K}$ & $0.687 \pm 0.032$ \\
$S_{\rho\rho,L}^{+-}$ & $-0.22 \pm 0.22$ \\
\hline
$A_{SL}$ & $-0.0026 \pm 0.0067$    \\
\hline
\hline
\end{tabular}
\end{center}
{\caption{Inputs used in the $\Delta F=2$ fits. The data is taken from
the August 1, 2005 update of Ref.~\cite{CKMfitter}, see 
http://www.slac.stanford.edu/xorg/ckmfitter/. \label{DF2}}}
\end{table}

\begin{figure}[htc]
\begin{center}
$\begin{array}{c@{\hspace{0.2in}}c}
\includegraphics[width=.5\textwidth]{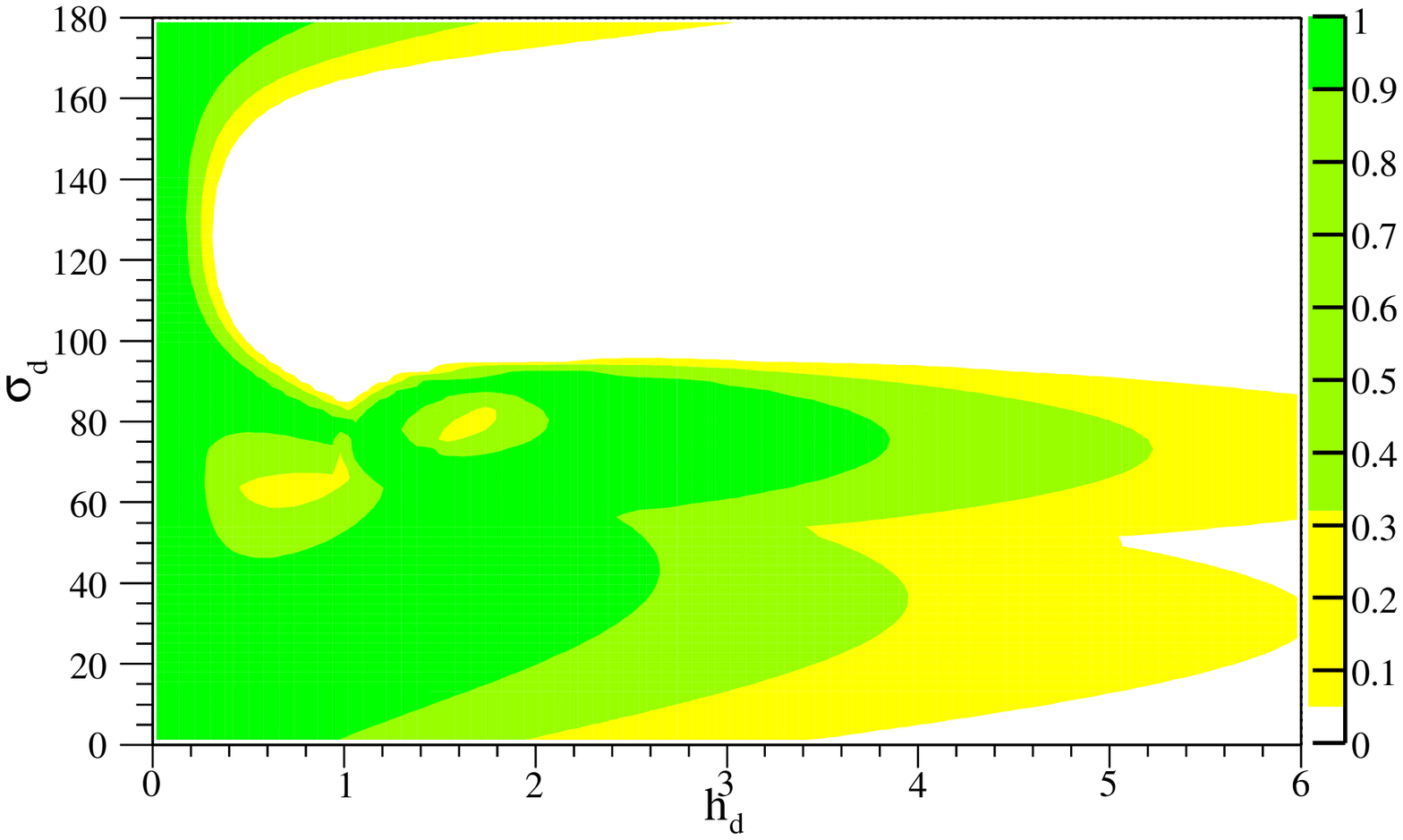} & 
\includegraphics[width=.5\textwidth]{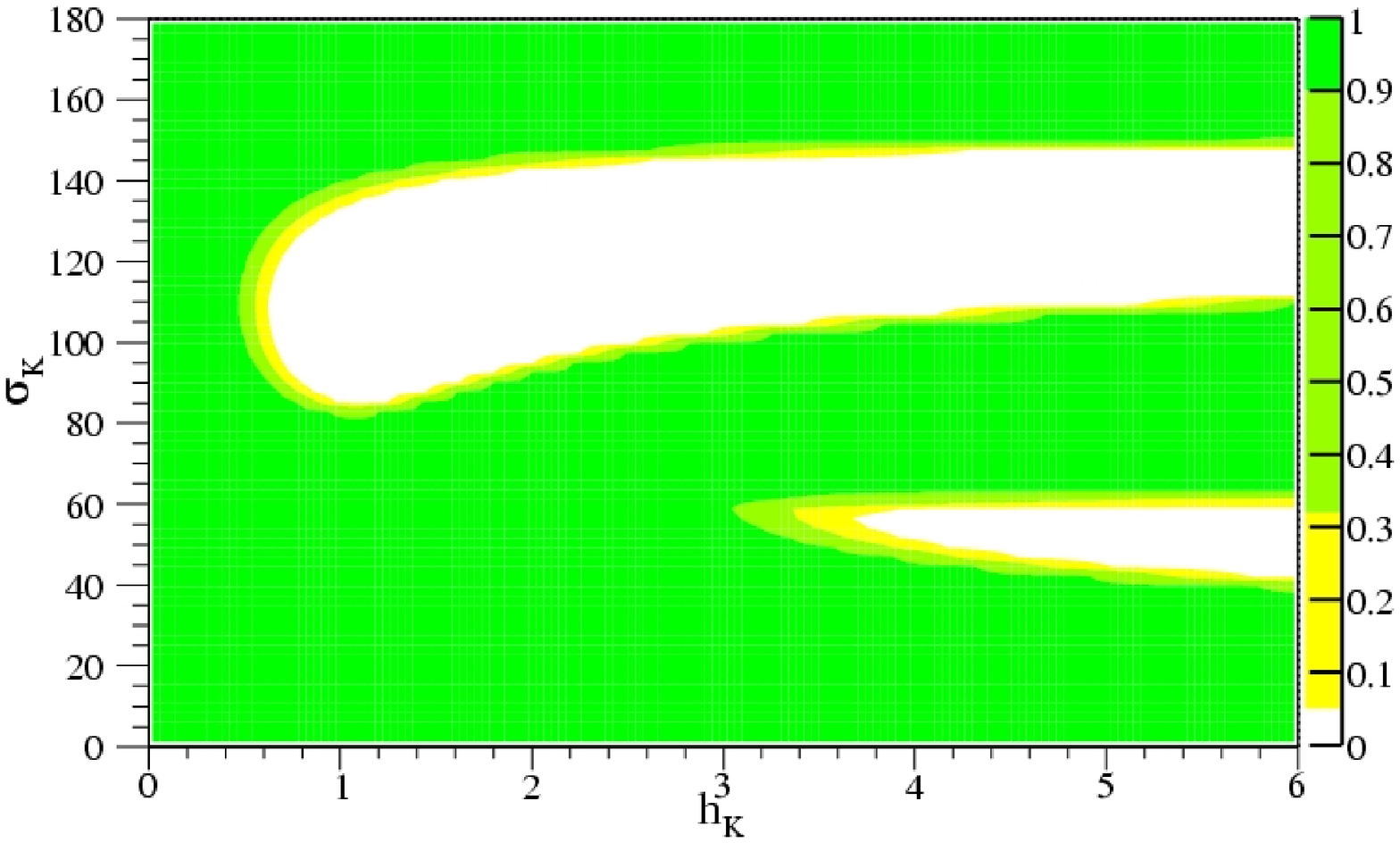} \\ 
\mbox{\bf (a)} & \mbox{\bf (b)}
\end{array}$
\end{center}
\caption{Left: The allowed range for $h_d$ and $\sigma_d$ before 2004
  from $V_{ub},\Delta m_d$ and $\spsi$.
Right: The allowed range for $h_K$ and $\sigma_K$ before 2004 from $V_{ub}$ and $\varepsilon_K$.}
\label{fighsigmaold}
\end{figure}

Next, we consider a more quantitative analysis.  We start our
discussion by considering the $B-\bar B$ system (here and below $B$
stands for $B_d$).  The required analysis, for this case, was
presented in~\cite{CKMfitter,Ligeti04} (see also~\cite{BBNR}). In that case the different
parameters $r_d ,\theta_d$ were used to constrain the NP
contributions, 
\beq 
r_d^2 e^{2i\theta_d} \equiv1+h_d e^{2i\sigma_d}\,,\label{rdtd}
\eeq 
The experimental data $\spsi^{\rm exp}$ and $\Delta m_d^{\rm  exp}$ 
yield the constraints
\beq 
\Delta m_d^{\rm SM} r_d^2=\Delta m_d^{\rm exp}\,,\qquad 
\sin(2\beta+2\theta_d)=\spsi^{\rm  exp}\,.  
\eeq 
It is remarkable that such an analysis, performed using data as before 2004, 
shows that the data only weakly constrained $h_d$ and $2\sigma_d$ to be in the
range: 
\beq 
h_d=0-6\,\ \ \ 2\sigma_d=0-\pi\,.  
\eeq 
This is
demonstrated in Fig. \mbox{\ref{fighsigmaold}(a)} where we plot 
the $h_d-\sigma_d$ plan at various CL. Here and in the following we
produced the plot using the CKMfitter code~\cite{CKMfitter}, suitably 
modified to accomodate our NP scenario.
Note that a {\em large} range of the NP phase $\sigma_d$ 
(roughly half the physical range) is allowed
for a given size of NP. This happens even in the case where NP
$\gtrsim$ SM, contrary
to the expectation from counting of parameters and data, since in the
above case there should be a correlation between $h_d$ and $\sigma_d$
(there is only one free parameter). 
This is due to the fairly large theory errors
in $\Delta m_d$ and $|V_{ ub }|$ (experimental errors are all small in comparison).
Thus there was at most a {\em mild} coincidence issue.

Improving the theory errors would have 
sharpened the coincidence issue, {\it i.e.},
large $h_d$ would have been allowed only for a {\em smaller} 
range of $\sigma_d$.
However, a sizable NP amplitude (up to $h_d\sim 3$) could not be ruled out
even if the theory errors are small since, with a $\sigma_d$ of a specific value,
the data could be always be fitted.
The point is that this bound on the NP size
is not dictated dominantly by errors, but rather its corresponding
strength is related to the counting of number of
parameters vs. data. Clearly, more independent observables were needed!

We move now to discuss the constraint from $\epsK\,$.  Since $\epsK$
is subject to a sizable hadronic uncertainty, the resulting constraint
is not very stringent.  This is demonstrated in
fig.~\mbox{\ref{figrhoetaold}(b)}~\cite{Lap} where the precise measurement of
$\epsK$ correspond to the light-blue band in the $\rho-\eta$ plane.
In order to find the allowed region for $h_K$ before the 2004 summer
results we use the relation (\ref{epsK}) and repeat the CKM fit
together with $h_K$
and $\sigma_K$. This is equivalent to asking what are the values of
$h_K$ and $\sigma_K$ for which the $\epsK$ hyperbola still
overlaps with the yellow
annulus of fig.~\mbox{\ref{figrhoetaold}(a)}, given by the $|V_{ub}/V_{cb}|$
constraint.  The resulting range is given by 
\beq h_K=0-0.6\,\ \ \
\sigma_K=0-\pi\,,  
\eeq 
although even for $h_K \gtrsim 1$
a large range of $\sigma_K$ (in fact, most of the physical
range except for near $\pi / 2$) is allowed.
This is demonstrated in fig.~\mbox{\ref{fighsigmaold}(b)}
which shows the region allowed by the measurements of $\epsK$ and
$V_{ub}/V_{cb}$ in the $h_K-\sigma_K$ plane.

We now briefly explain some of the features of this constraint
which will be useful in comparing in next section
with the status after the summer of 2004.
The cases $\sigma_K \sim 0, \pi /2$ are very 
simple to understand since the asymptotes of
the hyperbola of Eq. (\ref{epsK}) remain the same and only the 
intercept on the $\eta$ axis changes with $h_K$.
Consider first the case $\sigma_K \sim 0$. It is easy to
see that the $\eta$-intercept decreases as $h_K$ is increased,
becoming zero for $h_K \rightarrow \infty$, {\it i.e.}, it remains non-negative so that
the hyperbola always 
intersects with $V_{ub}$ circle, resulting in no constraint on $h_K$.
Next, consider $\sigma_K \sim \pi / 2$. In this case the $\eta$-intercept 
initially increases with $h_K$ 
(approaching
$+ \infty$ for $h_K \rightarrow 1^-$) so that the hyperbola goes outside of
the $V_{ ub }$ circle for some value of $h_K$ close to, but smaller than $1$. 
Whereas, for $h_K > 1$, the
hyperbola flips about the $\rho$-axis, {\it i.e.}, the $\eta$-intercept 
becomes negative (it approaches $-\infty$ for $h_K \rightarrow 1^+$).
The magnitude of the intercept decreases as $h_K$ increases 
(approaching zero  
as $h_K \rightarrow \infty$) so that 
eventually the hyperbola intersects 
$V_{ub}$ circle again for large $h_K$.
Thus, we find that only
a range of $h_K \sim 1$ (where 
the hyperbola does not intersect $V_{ub}$ circle at all) is excluded. 
For other values of $\sigma_K$, the asymptotes rotate precluding a simple analysis
as above.

\begin{figure}[htc]
\begin{center}
$\begin{array}{c@{\hspace{0.2in}}c}
\includegraphics[width=.5\textwidth]{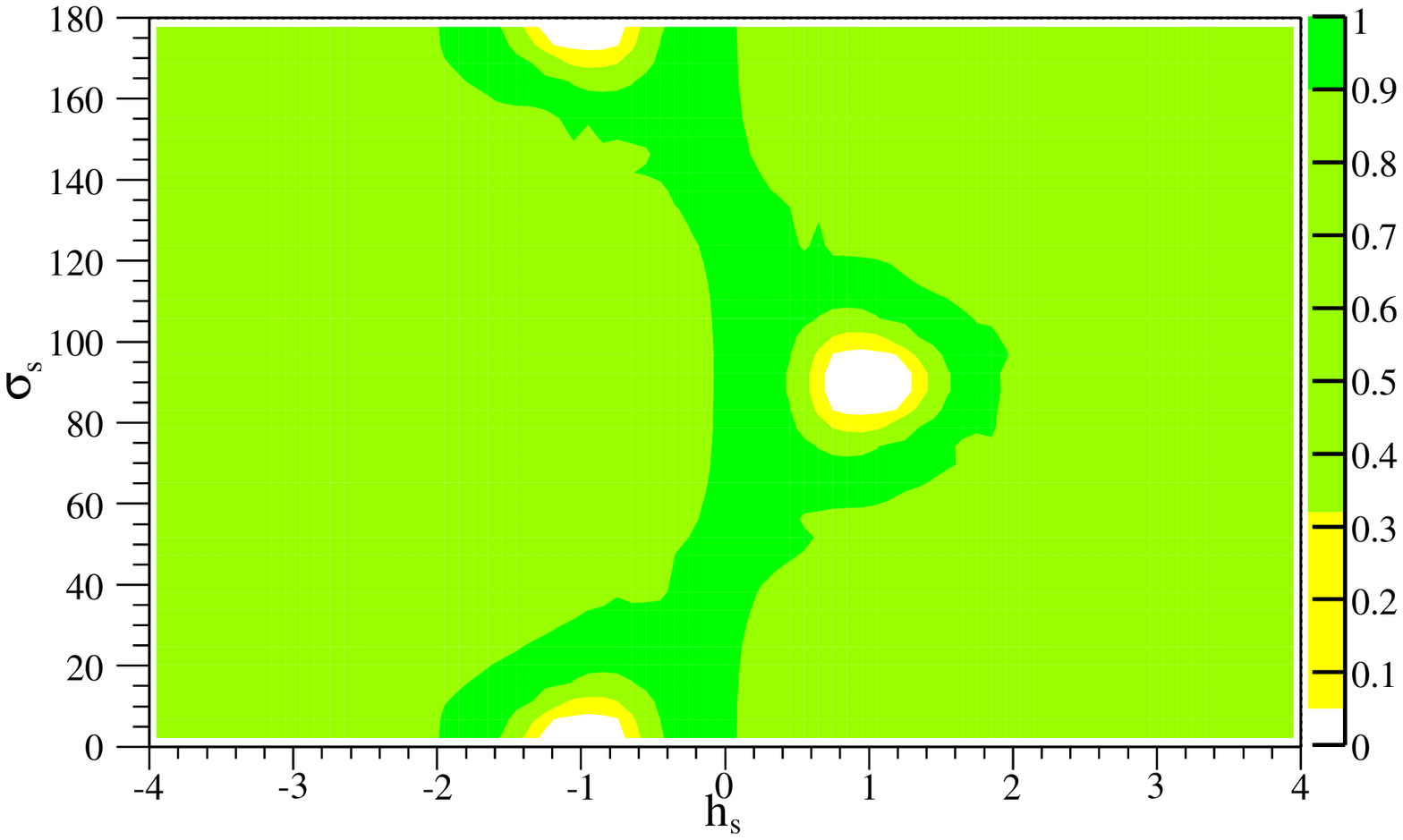} & 
\includegraphics[width=.5\textwidth]{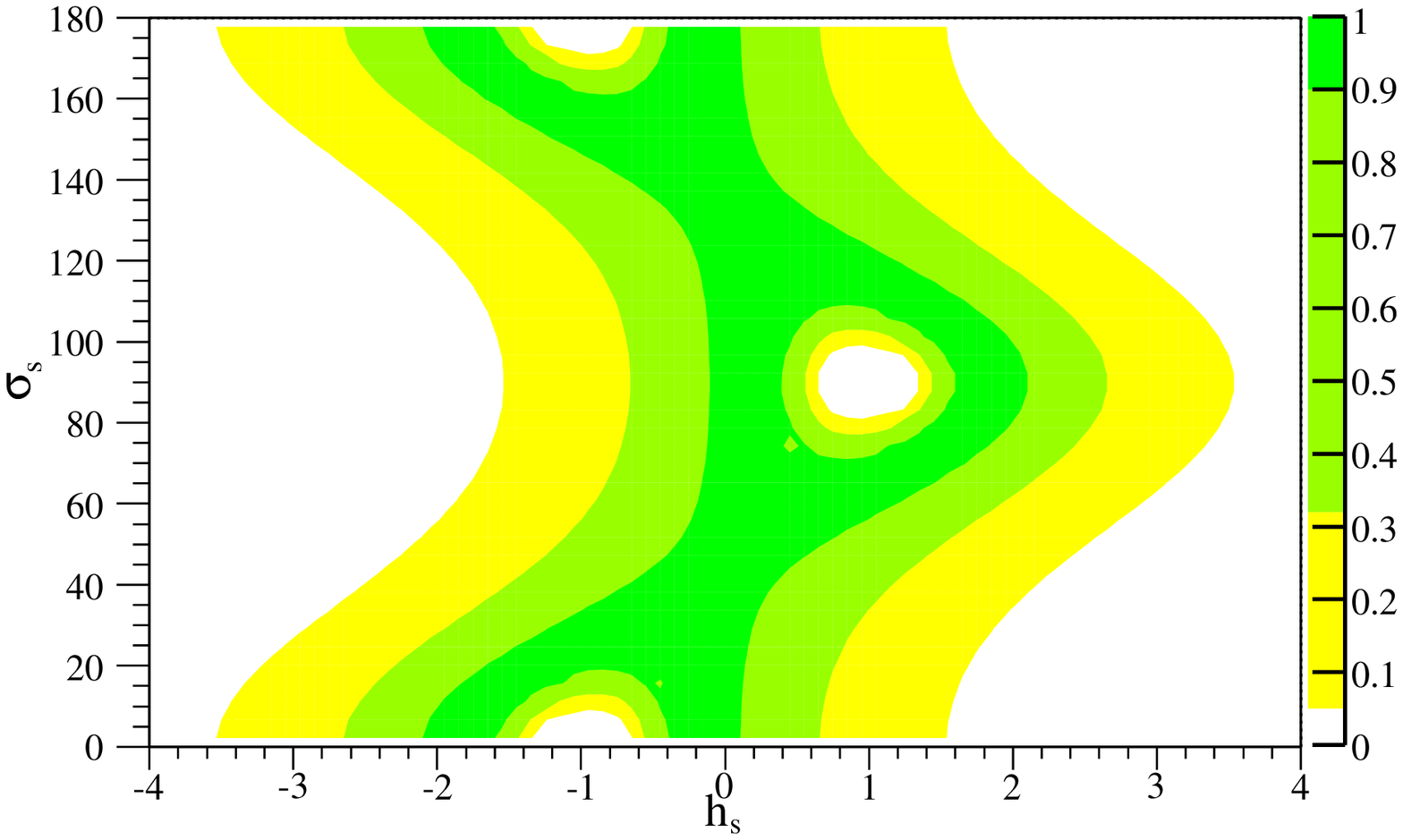} \\ 
\mbox{\bf (a)} & \mbox{\bf (b)}
\end{array}$
\end{center}
\caption{Left: The allowed range for $h_s-\sigma_s$ using the data on
  $\Delta m_s$. Right: the future projection for a measured $\Delta
  m_s=(18.3\pm 0.3) ps^{-1}$.}
\label{fighsss}
\end{figure}

The experimental lower value on the $B_s-\bar B_s$ mass difference,
$\Delta m_s^{\rm exp}$ yields the following constraint on
$h_s-\sigma_s$~\cite{hfag} 
\beq 
\left|1+h_s e^{2i\sigma_s}\right|
\Delta m_s^{\rm SM}\geq \Delta m_s^{\rm exp}\,.  
\eeq 
At present this
does not constrain the allowed range for $h_s$ 
\beq
h_s=0-\infty\,.\label{hsdms} 
\eeq 
The experimental lower bound on
$\Delta m_s$ yield exclusion regions in the $h_s-\sigma_s$ plane as
demonstrated in fig.~\mbox{\ref{fighsss}(a)}.\footnote{Since in our framework
  $\Delta m_d$ is affected by different NP which is only weakly constrained,
  employing the ratio $\Delta m_d/\Delta m_s$ to reduce the
  theoretical uncertainties on the hadronic matrix elements is of
  little use.}
It is easy to see that
$h_s \sim -1$, $\sigma_s \sim 0,\pi$ and $h_s \sim + 1$, $\sigma_s \sim
\pm\pi/2$ are excluded due to destructive interference between NP and SM
leading to $\Delta m_s$ below limit.\footnote{Here both signs of
  $h_s$ are kept since later we shall correlate the above result with
  the $b\to s$ transition.  In that case both transitions are governed by
  the same weak phase. Generically however the corresponding NP
  amplitudes could be of opposite sign so that both signs of $h_s$ are
  physical~\cite{LMP}.}

To conclude this discussion, the remarkable lesson we learnt from the
above analysis is that up to a year ago the NP contributions to
$\Delta F=2$ processes in models belonging to the class previously
defined could have been comparable in size or even dominate over the
SM ones.


\subsection{After 2005: coincidence issue $\rightarrow$ fine-tuning
  problem?}

The 2004 (and beyond) $B$ factory measurements dramatically changed the situation.
An obvious point from the above discussion is that a measurement of a
$4$th observable in $B_d$ system which depends on a different
combination of the same $4$ parameters leads to a determination of all
the $4$ parameters. NP is now constrained!  Precisely this happened in
the summer of 2004.  Specifically, the experimental 
results of 2004 (and
later) have provided us with
a direct probe of the SM flavor parameters ($\rho$ and $\eta$) through
new data related to SM tree level dominated processes (and hence
independent of NP) as follows.

We shall not try here to describe  all the relevant processes but list some
of them (for more details see {\it e.g}~\cite{CKMfitter} and refs.
therein).  The CP asymmetry in $B \rightarrow DK$, $\ADK$, which is governed by a
SM tree level transition and therefore unaffected by NP is: 
\beq
\ADK\sim \tan\gamma = \frac{ \eta }{ \rho } \,.  
\eeq 
The point is that $\ADK$ depends only on $\rho$, $\eta$ in a 
combination different than $V_{ ub }$.
The CP asymmetry in $B\to \rho\rho$, $\srr$, is given by\footnote{
  Note that in order to cleanly extract the SM parameters an isospin
  analysis is required. This is carried under the assumption that
  contributions from electroweak penguins (or any other ones with
  non-trivial isospin structure) can be safely neglected.  This is
  justified in the SM, but not model independently, i.e., not
in the presence of generic NP.
In most of the known NMFV examples however the above assumption holds
up to subdominant correction (see {\it e.g.}~\cite{APS}).  }
\beq \srr\propto\sin\left( 2 \gamma + 2 \beta + 2 \theta_d \right)\,.
\eeq
Thus, $S_{ \rho \rho }$ also depends only on $\rho$, $\eta$ {\em
  after} subtracting the phase of $B_d$ mixing (including the NP
phase) using $S^{ exp }_{ \psi K_s }$.

In short, we have a $2$nd direct measurement of $\rho$, $\eta$: $V_{
  ub }$ and $\ADK$ or $S_{ \rho \rho }$ are thus enough to fix
$\rho$, $\eta$ even in presence of NP!

\begin{figure}[htc]
\begin{center}
$\begin{array}{c@{\hspace{0.2in}}c}
\includegraphics[width=.5\textwidth]{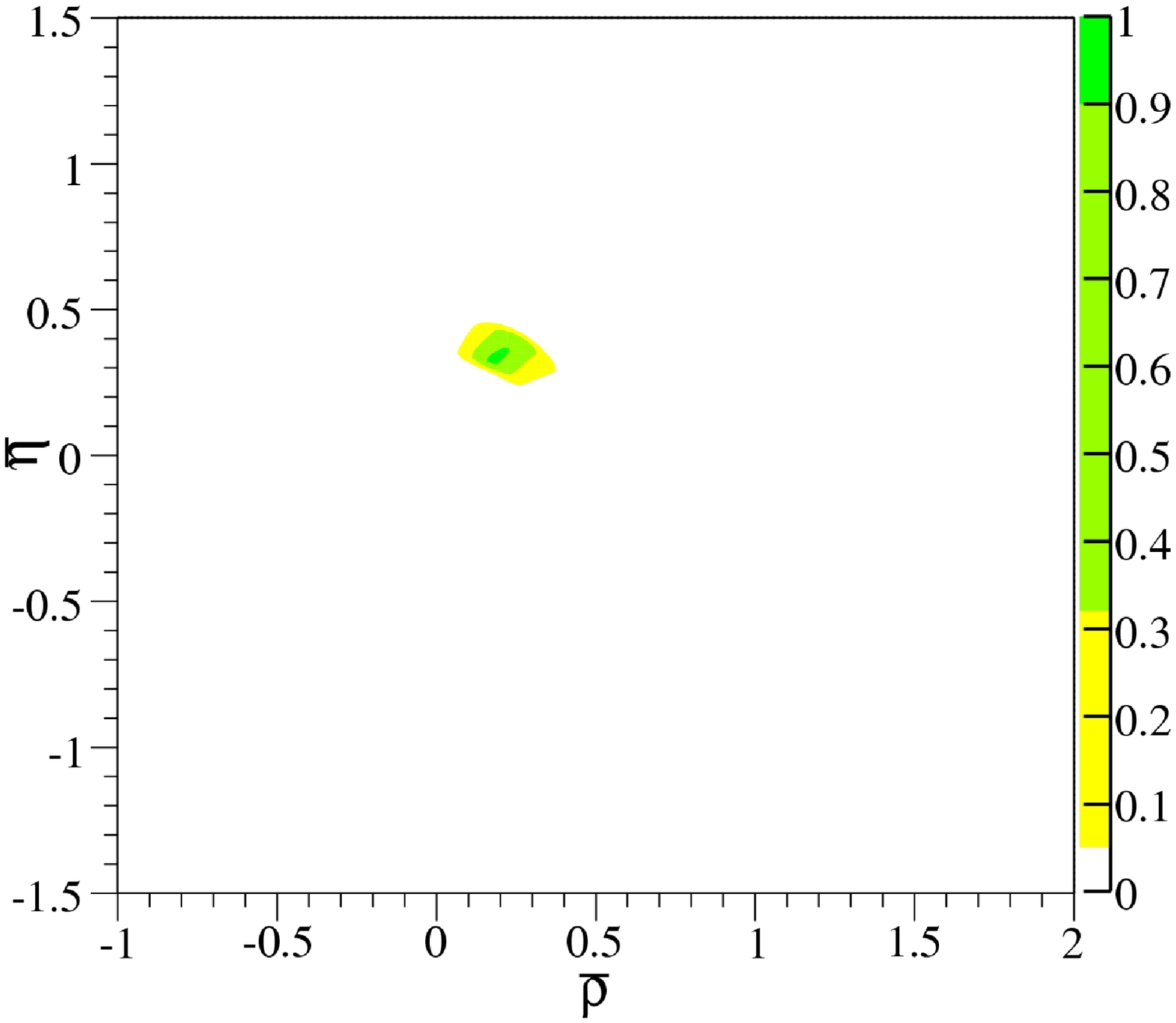} & 
\includegraphics[width=.5\textwidth]{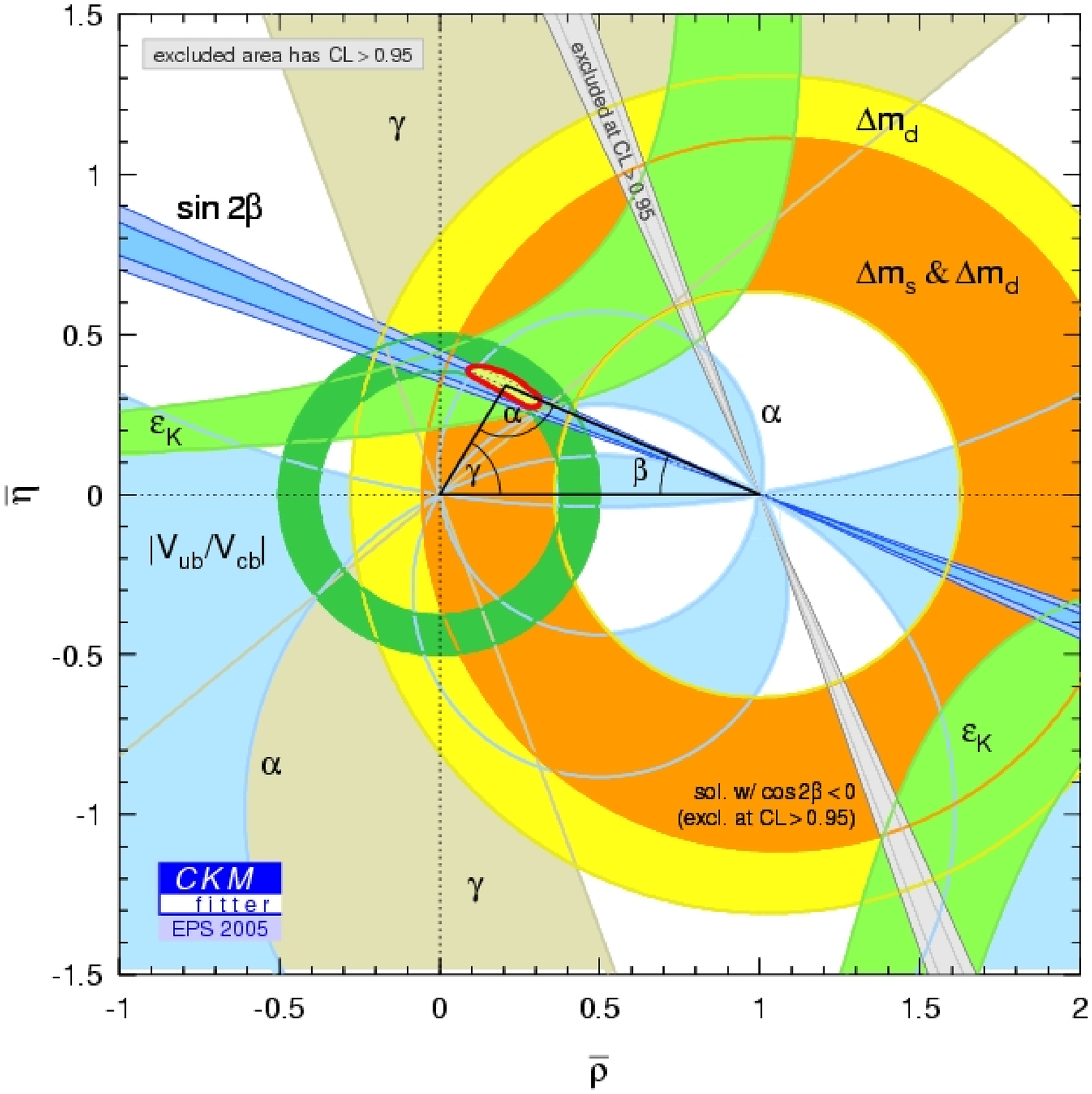} \\ 
\mbox{\bf (a)} & \mbox{\bf (b)}
\end{array}$
\end{center}
\caption{The constraint on the $\rho$ and $\eta$ 
Wolfenstein parameters after summer 2005. Left: allowing for NP in $\Delta F=2$~\cite{Ligeti04}.
Right: within the SM.}
\label{figrhoeta}
\end{figure}

The resulting values of $\rho$, $\eta$ are consistent with the SM
expectation, {\it i.e.} with $\rho$, $\eta$ coming from the SM fit {\em
  before} summer 2004 (see Fig. \mbox{\ref{figrhoetaold}(b)})~\cite{Ligeti04}.  This is
demonstrated in fig. \mbox{\ref{figrhoeta}(a)} which show that
with present data, even in presence of NP, the favored region in
the $\rho-\eta$ is now around the SM preferred region. The input
parameters used in this fit are listed in Table~\ref{DF2}.

This implies that the only solution for the $4$ parameters is one with
``small'' NP. To be explicit, the SM contribution to $B_d$ mixing
(which depends on $\rho$, $\eta$) is now known (even with NP present)
and it accounts for observed mixing (both $\Delta m_d$ and $S^{ exp
}_{ \psi K_s }$) so that there is not much room for NP.

Thus, the allowed NP went from $\gtrsim$ SM to $<$ SM -- in
fact, the ``naive'' expectation based on $\rho$, $\eta$ plane
(comparing Figs. \mbox{\ref{figrhoetaold}(a)} and \mbox{\ref{figrhoeta}(a)}) is that
NP $\ll$ SM! Consequently there is a potential
fine-tuning {\em problem} for NMFV models where the
{\em generic} expectation is that NP $\sim$ SM. It is no more just a
coincidence issue!

In short, the above measurements yield a
dramatic improvement in constraining the enlarged parameter space
relevant for the $\Delta F=2$ processes in presence of NP.

This dramatic effect on NP should be contrasted to the not-so-dramatic
effect on SM fit: an already non-trivial fit ($2$ parameters and $4$
data) became more so ($5$ data).  Moreover, the uncertainties in the
above new measurements are still rather large: for the SM the
corresponding improvement on the standard fit is not very significant
({\it i.e.}, errors or size of the region in $\rho$-$\eta$ plane didn't reduce much)
as one can see by examining fig.
\mbox{\ref{figrhoeta}(b)}~\cite{Ligeti04,CKMfitter} and comparing it with the
results from 2002 in fig. \mbox{\ref{figrhoetaold}(b)}.

\begin{figure}[htc]
\begin{center}
$\begin{array}{c@{\hspace{0.2in}}c}
\includegraphics[width=.5\textwidth]{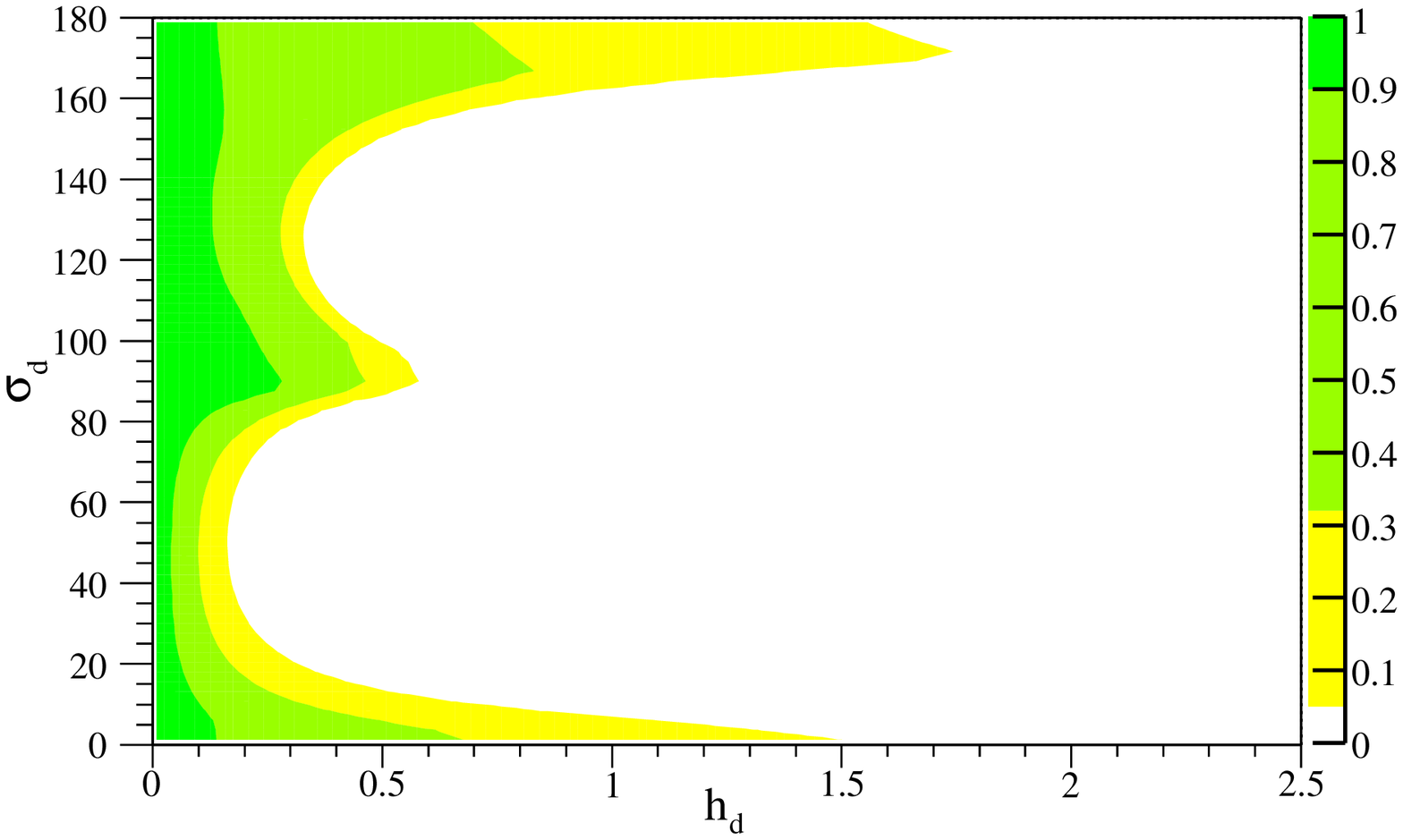} & 
\includegraphics[width=.5\textwidth]{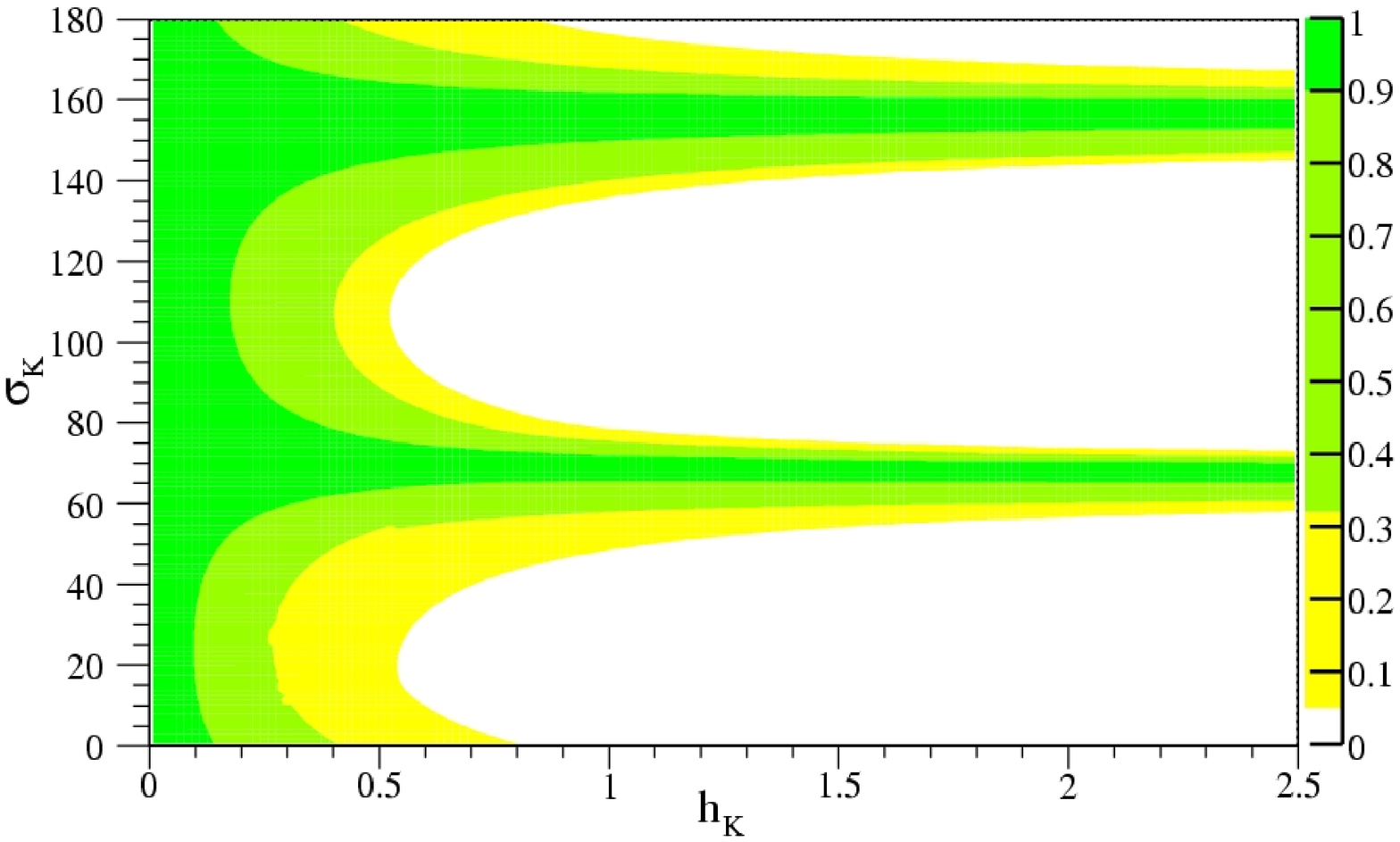} \\ 
\mbox{\bf (a)} & \mbox{\bf (b)}
\end{array}$
\end{center}
\caption{Left: The allowed range for $h_d$ and $\sigma_d$ after summer
  2005.
Right: The same for $h_K$ and $\sigma_K$.}
\label{fighsigma}
\end{figure}

We now quantitatively discuss the allowed size of NP after these new
results.
We start our discussion by considering the $B-\bar B$ system.  The
required analysis, for this case, was presented
in~\cite{CKMfitter,Ligeti04}.  We find the following range for $h_d$
\beq 
h_d=0-0.4\,\ {\rm for} \ \ 2\sigma_d=\pi-2\pi\,.  
\eeq 
This is demonstrated in fig. \mbox{\ref{fighsigma}(a)}
where we show the $h_d-\sigma_d$ allowed regions allowed
by the combined recent measurements.

\begin{figure}[tcbh]
\centering \includegraphics[width=.5\textwidth]{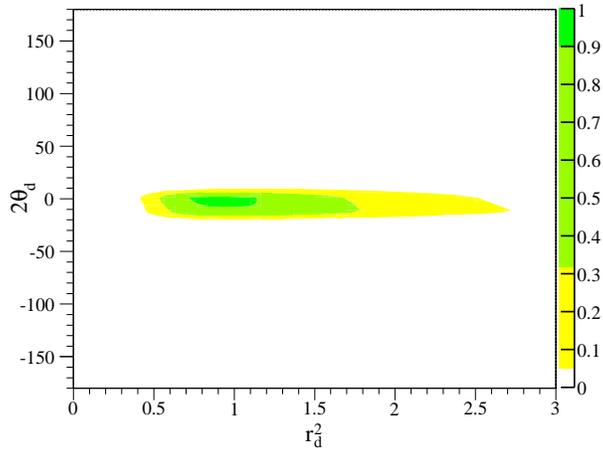}
\caption{The allowed range for $r_d$ and $\theta_d$ after 2005.} 
\label{figrdtd}
\end{figure}

Thus, in detail, the naive expectation mentioned above (based on
comparing Figs \mbox{\ref{figrhoetaold}(a)} and \mbox{\ref{figrhoeta}(a)}) is not
borne out, {\it i.e.}, surprisingly, NP $\sim 30-40 \%$ of SM (with almost any phase) is 
still allowed!
In view of above discussion, it is clear
that such a large size of NP being allowed is due to the (as yet) sizable errors
in 
the new data. The fact that
any phase being allowed for such a sizable NP contribution is due to theory errors in
data before summer, {\it i.e.}, $\Delta m_d$ and $V_{ ub }$ (since, just as before,
in the absence of these errors, $\sigma_d$ is fixed for given $h_d$).

At this point, it is worth comparing the two parameterizations for NP:
the $\left( r_d^2, 2 \theta_d \right)$ plot in Fig~\ref{figrdtd} shows
that\footnote{For more details, see {\it e.g.}~\cite{CKMfitter}. The
  differences with the plot shown here are due to the inclusion of
  $A_{SL}$ in the fit.} 
$| 2 \theta_d | \lesssim 20^{\circ}$. However, this  
does not directly imply that NP phase ($\sigma_d$) is
small since $2 \theta_d$ is
combination of $h_d$ and $\sigma_d$
(in particular, it is clear that for small $h_d$ large $\sigma_d$
still gives small $2 \theta_d$). In this sense,
$\left( h_d, \sigma_d \right)$ is a more transparent parametrization of NP than
$\left( r_d^2, 2 \theta_d \right)$.

Note that $h_d$ up to $1.5$ is allowed, but only 
for
$\sigma_d \sim 0$ 
since such NP 
only affects $\Delta m_d$ which has large theory errors. However, this implies
aligning of NP phase with SM and hence a fine-tuning.

Clearly, unlike before summer,
smaller experimental error (in $4^{th}$ data) will 
restrict the allowed $h_d$ (with {\em arbitrary} $\sigma_d$) to be less than
$\sim 30 \%$ and thus will sharpen the fine-tuning
problem for models where the generic expectation is
NP $\sim$ SM. Also, even if error in $4^{ th }$ data is not
improved, reducing theory errors in data
before summer ($\Delta m_d$ and $V_{ ub }$) is welcome since it will
restrict the range of 
$\sigma_d$ allowed for, say, $h_d \sim 30 \%$ (in addition to ruling out
$h_d \sim 1$ (with $\sigma_d \sim 0$)). Thus, the 
allowed $h_d$ (with arbitrary $\sigma_d$) will be smaller, again
leading to fine-tuning problem for NP models or in other words push
$\Lambda_{\rm NMFV}$ to a higher scale.

There is another potential fine-tuning problem related to the
Kaon system.
There is no new data in $K$-system, but $\rho$, $\eta$ and hence SM
contribution to $\epsK$ (even in presence of NP) is now known and
accounts for observed $\epsK$. This implies that $h_K$ cannot be
large: again, the naive expectation is that $h_K \ll 1$.
Let us discuss the constraint from $\epsK$ quantitatively. In
order to find the allowed region for $h_K$ after the 2004 summer
results we use the relation (\ref{epsK}) and scan over values of $h_K$
and $\sigma_K$ so that the resulting value is still within the
combined constraints in yellow shown in
fig.~\mbox{\ref{figrhoeta}(a)}
The resulting range is given by 
\beq
h_K=0-0.6\,\ \ \ \sigma_K=0-\pi\,,  
\eeq 
although (as was the case 
before summer of 2004) even for $h_K \sim 1$ a large range of $\sigma_K$
(roughly half the physical range)
is allowed (unlike for $h_d$, $\sigma_d$). This is demonstrated in
fig.~\mbox{\ref{fighsigma}(b)} which shows a plot of the region allowed by the
combined recent measurements.

The fact that the $2\sigma$ excluded region has not changed too much
before and after the summer can be easily understood from eq.
\ref{epsK}. 
For $\sigma_K \sim 0$, as explained before, the $\eta$-intercept decreases
(approaches zero) 
with increasing $h_K$ so that the hyperbola
does not intersect the $V_{ ub }$ circle in the allowed SM region at $1 \sigma$, although,
due to large theory errors, there is still an overlap with SM
region with a lower CL ($2 \sigma$). This explains the difference between
Figs. \mbox{\ref{fighsigmaold}(b)} and \mbox{\ref{fighsigma}(b)} for $\sigma_K \sim 0$.
Consider next $\sigma_K \sim \pi / 2$. For $h_K \gtrsim 1$, the
$\eta$-intercept is negative so that there is never an 
overlap with allowed (SM) region of $V_{ ub }$ circle\footnote{Although
the hyperbola does intersect the $
V_{ ub }$ circle for large $h_K$.}. Thus, $\sigma_K
\sim \pi /2$ is excluded not only for $h_K \sim 1$
(for which the hyperbola does not intersect the $V_{ ub }$ circle {\em
at all}), but also for
larger $h_K$ (cf. before summer 2004 where
$h_K > 1$ was allowed).

Thus, just as for $h_d$, the naive expectation for $h_K$ is not
realized. In fact, $h_K \sim 0.6$ is still allowed -- this is due to
large theory errors in hadronic matrix elements.
Thus it implies that
only a mild increase in the scale $\Lambda_{\rm NMFV}$ or a mild
suppression in the mixing angles (relative to the CKM angles) is required to
fit the data.

Also, we checked that the  correlation between $\left( h_d, \sigma_d \right)$
and $\left( h_K, \sigma_K \right)$ is weak at present due to the large errors.

During the last year no significant improvement in measurement related
to the $B_s-\bar B_s$ system were 
obtained (and the SM contribution is independent
of $\rho$, $\eta$ so that their direct measurement last year
does not affect
this analysis) . Consequently the
constraint on $h_s-\sigma_s$ is unmodified and is described by
fig.~\mbox{\ref{fighsss}(a)} 
\beq 
h_s=0-\infty\,.\label{hsdms1} 
\eeq
The same figure shows also how the constraints on $h_s$ and $\sigma_s$
will change when $\Delta m_S$ will be measured. The particular plot
in fig.~\mbox{\ref{fighsss}(b)} is for $\Delta m_s=(18.3\pm 0.3) ps^{-1}$.

We conclude this part with the following statement: even though the
data significantly improved in the last two years (in the sense
that new observables were measured), NP in $\Delta F=2$ amplitudes can still
fit the data with arbitrary phases and size comparable to the SM
contributions.
This implies that only recently we have started to constrain NMFV
models with $\Lambda_{\rm NMFV}=2-3\,$TeV.



\section{$\Delta F=1$ transitions}
\label{sec:deltaf1}

So far in the above we discussed how NP is constrained
using data only from $\Delta F=2$ processes. However,
as demonstrated in Eq. (\ref{NMFV1})  NP contributes not only to
$\Delta F=2$ processes but also
mediates $\Delta F=1$ transitions
like $b\to s$ or $s\to d$.
We shall analyze the implications of this scenario in the
context of recent measurements of the 
CP asymmetries 
in
$K\to \pi\bar\nu\nu$ (currently only an upper bound), 
$B\to\phi K_S,\eta^{ \prime } 
K_S$ and also $B \to K\pi$. 
As we will see, when dealing with the latter,
two main difficulties are found: (i) due to the presence of strong
phases, a computation of the hadronic
matrix elements is required, which suffers from theoretical
uncertainties.
In order to get an estimate of the size of the uncertainties
involved we shall obtain our results using two different hadronic
models.
For the $B\to K\pi$ transitions we shall use QCD
factorization~\cite{BBNS,Beneke:2003zv,BPRS}
and an SU(3) analysis (demonstrated here in complete form for the first time)
whereas for $B\to \phi,\eta' K$ case we shall use naive and QCD factorization.
(ii) We shall see that, generically, 
in the presence of strong phases the number of unknown UV
parameters
is too large and therefore does not allow us 
to obtain bounds on the weak phases.
Thus we are forced to make more assumptions which hold in a
narrower class of NMFV models.


\subsection{Model dependence of the $\Delta F=1$ processes}
\label{subsec:deltaf1}

We shall try to
analyze our framework in as model-independent fashion as possible. 
However, a completely model independent
analysis will be proven to be a non-trivial task given our present
experimental and theoretical knowledge.
Let us focus on the effective theory below the EWSB scale (or $M_W$) that
is obtained in our framework, in particular in the context
of $\Delta F=1$ processes. As we pointed out before, these have a much richer
structure than the $\Delta F=2$ ones and in order to
be able to obtain nontrivial results we therefore
must 
restrict
the definition of our framework. 
Let us now discuss in more detail the set of assumptions that 
define it. We consider these to be well motivated --
for example,
we will show in the next section 
that these assumptions hold in RS1 and various $Z'$ models and therefore provides us
with a way to test this framework. In addition we shall discuss how
one can use our results to constraint other subclasses of the NMFV framework.
\begin{itemize}
\item[(i)] NP induce only LH flavor-changing operators.
  This implies that our effective theory contains only
  operators which already exist in the SM one {\it i.e.} $O_{1-10}$
  (see {\it e.g.}~\cite{BBL} for definition of the standard operator basis).
\item[(ii)] The operators in the effective theory
  are obtained from integrating out color singlet particles~(see {\it e.g.}~\cite{Zp} 
and references therein for related discussions).
\end{itemize}
As already discussed in section~\ref{Over} the first assumption
implies that there is a correlation between the observables related to
$\Delta F=2$ and $\Delta F=1$ transitions. Note that this also clearly
implies that the $\Delta F=1$ processes are governed by a single
weak phase per transition.
Furthermore the assumption (ii) implies
that at the EWSB scale only the odd operators (since the color indices are
trivially contracted)\footnote{This is valid for
all the models in which the 
dominant effects come from  tree-level 
exchange of color singlet particles, since gluon
loops can generate non-trivial color structure even if singlet particles
are integrated out.}, 
$O_{3,5,7,9}$, are being
induced in the effective weak Hamiltonian.
Consequently the amplitude for each transition is characterized
by five UV parameters per transition, {\it i.e.} $C_{3,5,7,9}^{K,d,s}$ and a weak
phase, $\sigma_{K,d,s}$.
However, as we show next, even this reduction in the number of UV parameters
is not enough to enable us to
obtain a non-trivial constraint. 

For the $s\to d$ transition processes,
the only ones which are theoretically clean are the neutral and charge 
$K\to \pi \nu \bar{\nu}$
decays, which are related by isospin~\cite{GNR}. Thus we
basically have only a single measurement with too large an uncertainty
at this stage.
The $b\to d$ case would seem to be better, since
the B factories has provide us with more precise data on $B\to
\rho\rho,\rho\pi,\pi\pi$\footnote{There are some results on $B\to KK$ and
  other similar 
  decays but the corresponding data is presently much less precise and 
thus we shall not consider these processes
here.}.
Extracting the UV parameters from the data in a theoretically clean
way requires $SU(2)$ isospin analysis.
However, since the above processes are dominated by SM tree level
amplitudes, the resulting constraints on the above parameters are still
rather weak at present.
The $b\to s$ case is indeed better: there is data 
on the CP asymmetries in $B\to \phi,\eta^{ \prime} K_S$ 
decays. Furthermore, the data on various $B\to K\pi$
processes has reached a precision level. And, more importantly 
(unlike the former cases), these decays are 
penguin dominated.
Thus, with some
limited theoretical input and the use of $SU(2)$ isospin,
these decays can provide
non-trivial constraints on our UV parameters.
Below we show that without using a specific hadronic model for the matrix
elements, the data from $S_{\eta^{ \prime },
\phi K_S}$ and the $B\to K\pi$
system provides us with three non-trivial data point. Using a specific
hadronic model (like Naive factorization or QCD factorization)
increases the number of data points at our disposal, but introduces
additional theoretical uncertainties.
This is still not enough in order to constrain five fundamental
parameters, {\it i.e.}, $C_{3,5,7,9}$ and the weak phase $\sigma_s$.
The best we can do at this stage is to constrain frameworks with only
two NP Wilson coefficients and a single weak phase. 
However, instead of 
considering this most general possibility, we take another route
below, adding the following assumption.
\begin{itemize}
\item [ (iii) ]
We choose  a framework where NP affects only a single free NP Wilson coefficient
(which is a linear combination of $C_{3,5,7,9}^{s}$ discussed above), which we
find well motivated. Specifically,
we consider the class of models in which the NP $\Delta F=1$ operators
are aligned with the SM Z penguins.
\end{itemize}

This covers various 
$Z^{ \prime }$ models~(see~\cite{Zprecent} for recent related discussions), 
various little Higgs models and the RS1 framework.
Note that, given (iii), the fact that we have more observables than input parameters
implies that we can see whether these models are consistent with the
present data and if they pass their first non-trivial test.
The values of $C_{3,5,7,9}^s$ discussed above 
(at the $M_W$ scale) are
\begin{equation}
  \label{eq:2}
  C_{3,5,7,9}^s (m_W) \equiv C_{3,5,7,9}^Z (m_W) h_s^1 e^{i \sigma_s}
\end{equation}
in such a way that when $h_s^1=1$ and $\sigma_s=0$ they coincide with
the SM Z-penguin contribution.

We are led thus to consider the following extension of the weak
$\Delta B=1$ Hamiltonian, including terms introduced by NP:
\begin{eqnarray}\label{Heff}
{\cal H}_W^{SM} = \lambda_u^{(f)} [C_1 O_1^u + C_2 O_2^u]
  + \lambda_c^{(f)} [C_1 O_1^c + C_2 O_2^c] - \lambda_t^{(f)}
  \sum_{i=3}^{10} [C_i^{SM}(\mu) + h_s^1 e^{i \sigma_s} C_i^{(Z)}(\mu)] O_i 
\nonumber \\
\end{eqnarray}
with $f=d,s$. The operators $O_i$ are defined as in Ref.~\cite{BBL} and include
the tree operators $O_{1,2}^{u,c}$, the QCD penguin operators $O_{3-6}$, and the
electroweak penguin operators $O_{7-10}$.

The Wilson coefficients $C^{SM}_{3-10}(\mu)$ and $C^{(Z)}_{3-10}(\mu)$
are found using the standard method.
First, one integrates out the $W$ and top quark at the $\mu =M_W$ scale, 
followed by running down to $\mu \sim m_b$. At the matching scale, the Wilson
coefficients have an expansion in $\alpha_s(M_W)$ of the form
\begin{eqnarray}\label{Wil}
  \vec C(M_W) \equiv \vec C^{SM} + \vec C^{(Z)}&=& \vec C_s^{(0)}(M_W) +
  \frac{\alpha_s(M_W)}{4\pi} 
\vec C_s^{(1)}(M_W) +
\frac{\alpha_{em}}{4\pi} [\vec C_{\rm ew}^{\gamma+b}(M_W) \nonumber\\
&&+
  \left(1+h_s^1 e^{i \sigma_s}\right) \vec C_{\rm ew}^Z(M_W)]\,.
  \label{NPEWP}
\end{eqnarray}
In the matching condition we
distinguish between the contributions from the photon penguin and box
graphs $C^{\gamma+b}$ and from the $Z$ penguins $C^Z$. 
The Wilson coefficients at a low scale $\mu \sim m_b$ are obtained by
running down with the $10\times 10$ anomalous dimension matrix of the
operators $O_{1-10}$.

We quote below the values of the Wilson coefficients in Eq.~(\ref{Heff})
at a low scale $\mu=4.25$ GeV. We used the following
parameters in obtaining these results 
$\Lambda = 225 $ MeV, $m_t=170$ GeV, $\sin^2\theta_W=0.231$.
\begin{eqnarray}\label{WCi}
\begin{array}{c|cc|cccc|cccc}
 & C_1 & C_2 & C_3 & C_4 & C_5 & C_6 & C_7 & C_8 & C_9 & C_{10} \\
\hline
\vec C^{(SM)} & 1.086 & -0.192 & 0.014 & -0.036 & 0.01 & -0.043 & -0.0005 & 0.0004 & -0.009 & 0.0017 \\
\hline
10^3\times \vec C^{(Z)} & 0 & 0 & 1.80 & -0.64 & 0.050 & -0.298 & 
1.182 & 0.446 & -4.539 & 1.040 \\
\hline
\end{array}
\nonumber \\
\end{eqnarray}
In the evaluation of the electroweak penguin coefficients we used the
electromagnetic coupling $\alpha = 1/129$.

A similar approach has been followed in section \ref{intro} to include
the NP effects in the $\Delta S=2$, $\Delta B = 2$ effective
Hamiltonian.  The weak phase appearing there, $2 \sigma_s$, is related
to the one in the $\Delta S = \Delta B = 1$ effective Hamiltonian
($\sigma_s$).  However, the magnitude 
of the NP in mixing will be in
general different $h_s \neq h_s^1$.

\subsection{Analysis of $\Delta F=1$ transitions in $K$ system}
\label{subsec:deltaf1anal}

We first briefly discuss
the $K\to\pi \nu \bar \nu$ decay process which involves the
$s\to d$ transition.
In our framework the NP contributions are governed by $\sigma_K$, {\it i.e.}, 
the same phase which controls the NP in $\varepsilon_K\,.$
The measurement of the charged mode has large errors so that the 
resulting constraint on $\sigma_K$ is rather weak. For the neutral mode
$K_L\to\pi\nu\bar\nu$ there is presently only an upper bound so that the
resulting constraint is much weaker. 
However, we still show the relation between these BR's and the NP model 
parameters for future reference since the situation can dramatically change
once more data is collected.
For the charged mode we can relate the branching ratio to our NP
parameters as follows
\beq \label{bkpn}
{\rm BR}(K^+ \to \pi^+ \nu \bar\nu)/\kappa_+ \propto
\left\{ X^2 \left(\eta ^2+(\rho -1)^2\right) \left((h_k^1)^2+2 \cos
   \left(\sigma _K\right) h_k^1+1\right) A^4+\right. \nonumber\\
\left.  2 X \left(\eta h_k^1 
    \sin \left(\sigma _K\right)-(\rho -1) \left(h_k^1 \cos
   \left(\sigma _K\right)+1\right)\right) P_0 A^2+P_0^2 \right\} \,,
\eeq
(see {\it e.g.}~\cite{kpn} and refs. therein) where
$X\sim 1.5$ and $P_0\sim0.4$~\cite{BBL}. A similar relation is
obtained for the neutral decay mode.
In the future, measuring the above processes will 
directly constrain $\sigma_K$, allowing for a comparison with the
$\varepsilon_K$ result.

We now move to discuss in more details the constraints related to
$b\to s$ transitions.


\subsection{$B\to\phi K_S,\eta' K_S$ transitions}

An important source of information about the new physics in $b\to s$
transitions comes from time-dependent CP violation in $B$ decays into
CP eigenstates. 
The relevant parameters for the $B^0(t)\to f$ transition are 
defined in terms of the ratio of amplitudes
\begin{eqnarray}
\lambda_f = e^{-i\phi_M} \frac{\bar A_f}{A_f}
\end{eqnarray}
where $\phi_M$ is the $B^0 -\bar B^0$ mixing phase, and $A_f, \bar A_f$ are the
$B^0\to f$ and $\bar B^0\to f$ decay amplitudes, respectively. 
The time-dependent CP asymmetry parameters $S_f$ and $C_f$ are given in terms of
these parameters by
\begin{eqnarray}
S_f = \frac{2\mbox{Im } \lambda_f}{1+|\lambda_f |^2}\,,\qquad
C_f = \frac{1-|\lambda_f|^2}{1+|\lambda_f |^2}
\end{eqnarray}

Assuming that the amplitudes $A_f, \bar A_f$ are dominated by
one single weak phase, the $S_f$ parameter gives a direct measurement
of the interference of this phase with the $B^0 -\bar B^0$ mixing amplitude.
One such case is the tree mediated decay $B\to \psi K_S$, which measures
the $B^0-\bar B^0$ mixing phase. 
Allowing for NP
in the $B^0 -\bar B^0$ mixing amplitude, this relation is modified in our
framework as mentioned in Sec.~III
\begin{eqnarray}
S_{\psi K_S} = \sin (2\beta + 2\theta_d)
\end{eqnarray}
with $1 + h_d e^{i\sigma_d} = r_d^2 e^{2i\theta_d}$. The experimental
result for $S_{\psi K_S}$ is shown in Table 1.

Next we consider the penguin-mediated decays $b\to s\bar qq$, which have in 
general a more complicated structure. Allowing for NP as described by the effective
Hamiltonian Eq.~(\ref{Heff}),
the general form of the decay amplitude can be written as
\begin{eqnarray}\label{b2samp}
A_f = \lambda_c^{(s)} P_f (1 + e^{i\gamma} d_f e^{i\theta_f} +
h_s^1 e^{-i\sigma_s} q_f^c e^{i\phi_f^c} (1 + R_b \lambda^2 e^{i\gamma}))
\end{eqnarray}
where $R_b=\sqrt{{\bar \rho}^2+ {\bar \eta}^2}$. The different terms in this formula 
arise from the operators in the
weak Hamiltonian Eq.~(\ref{Heff}) as follows: $P_f$ is contributed
by the matrix elements of $O_{1,2}^c$ and $O_{3-10}$ with the SM
Wilson coefficients; the $d_f e^{i\theta_f}$ arises from the
operators $O_{1,2}^u$; finally, the terms proportional to $q_f e^{i\phi_f^c}$
are contributed by the operators $O_{3-10}$ with the Wilson coefficients
$C_i^Z(\mu)$. The small correction $\sim \lambda^2 R_b \sim O(1\%)$ 
is introduced by the fact that NP appears in Eq.~(\ref{Heff}) multiplying 
the CKM coefficient $\lambda_t^{(s)}$.

In the absence of the $d_f$ term (usually called the `SM contamination'), 
and assuming the validity of the SM, the amplitude Eq.~(\ref{b2samp}) gives a CP
asymmetry parameter $S_f = -\eta_{CP} \sin (2\beta + 2\theta_d)$, where
$\eta_{CP}$ is the CP eigenstate of the final state. In particular,
$S_f$ measured in these decays must be directly related to the
corresponding parameter in $B\to \psi K_S$. Any significant deviation from
zero of the difference $-\eta_{CP} S_f - S_{\psi K_S}$ can 
be therefore interpreted as new physics~\cite{London:1997zk,Grossman:1996ke}.

We show in Table \ref{table1} the current
measured values of the CP violating parameters $S_f$ and $C_f$ for several 
decays mediated by the $b\to s$ penguin. The individual results
for $S_f$ from BABAR and BELLE are listed, together with their world
average. For reference, we give also the corresponding result for
$B\to J/\psi K_S$, to which they are equal in the SM in the limit of the 
decay amplitude being dominated by the penguin. The results display a 
deviation from the naive SM expectation  $-\eta_{CP} S_f = S_{\psi K_S}$ 
of roughly $2\sigma$. The agreement
improved after the most recent results reported at the Lepton-Photon 
2005 conference~\cite{LP05}.

A nonzero difference $-\eta_{CP} S_f - S_{\psi K_S}$ can be
introduced from terms in the decay amplitude with a weak phase different from 
that of the penguin. Such terms are present in the SM, where they 
originate from the matrix elements of $O_{1,2}^u$, which are multiplied with the 
CKM coefficient $\lambda_u^{(s)} = e^{-i\gamma} |\lambda_u^{(s)}|$
(parameterized by  $\sim d_f e^{i\phi_f}$ in Eq.~(\ref{b2samp})). 
Since such contributions are CKM- and loop-suppressed, they are
expected to be small. 

The SM contamination in the differences $\Delta_f =
-\eta_{CP} S_f - S_{\psi K_S}$  has been studied using several methods.
In Refs.~\cite{Grossman:2003qp,Gronau:2003kx}, these effects have been
bound using SU(3) flavor symmetry, and inputs from the measured branching
fractions of $b\to d$ modes. This method allows also the study of 
correlations with data on $C_f$ and has been recently extended also to
3-body modes~\cite{3body}. Further applications of these bounds, with
additional dynamical input, have been presented in Ref.~\cite{boundsplus}.

A different approach makes use of QCD factorization in the heavy
quark limit, to compute the matrix elements of the operators in the
weak Hamiltonian.
Such computations were performed in~\cite{Beneke:2003zv,Beneke:2005pu}
and found to give small positive results for the differences 
$-\eta_{CP} S_f - S_{\psi K_S}$ (see Ref.~\cite{Beneke:2005pu} for a recent update).
On the other hand, the observed sign of this difference in the experimental 
data is predominantly negative, although with significant errors. Therefore it is
natural to attempt an explanation of the data in terms of new physics contributions
to the weak effective Hamiltonian.

\begin{table}\label{table1}
  \begin{center}
    \begin{tabular}{|c|ccc|c|}
\hline
 $f$ &  \multicolumn{3}{c}{$-\eta_{CP}S_f$} & $C_f$ \\
\cline{2-5}
 & BABAR & BELLE & WA & WA \\
\hline
\hline
$\phi K_S$ & $0.50 \pm 0.25 ^{+0.07}_{-0.04}$ & $0.44 \pm 0.27 \pm 0.05$ & $0.47\pm 0.19$ & $-0.09 \pm 0.14$\\
$\eta' K_S$ & $0.36 \pm 0.13 \pm 0.03$ & $0.62 \pm 0.12 \pm 0.04$ & $0.50\pm 0.09$ & $-0.07 \pm 0.07$ \\
$\pi^0 K_S$ & $0.35 ^{+0.30}_{-0.33}\pm 0.04$ & $0.22 \pm 0.47 \pm 0.08$ & $0.31\pm 0.26$ & $-0.02 \pm 0.13$ \\
\hline
$J/\psi K_S$ &  &  &  $0.687\pm 0.032$ & \\
\hline
\end{tabular}
\end{center}
{\caption{Experimental results \cite{BaBe} for the CP asymmetry parameters in neutral $B$ decay into
CP eigenstates mediated by the $b\to s \bar qq$ transition. For reference we
show also the corresponding parameter measured in $B\to \psi K_S$ decays.}}
\end{table}

In the remainder of this section, we will study the implications of these
data for the NP framework considered here, in particular the constraints
on the parameters $(h_s^1, \sigma_s)$. Some aspects of our analysis have
been partially considered in previous work, so a brief review is in order.

An important question concerns the assumptions built into the NMFV, in particular
assumption (i) of the NP inducing only LH flavor-changing operators.
Due to the different parity of the final state in $\sphi,\seta$ the
resulting asymmetries are sensitive to the chiral structure of the NP
contributions~\cite{Kagan}.  In~\cite{Endo:2004dc}, Endo, Mishima, and
Yamaguchi demonstrated that a possible difference between the mixing-induced
CP-asymmetries in $B\to \phi K_S$ and $B\to \eta' K_S$
can be easily explained by NP-induced operators with LH
chiralities for a wide range of weak phases and
amplitudes.\footnote{This holds as long as the relative sizes of the
  corresponding matrix elements between the two final states are close
  to each other and the NP contributions are subdominant.
  The status of RH NP in view of these measurements is
  discussed in~\cite{LMP}.}  This observation validates the assumption (i) of the
NMFV approach (see Sec. IV.A).

Another important information yielded by
the recent measurements is that, with left handed NP
operators~\cite{LMP}, the data can be accounted even if the NP
contributions are subdominant.
Thus it is interesting to examine whether the data can be explained by
$O(1)$ modification of the SM electroweak operators (as in
our framework). Such a scenario
occurs in many models with NP mediated by a $Z'$ gauge
boson~\cite{Zp}. 
Our main motivation however to discuss this is that 
it was recently shown (before the summer results) in~\cite{APS}
that this is exactly the case in RS1 models with low KK masses and
bulk custodial isospin.

\begin{figure}[!htp]
\begin{center}
$\begin{array}{c@{\hspace{0.2in}}c}
\includegraphics[width=.5\textwidth]{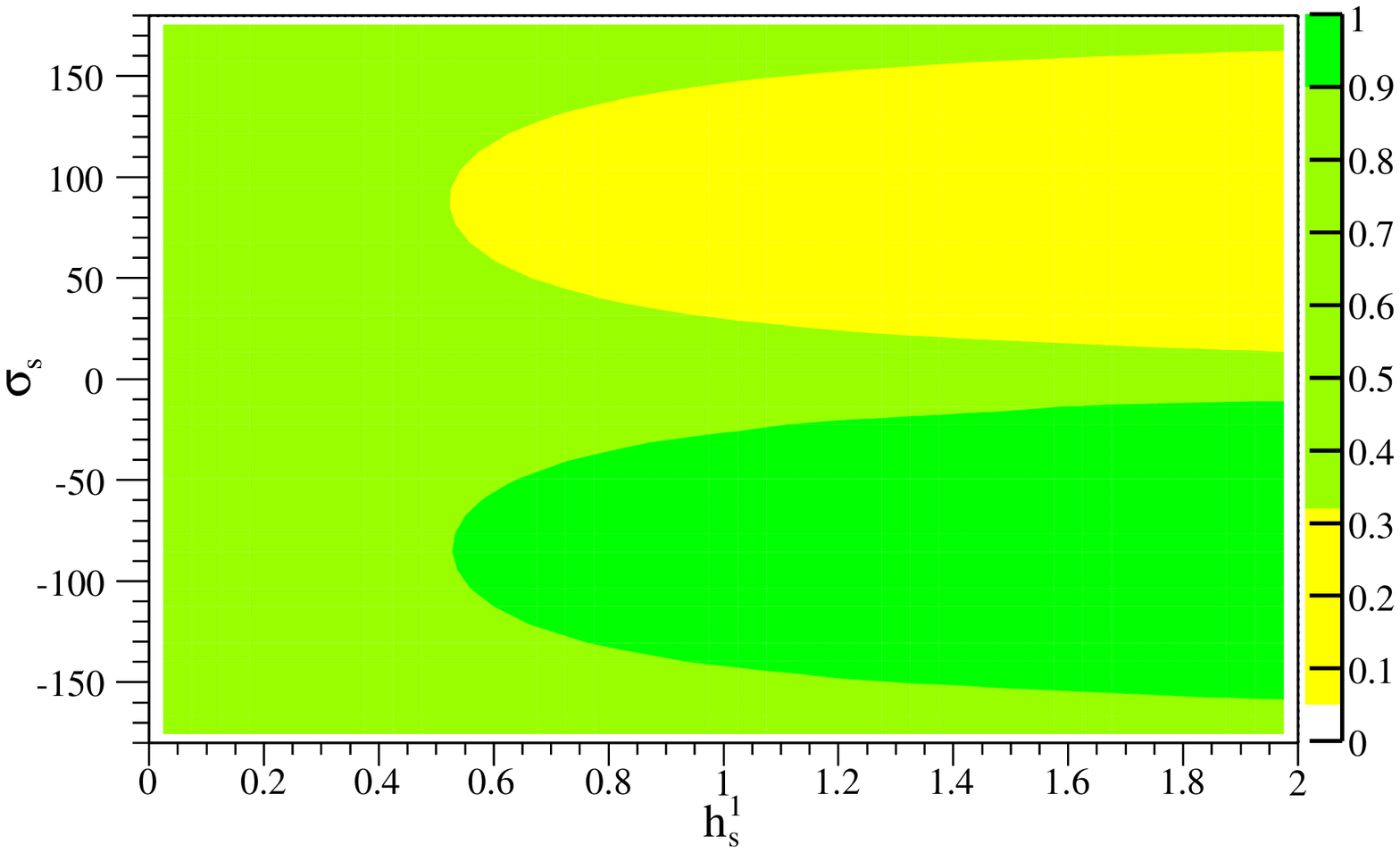} & 
\includegraphics[width=.5\textwidth]{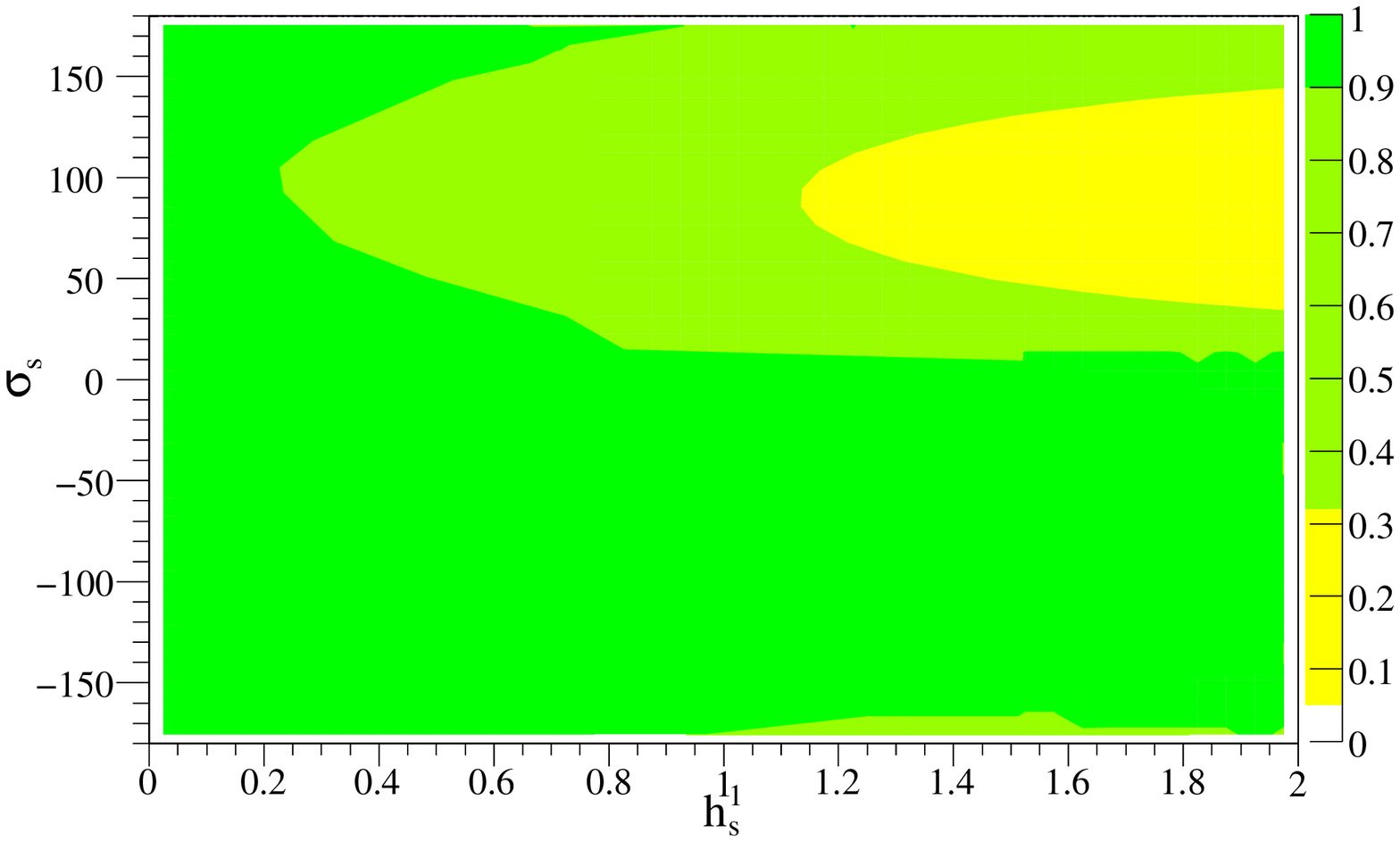} \\ 
\mbox{\bf (a)} & \mbox{\bf (b)} \\
\includegraphics[width=.5\textwidth]{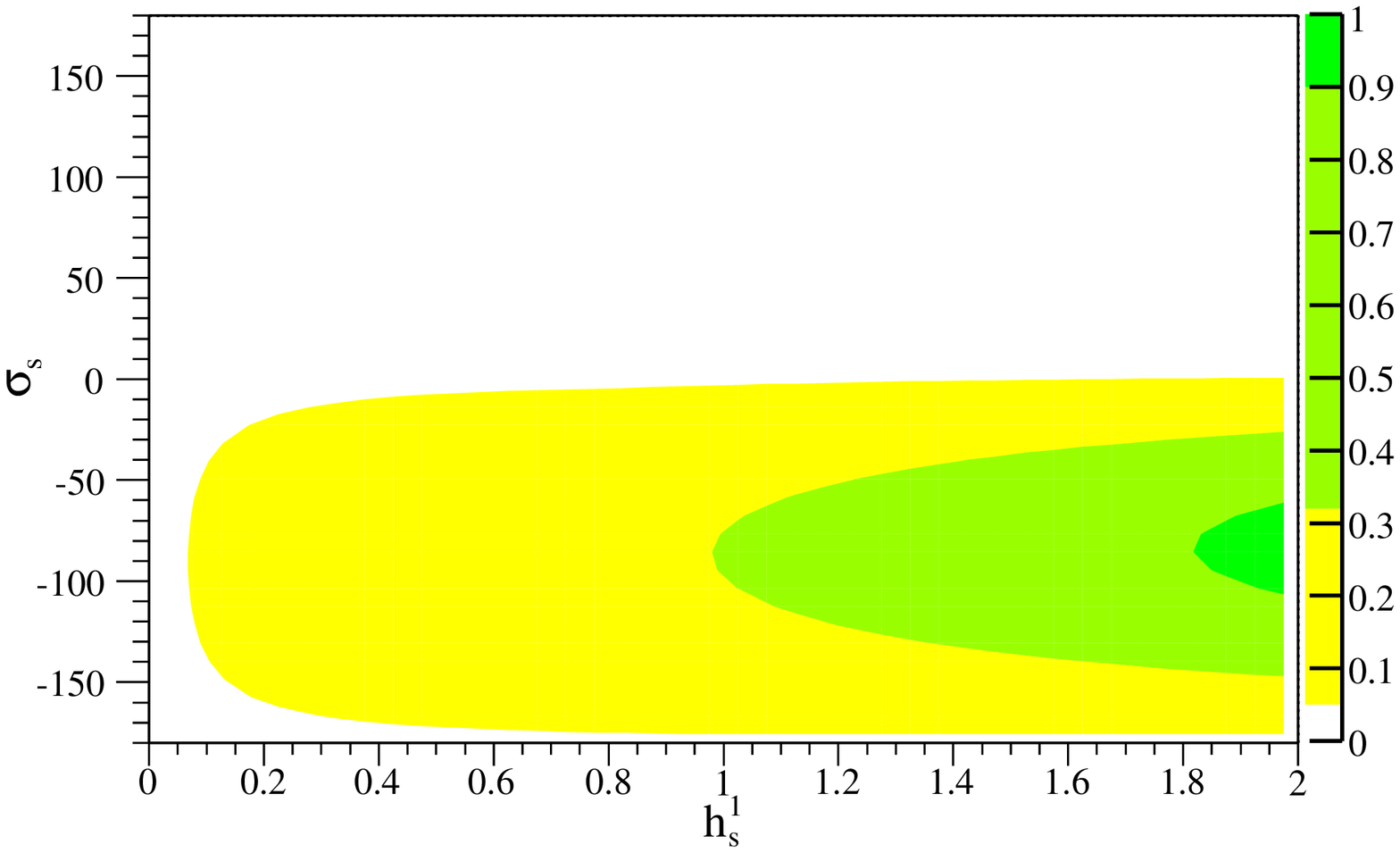} & 
\includegraphics[width=.5\textwidth]{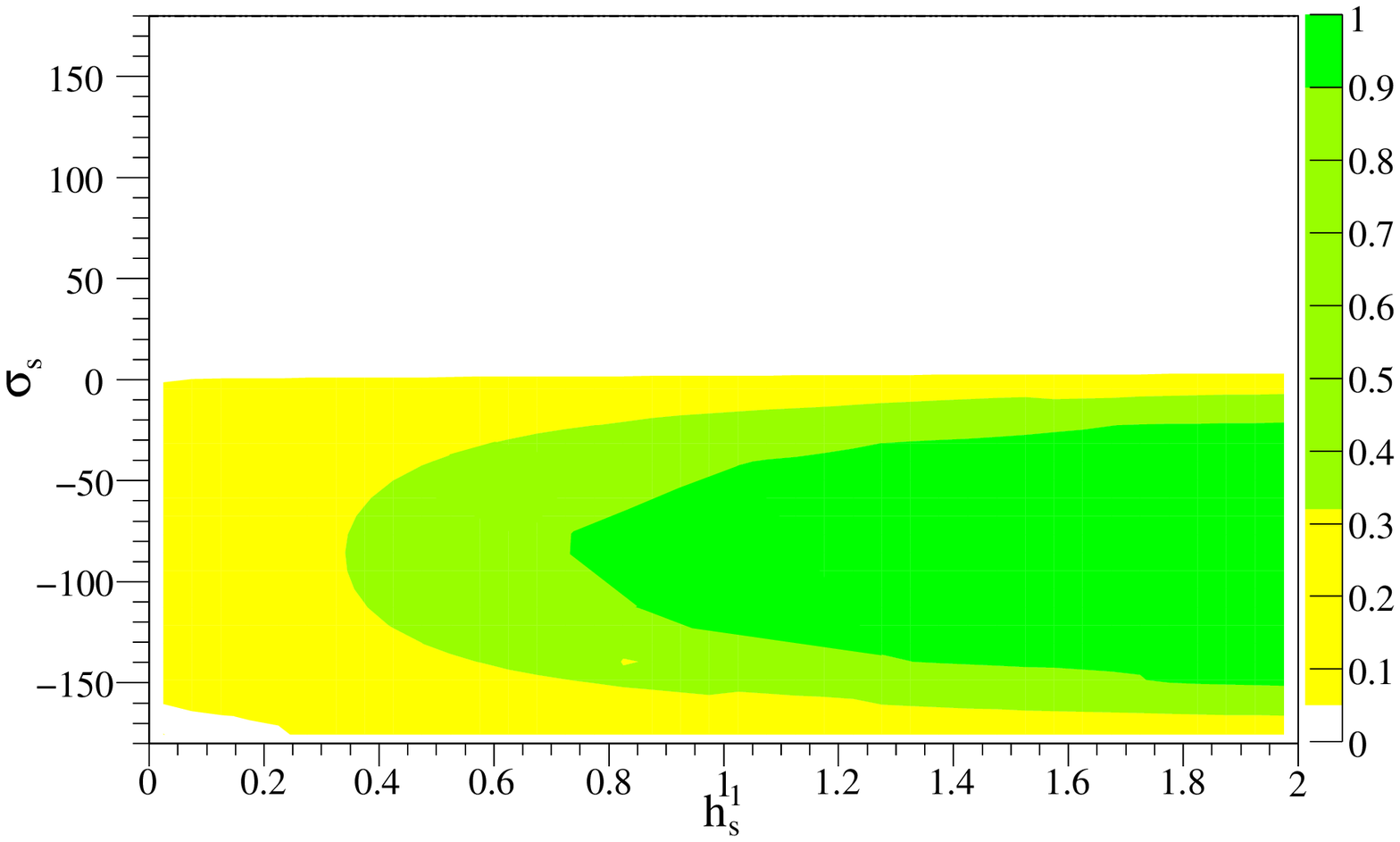} \\ 
\mbox{\bf (c)} & \mbox{\bf (d)} \\
\includegraphics[width=.5\textwidth]{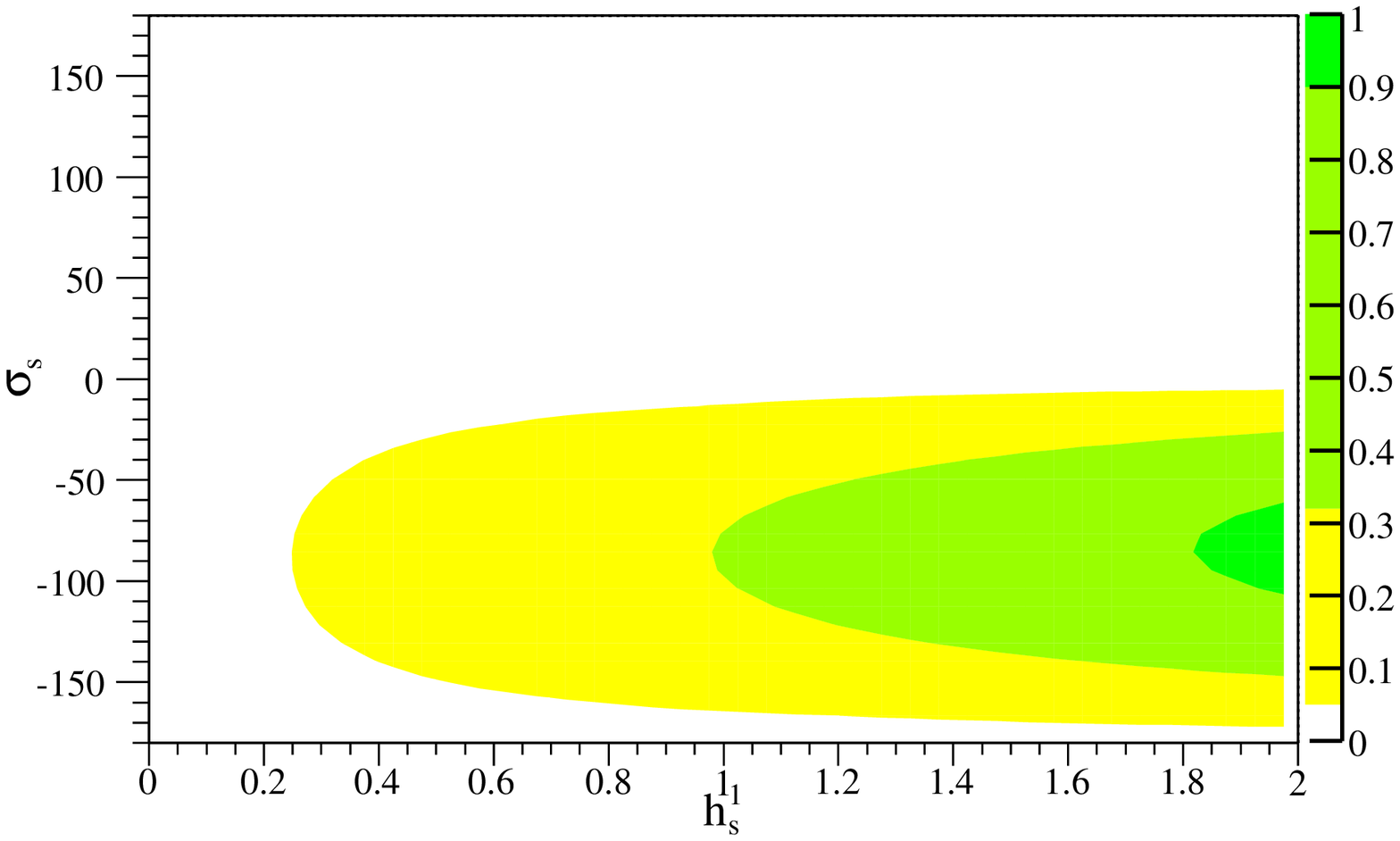} & 
\includegraphics[width=.5\textwidth]{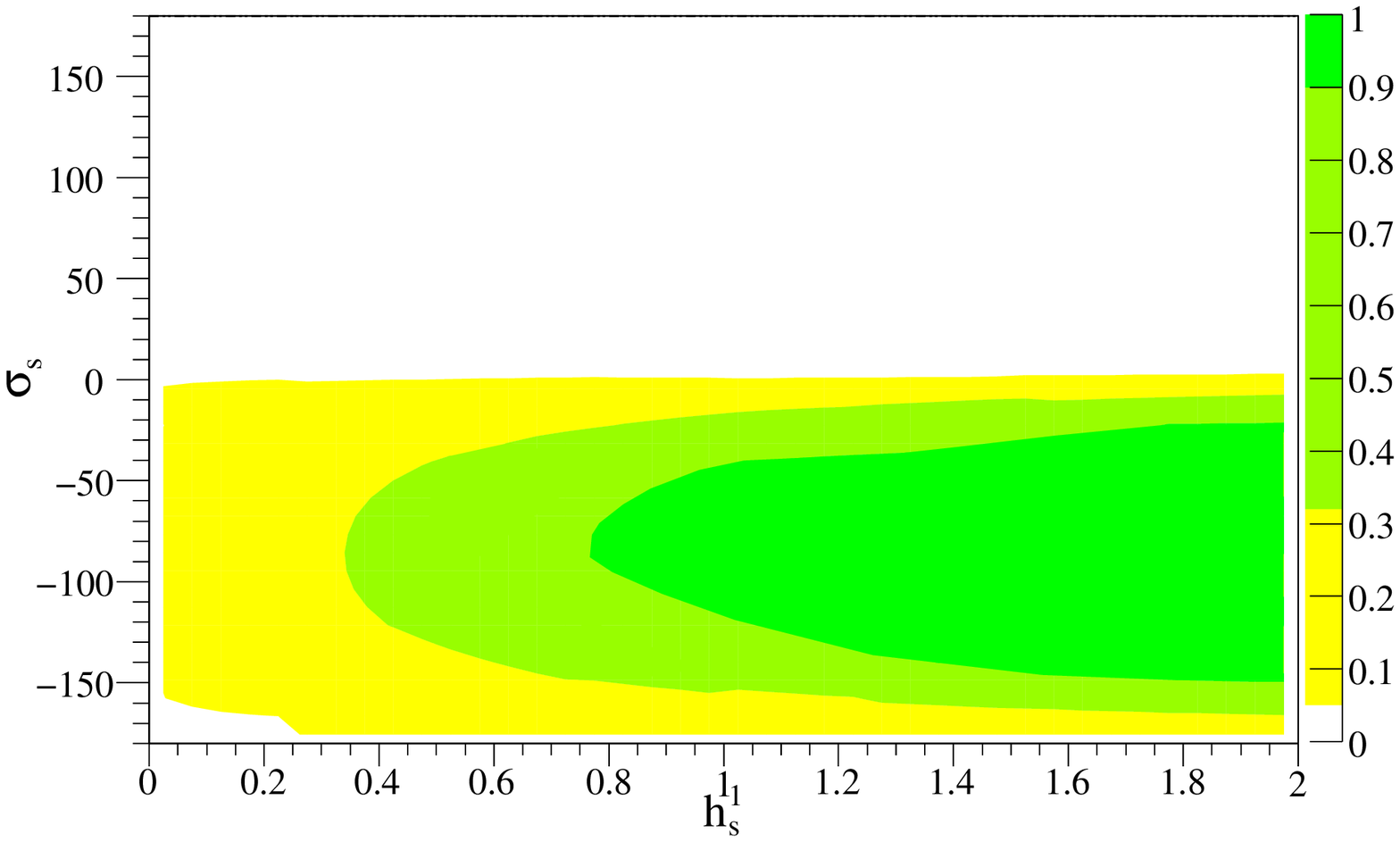} \\ 
\mbox{\bf (e)} & \mbox{\bf (f)}
\end{array}$
\end{center}
\caption{The constraints on $h_s^1$ and $\sigma_s$ from various
  $\Delta F=1$ processes. Left column: naive factorization
  results. Right column: QCD factorization results. The various rows
  are respectively for (from top to bottom): $B\rightarrow \phi K_S$,
  $B\rightarrow \eta' K_S$, the two combined.}
\label{figphietaK}
\end{figure}

We describe next the details of our analysis. We add
new terms $q_f e^{i\phi_f}$ in the $b\to s$ amplitude in Eq.~(\ref{b2samp})
induced by the NP terms in the weak effective Hamiltonian (\ref{Heff}). 
The coefficient $q_f e^{i\phi_f}$ is related to matrix elements of the operators 
$O_{3-10}$ with appropriate Wilson coefficients $C_i^Z$. We 
computed these matrix elements for several final states $f$ of interest
using: a) naive factorization~\cite{Naive} and b) the QCD factorization relations
using the heavy quark limit~\cite{BBNS},
including the leading radiative corrections and chirally enhanced terms.
The results of this analysis are shown in the plots of
Fig.~\ref{figphietaK}, where the left column is for the naive
factorization case and the right one is for the QCD factorization case.
In the numerical evaluation of the factorization formulas we used the 
hadronic parameters quoted in Ref.~\cite{Beneke:2003zv} with the exception
of the quark masses, for which we take $m_b(m_b) = 4.8 \pm 0.1$ GeV
and $m_s(2\, GeV) = 0.11 \pm 0.025$ GeV.

This type of analysis overlaps partially with previous work done in
Ref.~\cite{BFRS}, where constraints on several NP scenarios 
were obtained from mixing-induced CP violation in $b\to s$ decays.
The parameters $\varepsilon_z e^{i\theta_z}$ introduced in~\cite{BFRS} describing  
NP induced through Z$-$penguins are similar to the parameters $(h_s^1,\sigma_s)$ used here.
Ref.~\cite{BFRS} used leading order factorization to compute the
matrix elements of the relevant weak Hamiltonian operators. There are 
important differences between our analysis and that presented in Ref.~\cite{BFRS}, 
which we discuss next. 
First, we take into account
also the NP present in $B^0 - \bar B^0$ mixing, parameterized by the
$(h_d, \sigma_d)$. This is done by performing a correlated
analysis of these two types of processes. Second, the analysis of
\cite{BFRS} does not take into account experimental information on
direct CP asymmetries. The reason for this is that at leading order
in QCD factorization, the strong phases vanish. This implies that the
direct CP asymmetries are predicted to vanish, and thus no information
is gained at this order. In our fits the $A_{CP}$ data is included as 
inputs for the QCD factorization case.

\subsection{$B\to K\pi$ transitions}
\label{Kpi}

The $B\to K\pi$ decays are penguin-dominated processes, with subleading
contributions coming from tree and electroweak penguin amplitudes. 
In this respect they are similar to the $B\to \phi K_S$ and $\eta' K_S$
decays considered previously, however with a larger ``up-type'' contamination.

In the previous subsection we showed that the time-dependent CP asymmetry 
parameter $S_{f}$ in $B^0(t) \to f$ can be
used to constrain the NP parameters $(h_s^1,\sigma_s)$, provided that the
hadronic parameters determining the NP contribution
are known. Writing the amplitude for a generic $B\to f$ mode as
\begin{eqnarray}\label{b2sgeneric}
A(\bar B^0 \to f) = P_f (1 + d_f e^{i\psi_f} e^{-i\gamma} + h_s^1 e^{i\sigma_s} 
q_f e^{i\phi_f})\,,
\end{eqnarray}
the NP parameters $(h_s^1,\sigma_s)$ have been constrained using dynamical
computations of the hadronic parameters $(d,\psi)$ and $(q,\phi)$.

In this subsection we show that in the particular case of the $B^0 \to K_S \pi^0$
decay, these parameters can be determined from data on other $B\to K\pi$ decays,
using only flavor SU(3) symmetry and minimal assumptions about the smallness of
certain contributions. Some of these assumptions are satisfied at leading order 
of the heavy mass expansion in $\Lambda/m_b$, and are explicitly checked in
dynamical computations of the nonleptonic decay amplitudes in 
factorization~\cite{BBNS,Beneke:2003zv,BPRS}. We
estimate the corrections introduced by the remaining assumptions by comparing
with the results of such dynamical computations.
In principle, the output of such an analysis allows also a
direct test of the QCD factorization computation of these hadronic parameters.
The current precision of the data is however not sufficient for performing
a significant test. Tests for NP in these modes using related 
methods were also presented in Refs.~\cite{xx,xx1,xx2,xx3,xx4}.

We give also the results for the constraints on $(h_s^1 , \sigma_s)$ using
as input the QCD factorization results for the hadronic parameters
$(d,\psi)$ and $(q,\phi)$ in Eq.~(\ref{b2sgeneric}).

Before proceeding with the details of our analysis we list our data points.
Altogether we have  9 (10 when we include the
constraint on the CKM phase as we shall do eventually using the results
of our $\Delta F=2$ analysis) data points.
These are given by four branching ratios $B^{-,0}\to \pi^-\bar
K^0,\pi^0\bar K^-,\pi^+ K^-,\pi^0 K^0\,$\footnote{In our actual analysis
we just use the ratios between the above branching fractions and use also
the one from $B^-\to\pi^-\pi^0$.}, four direct CP asymmetries 
and a single time-dependent CP asymmetry $S_{K_S \pi^0}$.
The corresponding experimental values are listed in Table~\ref{data:Kpi}.

We start by neglecting the NP and discuss the most general form for these
decay amplitudes in the SM. 
The structure of the $B\to K\pi$ decay amplitudes, assuming only
isospin symmetry, can be written in terms of graphical amplitudes as 
(see, e.g.~\cite{GHLR})
\begin{eqnarray}\label{Kpigeneral}
&& A(B^-\to \pi^- \bar K^0) = \lambda_u^{(s)} (P_{uc} + A) +
\lambda_t^{(s)} (P_{tc} + EW_P + \frac12 EW_C - EW_E) \\
&& \sqrt2 A(B^-\to \pi^0 K^-) =
\lambda_u^{(s)} (-P_{uc} -A-T-C) +
\lambda_t^{(s)} (-P_{tc} - EW_P + EW_T + EW_C + EW_E) \nonumber \\
&& A(\bar B^0\to \pi^+ K^-) =
\lambda_u^{(s)} (-P_{uc} - T )
 + \lambda_t^{(s)} (-P_{tc} - EW_P + EW_C  - \frac12 EW_E)\nonumber\\
&& \sqrt2 A(\bar B^0\to \pi^0 \bar K^0) =
\lambda_u^{(s)} (P_{uc}-C )  + \lambda_t^{(s)} (P_{tc} + EW_P + EW_T +
\frac12 EW_C + \frac12 EW_E)
\nonumber
\end{eqnarray}
The notation adopted implies the use of the unitarity of the CKM matrix to 
eliminate the
$c$-quark CKM factor as $\lambda_c^{(s)} = -\lambda_u^{(s)}-\lambda_t^{(s)}$.
We combined the QCD penguin amplitudes as $P_{ut} = P_u-P_t$ and
$P_{ct} = P_c-P_t$. 

The graphical amplitudes appearing in Eqs.~(\ref{Kpigeneral}) arise
as matrix elements of specific operators in the weak Hamiltonian, as follows: 
the operators $O_{1,2}^{u,c}$ give rise to the tree $T$, color-suppressed
$C$, weak annihilation $A$ and $u,c$-penguin $P_{u}, P_c$ amplitudes. The 
matrix elements of the electroweak penguin operators $Q_{7-10}$ appear in 
several possible combinations:  color-allowed $EW_T$, color-suppressed $EW_C$, 
penguin-type contractions $EW_P$ and weak annihilation $EW_E$. 

In the presence of new physics as described by the weak Hamiltonian
Eq.~(\ref{Heff}), the QCD and EW penguin amplitudes can be split into
contributions which contain, respectively do not contain, the NP
factor $h_s^1 e^{i\sigma_s}$ as follows
\begin{eqnarray}\label{SM2NP}
&& P_t \to P_t + h_s^1 e^{i\sigma_s} P_Z \\
&& EW_i \to EW_i + h_s^1 e^{i\sigma_s} EW_i^{(Z)}
\end{eqnarray}
The new penguin amplitude $P_Z$ is proportional to the Wilson coefficients
$C_{3-6}^Z$, and the new EWP amplitudes are proportional to the corresponding
Wilson coefficients $C_{7-10}^Z$. In the remainder of this part we will
assume the SM form of the $B\to K\pi$ amplitudes. Their
modification to include NP effects using Eq.~(\ref{SM2NP}) is straightforward and 
will be included below.

Isospin symmetry gives one relation among these amplitudes
\begin{eqnarray}\label{Kpiisospin}
 A(B^-\to \pi^- \bar K^0) + \sqrt2 A(B^-\to \pi^0 K^-) =
A(\bar B^0\to \pi^+ K^-) + \sqrt2 A(\bar B^0\to \pi^0 \bar K^0)\,.
\end{eqnarray}
Counting hadronic parameters gives 6 independent complex amplitudes
(8 coefficients of the $\lambda_{u,t}$ CKM factors minus two relations 
from (\ref{Kpiisospin})). Taking into account that one unphysical phase can be
eliminated, this gives 11 independent real hadronic parameters.
Together with the CKM phase, this counting shows that the $K\pi$ system is 
parameterized by 12 unknown
parameters, which can not be fixed with only the help of the 9 data points in
Table 2 (even if the CKM phase is added as an input).
In other words, the most general form Eq.~(\ref{Kpigeneral}) contains too 
many amplitudes to be predictive, and additional input is necessary.
We will be very explicit about the assumptions made, and discuss possible 
tests for their validity.
\vspace{0.5cm}

{\em Assumption A.} In the heavy quark limit, weak annihilation diagrams
are power suppressed by $\Lambda/m_b$ relative to the tree and penguin 
amplitudes~\cite{BBNS,Beneke:2003zv,BPRS}. This is a theoretically clean 
approximation, although
the numerical size of the power corrections $O(\Lambda/m_b)$ could be significant.
In this limit we can neglect the weak annihilation amplitudes $A, EW_E$ appearing
in Eqs.~(\ref{Kpigeneral}).
This approximation by itself does not reduce the number of $K\pi$ independent 
amplitudes, but will be required in conjunction with the approximation
{\em B.} introduced below.
\vspace{0.5cm}

{\em Assumption B.}
Flavor SU(3) symmetry\footnote{The analysis presented here neglects all the
  $SU(3)$ breaking effects, but they could be included in
  factorization.} and the neglect of the matrix elements related
to electroweak penguins $Q_{7,8}$. 
This approximation is justified in the SM because of the smallness of the
Wilson coefficients $C_{7,8}$ relative to $C_{9,10}$ (see Eq.~(\ref{WCi})). In the
presence of new physics mediated by the $Z$ coupling (\ref{Heff}), this inequality 
still holds $C_{7,8}^{(Z)} < C_{9,10}^{(Z)}$,
but the Wilson coefficients $C_{7,8}^{(Z)}$ are not negligible compared with the 
QCD penguin coefficients $C_{3-6}^{(Z)}$. Their effects must be therefore included
in our analysis. 

The latter assumption allows a reduction in the number of electroweak penguin 
amplitudes, and gives four relations for these amplitudes. These relations follow
from SU(3) symmetry, and were derived in Refs.~\cite{NeRo,BuFl,GPY}. 
The first relation does not require the approximation {\em A.} and relates the
combination $EW_T + \frac32 EW_C$ as~\cite{NeRo}
\begin{eqnarray}\label{EWPT+C}
&& EW_T + \frac32 EW_C = \frac32 \kappa_+(T+C) 
\end{eqnarray}
where $\kappa_\pm$ are given by ratios of Wilson coefficients of  the
dominant EW penguins $O_{9,10}$ as
\begin{eqnarray}
\kappa_\pm \equiv \frac{C_9 \pm C_{10}}{C_1 \pm C_2}
\end{eqnarray}

Adopting also the approximation {\em A.}, two additional identities can be
written down, which fix all individual EW penguin amplitudes in terms of the
$T,C$ and $P_{uc}$ amplitudes~\cite{GPY}. The first relation holds for 
any values of $\kappa_\pm$
\begin{eqnarray}\label{EWPinSU3}
EW_C = \frac12\kappa_+(T+C) + \frac12 \kappa_-(C-T)\nonumber\,.
\end{eqnarray}
and the second relation 
\begin{eqnarray}\label{EWPP}
EW_P = \kappa P_{u}  \,.
\end{eqnarray}
assumes furthermore the equality 
$\kappa_+ \simeq \kappa_- \equiv \kappa$, which holds to a good precision in
the SM. It continues to hold in any new physics model for which 
$C_{10}(M_W)$ is subdominant, including the class of models considered here.

The approximations {\em A.} and {\em B.} taken together reduce the number of 
individual hadronic parameters to 4 complex amplitudes: $P_{ut}, P_{ct}, T,C$ 
(minus an overall phase).  
Counting in also $\gamma$, this gives 8 unknown real parameters, which 
can be extracted from data.
This is essentially the approach proposed in Ref.~\cite{London} for 
determining $\gamma$ from $B\to K\pi$ data alone.

Including also the NP effects through the substitutions Eq.~(\ref{SM2NP})
introduces several new parameters:
$(h_s^1,\sigma_s)$ and the NP penguins $P_Z$ and $EW_i^Z$. 
This renders the system again undetermined. The $EW_i^Z$ amplitudes can be 
expressed in terms of the tree amplitudes, using relations similar to those
for $EW_i$. The relation similar to Eq.~(\ref{EWPT+C}) is
\begin{eqnarray}\label{EWPT+CNP}
&& EW_T^Z + \frac32 EW_C^Z = \frac32 \kappa_+^Z(T+C) + \frac32 \Delta_1 (T+C)
\end{eqnarray}
where $\kappa_\pm^Z$ are given by
\begin{eqnarray}
\kappa_\pm^Z \equiv \frac{C_9^Z \pm C_{10}^Z}{C_1 \pm C_2}
\end{eqnarray}
and $\Delta_1$ is a correction proportional to $C_{7,8}^Z$. Inspection of the
Wilson coefficients in Eq.~(\ref{WCi}) shows that their effects can be
significant, and have to be included.
In QCD factorization this coefficient is given explicitly by
\begin{eqnarray}
\Delta_1 = \frac{r_\chi^M a_{8Z}^c - a_{7Z}^c}{a_1+a_2} = 
-\frac{C_7^Z + C_8^Z/N_c}{(C_1+C_2)(1+1/N_c)} +
r_\chi^K \frac{C_8^Z+C_7^Z/N_c}{(C_1+C_2)(1+1/N_c)} + O(\alpha_s)
\end{eqnarray}
where we used the notations of \cite{Beneke:2003zv} for the coefficients $a_i$.
A subscript $Z$ means that $a_{iZ}$ has to be computed using the $C_i^Z$ Wilson
coefficients. 
We retained here also the $a_{8Z}^c$ term, although it is formally power suppressed
by the factor
$r_\chi^K = 2M_K^2/(m_b( m_q+m_s))$. However, numerically it is comparable with 
the $a_{7Z}^c$ term, and it partially cancels it.
Using the Wilson coefficients in Eq.~(\ref{WCi}), the effect of $\Delta_1$ is a
6\% increase in the value of $\kappa_+$ (in absolute value). This shows that
the contributions from $C_{7,8}^Z$ can be neglected to a very good approximation in 
the relation Eq.~(\ref{EWPT+CNP}).

The relations analogous to Eqs.~(\ref{EWPinSU3}), (\ref{EWPP}) read
\begin{eqnarray}\label{EWPinSU3NP}
&& EW_C^Z = 
\frac12\kappa_+^Z (T+C) + \frac12 \kappa_-^Z (C-T) + \Delta_2 (T+C)\\
&& EW_P^Z = \kappa^Z P_{u} + \Delta_3 \,.
\end{eqnarray}
with $\kappa^Z \equiv \kappa_+^Z \sim \kappa_-^Z$, and the correction terms 
$\Delta_{2,3}$ contain again the contributions proportional to $C_{7,8}^Z$.
We will need only $\Delta_2$, which is given in QCD factorization by \cite{BBNS,Beneke:2003zv}
\begin{eqnarray}
\Delta_2 = \frac{r_\chi^K a_{8Z}^c}{a_1+a_2} = r_\chi^K \frac{C_8^Z + C_7^Z/N_c}{(C_1+C_2)(1+1/N_c)}
+ O(\alpha_s)
\end{eqnarray}
This gives that the $C_{7,8}^Z$ contributions to the relation Eq.~(\ref{EWPinSU3NP})
are both power suppressed and color suppressed. However, they appear multiplied with the 
chirally enhanced
coefficient $r_\chi^K$, so for consistency with Eq.~(\ref{EWPT+CNP}) we will include them
in the following. Numerically the effect of $\Delta_2$ is a negative shift (again in absolute
value) in $\kappa_+$ of $\sim -40\%$. However, when substituted in the
$\bar B\to K\pi$ amplitudes, the overall contribution
from $Q_{7,8}$ is negligible (below 1\% of the total amplitude) over most
of the parameter space. Therefore we will neglect these contributions in our fit.

Counting the number of parameters, we have now in addition to the 7 SM
parameters, $P_{ut},P_{ct},T,C$, another 4 unknown parameters $h_s^1,\sigma_s,
P_Z$. This renders the system undetermined. To be able to proceed, 
we make one last approximation.
\vspace{0.5cm}

{\em Assumption C.} Our final approximation here is to neglect the amplitude
$P_{ut}$ which is doubly Cabibbo suppressed by the small CKM coefficient
$\lambda_u^{(s)}$.

With this approximation, the $K\pi$ system is described in the presence of NP 
by 4 complex hadronic amplitudes $P_{ct},T,C, P_Z$
and the two NP parameters $(h_s^1,\sigma_s)$, altogether 9 unknowns
(the CKM phase  will be taken as an input from our $\Delta F=2$
analysis).
The number of hadronic parameters can be reduced by one if we consider
also the decay $B^+ \to \pi^0\pi^+$. As this mode is dominated by a SM tree level
transition, it receives a negligible NP
contribution. Neglecting small electroweak penguin contributions, the
amplitude for this decay is
\begin{eqnarray}
\sqrt{2} A(B^- \to \pi^0 \pi^-) = \lambda_u^{(d)} (T+C)
\end{eqnarray}
with branching fraction BR$(B^+ \to \pi^0 \pi^+) = (5.5 \pm 0.6)
\times 10^{-6}$~\cite{hfag}.
Using the branching fraction for this mode as an input eliminates the
hadronic amplitude $|T+C|$, and reduces the number of unknown parameters
to 8.
\vspace{0.5cm}

\begin{table}[t!]
\begin{center}
\begin{tabular}{ccc}
\hline
 & $Br\, (\times 10^{-6})$ & $A_{CP}$ \\
\hline
$B^-\to \pi^-\bar K^0$ & $24.1\pm 1.3$ &  $-0.02\pm 0.04$ \\
$B^-\to \pi^0 K^-$ & $12.1\pm 0.8$ &  $0.04\pm 0.04$ \\
$B^0\to \pi^+ K^-$ & $18.9\pm 0.7$ & $-0.11\pm 0.02$ \\
$B^0\to \pi^0 \bar K^0$ & $11.5\pm 1.0$ & $0.01\pm 0.16$ \\
\hline
\hline
$S_{K_S\pi^0}$ & $0.34\pm 0.28$ & \\
\hline
\end{tabular}
\end{center}
{\caption{\small World averages of $B\to K\pi$ branching fractions, direct CP asymmetries
and the $S_{K_S\pi^0}$ parameter~\cite{hfag}.}
\label{data:Kpi} }
\end{table}

The most general parameterization of the $\bar B\to K\pi$
amplitudes compatible  with the  assumptions {\em A.,B.,C.} can be written as
\begin{eqnarray}\label{KpiepsNP}
&& A(B^-\to \pi^- \bar K^0) =
\lambda_c^{(s)} P_{ct} (1 + \frac13 \delta_{EW} \varepsilon_C e^{i\phi_C} + h_s^1 e^{i\sigma_s}
[p_Z e^{i\phi_Z} + \frac13 \delta_{EW}^Z \varepsilon_C e^{i\phi_C}]) \\
&& \sqrt2 A(B^-\to \pi^0 K^-) =
-\lambda_c^{(s)} P_{ct} (1 - \delta_{EW} \varepsilon e^{i\phi}
+ \frac13 \delta_{EW} \varepsilon_C e^{i\phi_C}\\
&& \hspace{3cm} + h_s^1 e^{i\sigma_s}
[p_Z e^{i\phi_Z} - \delta_{EW}^Z \varepsilon e^{i\phi} + \frac13 \delta_{EW}^Z \varepsilon_C
e^{i\phi_C}] + e^{-i\gamma}  \varepsilon e^{i\phi} )\nonumber\\
&& A(\bar B^0\to \pi^+ K^-) =
-\lambda_c^{(s)} P_{ct} (1  - \frac23 \delta_{EW} \varepsilon_{C} e^{i\phi_C}
+ h_s^1 e^{i\sigma_s} [p_Z e^{i\phi_Z} - \frac23 \delta^Z_{EW} \varepsilon_C e^{i\phi_C}]
  +  e^{-i\gamma} \varepsilon_{T} e^{i\phi_T} )\nonumber\\
&&\sqrt2 A(\bar B^0\to \pi^0 \bar K^0) =
\lambda_c^{(s)} P_{ct} (1  + \delta_{EW} \varepsilon_{T} e^{i\phi_T}
+ \frac13 \delta_{EW} \varepsilon_{C}  e^{i\phi_C}\\
&& \hspace{3cm} + h_s^1 e^{i\sigma_s} 
[p_Z e^{i\phi_Z} + \delta^Z_{EW} \varepsilon_T e^{i\phi_T} + \frac13 \delta^Z_{EW} \varepsilon_C e^{i\phi_C}]
  -  e^{-i\gamma} \varepsilon_{C}  e^{i\phi_C} )\nonumber
\end{eqnarray}

We introduced here the tree/penguin ratios $(\varepsilon_i, \phi_i)$, and the reduced
NP penguin parameters $(p_Z, \phi_Z)$, defined as
\begin{eqnarray}
 \varepsilon_T e^{i\phi_T} = \frac{|\lambda_u^{(s)}|}
{|\lambda_c^{(s)}|}\frac{T}{P_{ct}}\,,\quad
\varepsilon_C e^{i\phi_C} = \frac{|\lambda_u^{(s)}|}{|\lambda_c^{(s)}|}
\frac{C}{P_{ct}}\,,\quad
 \varepsilon e^{i\phi} = \frac{|\lambda_u^{(s)}|}
{|\lambda_c^{(s)}|}\frac{T+C}{P_{ct}}\,, \quad
p_Z e^{i\phi_Z} = \frac{P_Z}{P_{ct}}\,.
\end{eqnarray}
The parameters $\delta_{EW}$ and  $\delta_{EW}^Z$ describe the 
contributions of
the two types of electroweak penguin operators $C_{9,10}$ and $C_{9,10}^Z$, 
respectively. Their numerical values can be obtained from Eq.~(\ref{WCi})
\begin{eqnarray}
&& \delta_{EW}  = -\frac32 \frac{|\lambda_c^{(s)}|}
{|\lambda_u^{(s)}|}\frac{C_9 + C_{10}}{C_1 + C_2} \simeq 0.65 \,,\qquad
\delta_{EW}^Z  = -\frac32 \frac{|\lambda_c^{(s)}|}
{|\lambda_u^{(s)}|} \frac{C_9^Z + C_{10}^Z}{C_1 + C_2} \simeq 0.287 \,.
\end{eqnarray}

We will use as hadronic inputs the following ratios of CP-averaged 
branching fractions
\begin{eqnarray}\label{Kpidata}
&& R_\pi = \frac{2 Br(B^+ \to \pi^0 \pi^+)}{Br(B^+ \to K^0 \pi^+)} = 
0.456 \pm 0.055\,,\qquad
R_c = \frac{2 Br(B^+ \to \pi^0 K^+)}{Br(B^+ \to K^0 \pi^+)} = 1.004 \pm 0.086\\
&& R_n = \frac{Br(B^0 \to K^+\pi^-)}{Br(B^+ \to K^0 \pi^+)}
\frac{\tau_{B^+}}{\tau^{B^0}}
 = 0.816 \pm 0.057\,,\qquad
  R_0 = \frac{2Br(B^0 \to K_S\pi^0)}{Br(B^+ \to K^0 \pi^+)}
\frac{\tau_{B^+}}{\tau^{B^0}}
 = 1.032 \pm 0.106
\nonumber \\
\end{eqnarray}
together with the direct CP asymmetries and the $S_{K_S\pi^0}$ parameter in Table 2.
We will use the weak phase $\gamma$ as determined from the global CKM
fit ~\cite{CKMfitter} 
\begin{eqnarray}
\gamma = (57^{+13}_{-8})^\circ  
\end{eqnarray}
This gives a total of 9 data points which can be used to constrain the 8 hadronic parameters 
$(\varepsilon, \phi), (\varepsilon_C,\phi_C), (p_Z,\phi_Z)$ and  
the NP parameters $(h_s^1, \sigma_s)$ from data. 

For the purposes of a numerical analysis, it is convenient to simplify the
theoretical expressions for the decay amplitudes, by introducing a new
penguin amplitude
\begin{eqnarray}
P \equiv \lambda_c^{(s)} P_{ct} (1 + \frac13 \delta_{EW} \varepsilon_C e^{i\phi_C})
\end{eqnarray}
and redefining the ratios of amplitudes as
\begin{eqnarray}
&& \frac{\varepsilon_i e^{i\phi_i}}{1+\frac13 \delta_{EW} \varepsilon_C
  e^{i\phi_C}} \to \varepsilon_i e^{i\phi_i}\,,\qquad i = T,C \\
&&  \frac{p_Z e^{i\phi_Z} + \frac13 \delta_{EW}^Z \varepsilon_C e^{i\phi_C}}
{1+\frac13 \delta_{EW} \varepsilon_C
  e^{i\phi_C}} \to p_Z e^{i\phi_Z}\,.
\end{eqnarray}
Expressed in terms of the redefined parameters, the $B\to K\pi$ amplitudes in
Eqs.~(\ref{KpiepsNP}) are given by
\begin{eqnarray}\label{Kpisimple}
&& A(B^-\to \pi^- \bar K^0) =
P (1 + h_s^1 e^{i\sigma_s} p_Z e^{i\phi_Z} ) \\
&& \sqrt2 A(B^-\to \pi^0 K^-) =
- P (1 - \delta_{EW} \varepsilon e^{i\phi} + e^{-i\gamma}  \varepsilon e^{i\phi} 
+ h_s^1 e^{i\sigma_s}
[p_Z e^{i\phi_Z} - \delta_{EW}^Z \varepsilon e^{i\phi} ]  )\\
&& A(\bar B^0\to \pi^+ K^-) =
- P (1  -  \delta_{EW} \varepsilon_{C} e^{i\phi_C} + e^{-i\gamma} \varepsilon_{T} e^{i\phi_T} 
+ h_s^1 e^{i\sigma_s} [p_Z e^{i\phi_Z} - \delta^Z_{EW} \varepsilon_C e^{i\phi_C}] )
\nonumber\\
&&\sqrt2 A(\bar B^0\to \pi^0 \bar K^0) =
P (1  + \delta_{EW} \varepsilon_{T} e^{i\phi_T} - e^{-i\gamma} \varepsilon_{C}  e^{i\phi_C} 
 + h_s^1 e^{i\sigma_s} 
[p_Z e^{i\phi_Z} + \delta^Z_{EW} \varepsilon_T e^{i\phi_T} ])\nonumber
\end{eqnarray}

\begin{figure}[!htp]
\begin{center}
$\begin{array}{c@{\hspace{0.2in}}c}
\includegraphics[width=.5\textwidth]{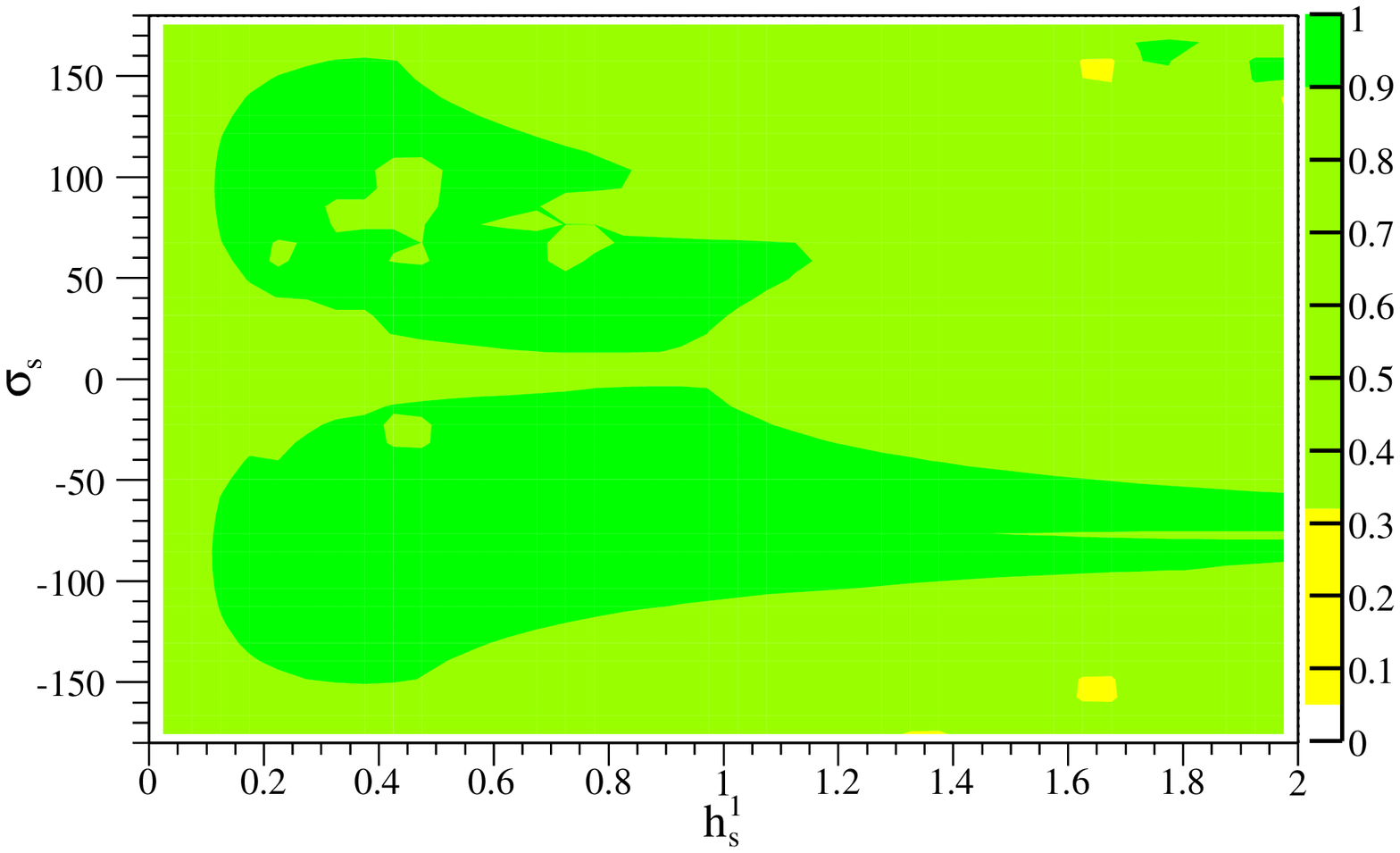} & 
\includegraphics[width=.5\textwidth]{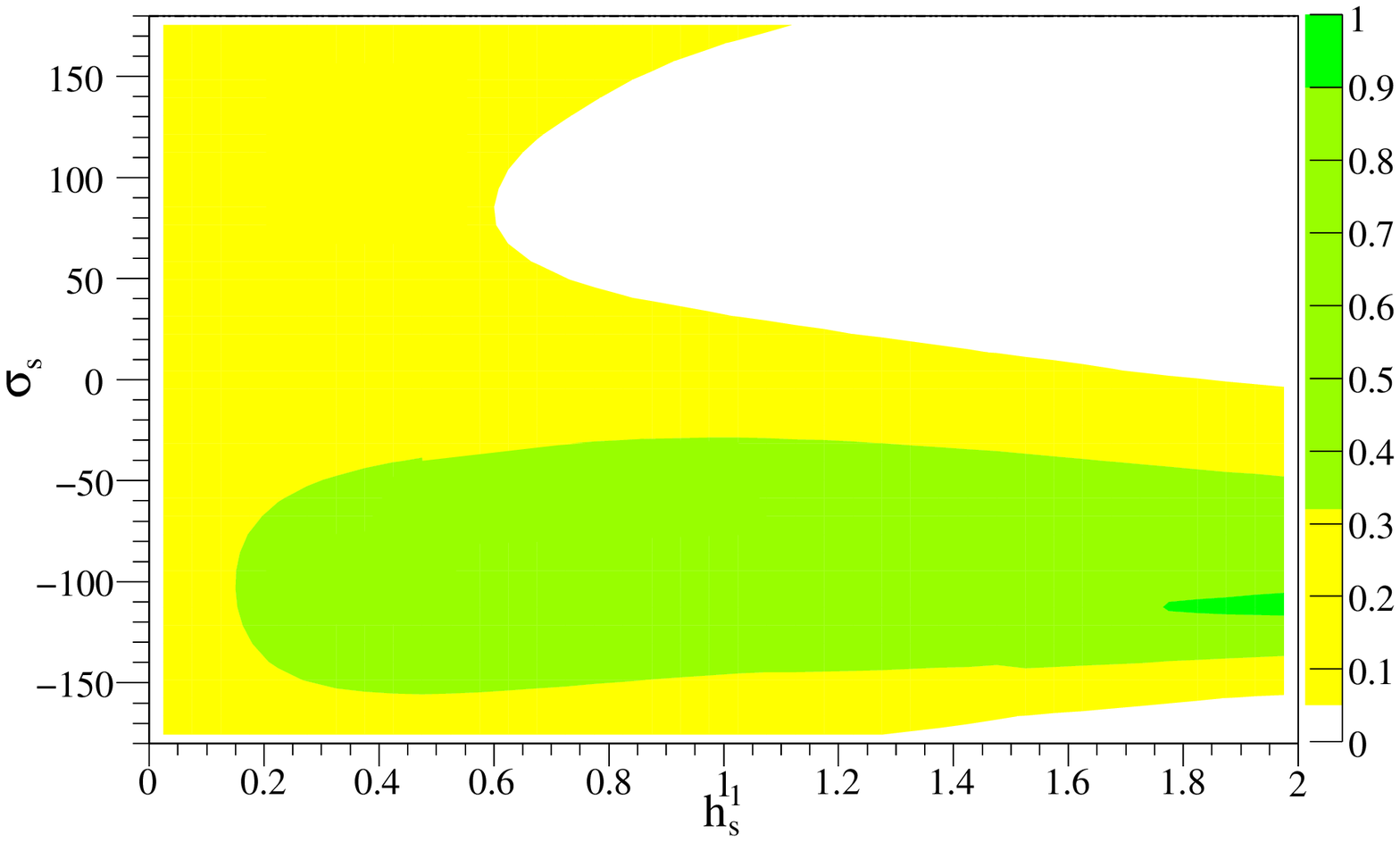} \\ 
\mbox{\bf (a)} & \mbox{\bf (b)} \\
\includegraphics[width=.5\textwidth]{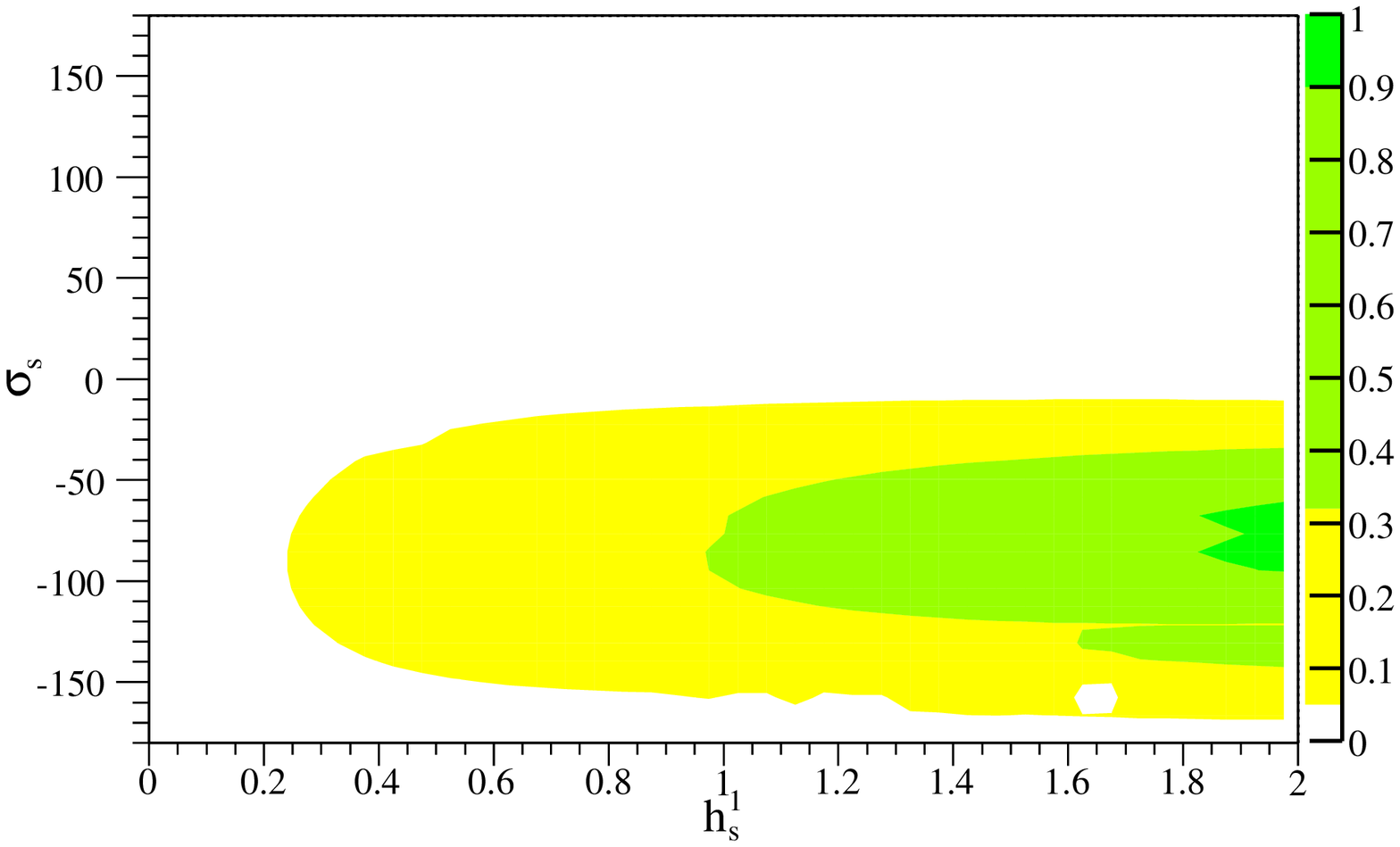} &
\includegraphics[width=.5\textwidth]{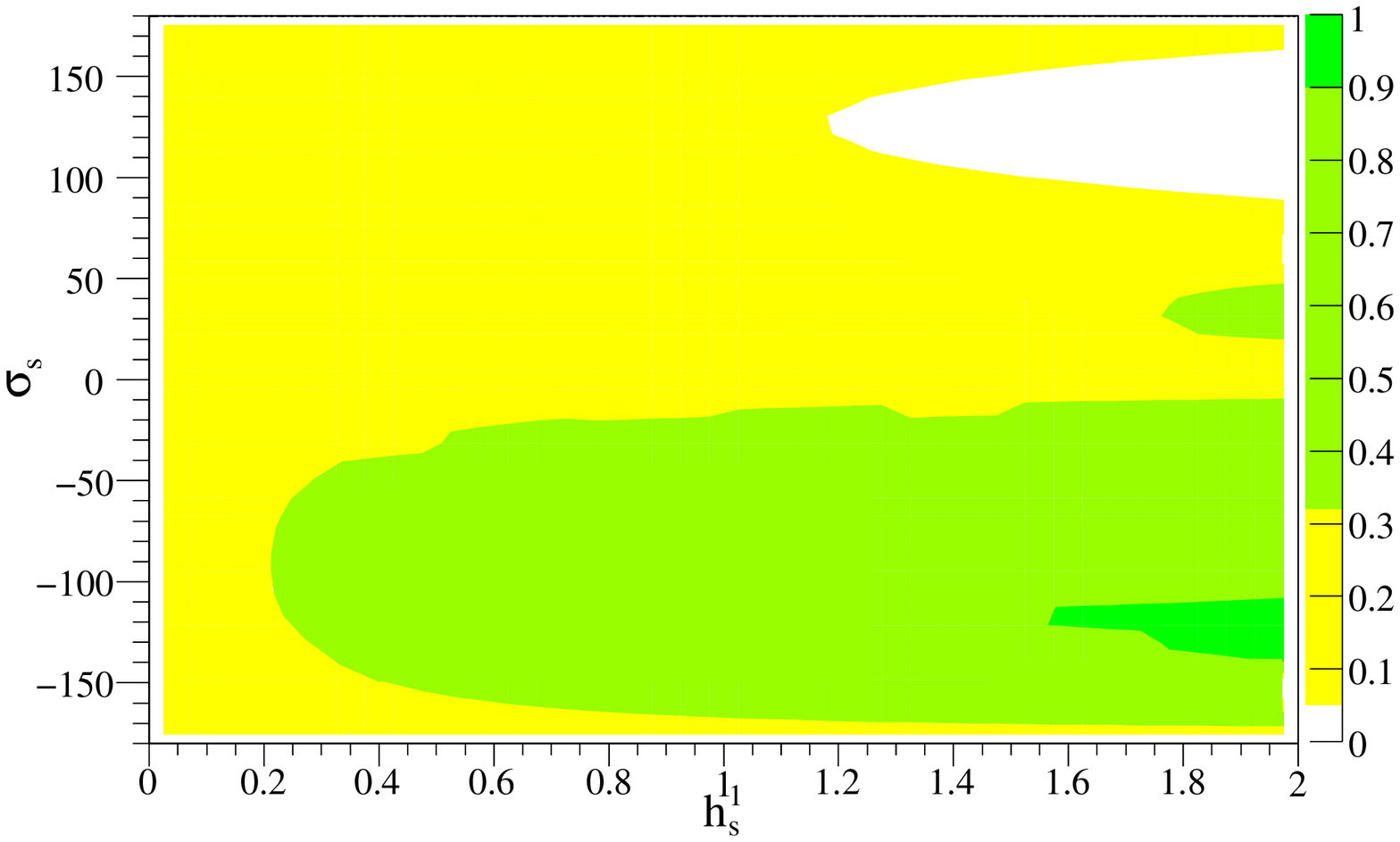} \\ 
\mbox{\bf (c) } & \mbox{\bf (d)} \\
\includegraphics[width=.5\textwidth]{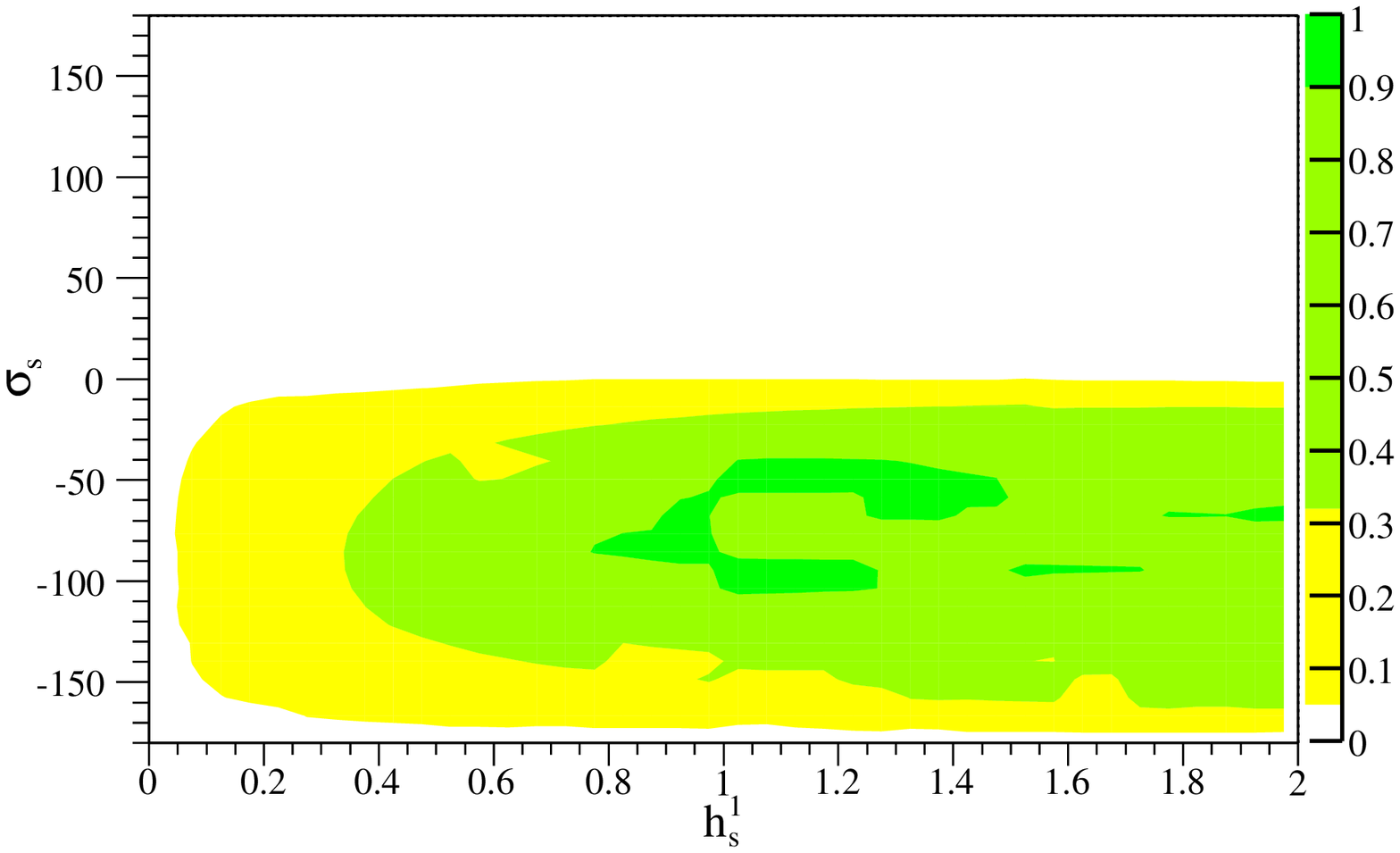} & 
\includegraphics[width=.5\textwidth]{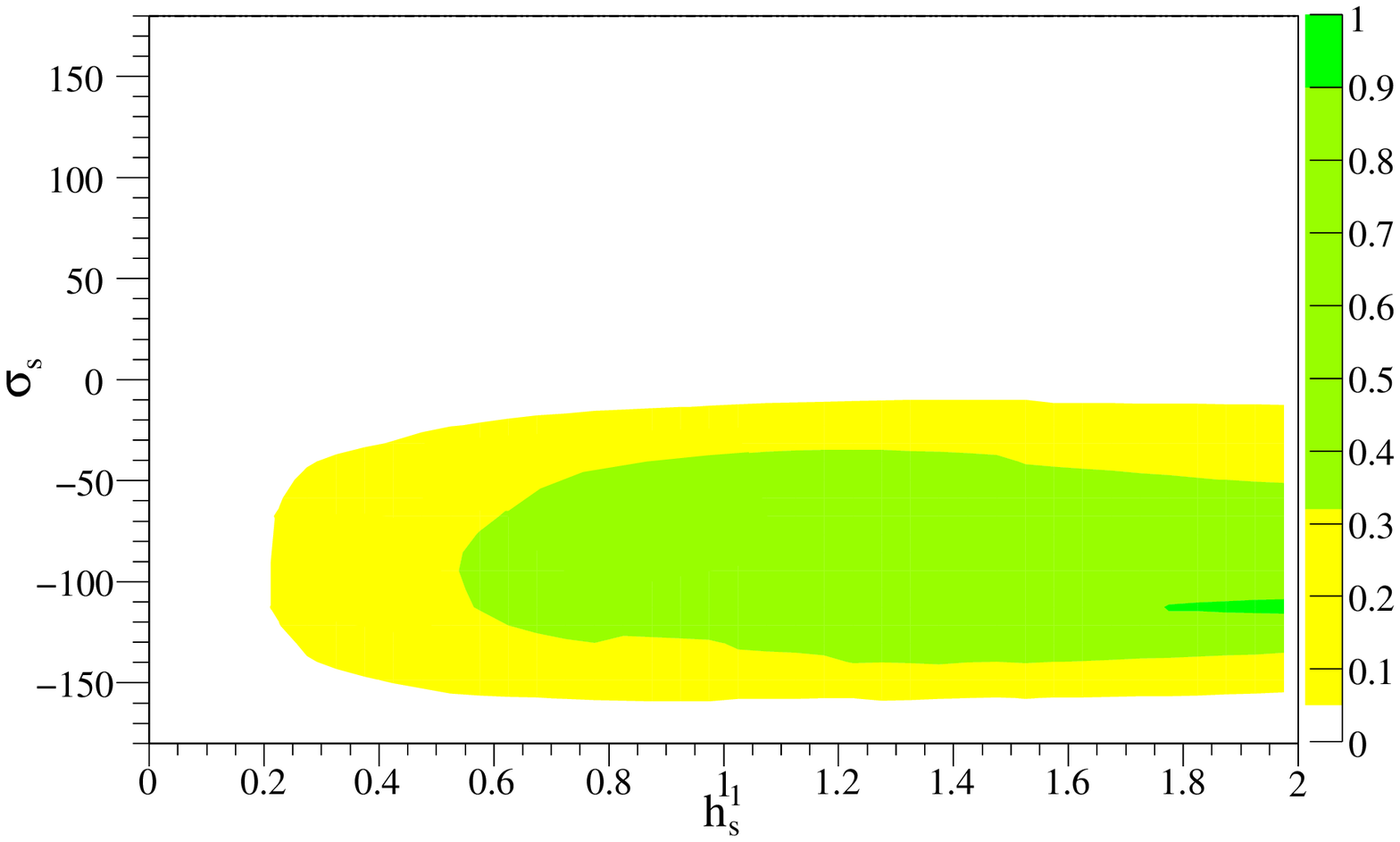} \\ 
\mbox{\bf (e)} & \mbox{\bf (f)}
\end{array}$
\end{center}
\caption{The constraints on $h_s^1$ and $\sigma_s$ from the $K \pi$
  system. Left column: $SU(3)$ analysis. Right column: QCD
  factorization. Plots (a) and (b) are obtained taking into
  account BR information. Plot (d) is obtained from CP asymmetries
  only. Plots (c) and (e) are obtained combining (a)  with
  $B\rightarrow \phi K_S,\eta' K_S$ in naive and QCD factorization
  respectively. Plot (f) is obtained combining (b) with 
  $B\rightarrow \phi K_S,\eta' K_S$ in QCD factorization.}
\label{figKpi}
\end{figure}

We performed a fit to the hadronic parameters $\varepsilon_T, \phi_T, \varepsilon_C, \phi_C,
p_Z, \phi_Z$ and the NP parameters $(h_s^1, \sigma_s)$, using as input the 
data in Eq.~(\ref{Kpidata}). 
We allowed $\varepsilon_T$ and $\varepsilon_C$ to vary in the range
$[0,1]$ while $p_Z$ in $[0,0.6]$.
We also impose the following constraints on the strong phases, motivated by
factorization predictions in the heavy quark limit~\footnote{Without imposing these
constraints, we find that the most favored region is 
$h_s^1 \sim 2$ and $\sigma_s=0$. The reason for a good fit in this
case is that the NP effectively 
enhances the $\gamma$ contribution by interfering destructively with the SM 
QCD penguins, while having $ \phi_C \sim 0$ 
(opposite to the QCD factorization prediction) 
allows the $\gamma$ contribution to decrease $S_{K_S \pi^0}$ relative to $S_{J/\psi K_S}$.}:
\beq
\phi_C, \, \phi_T \in [90^\circ,270^\circ] \qquad \phi_Z \in [-90^\circ,90^\circ]
\eeq
We present in Fig.~\mbox{\ref{figKpi}(a)} the constraints on the NP
parameters $(h_s^1,\sigma_s)$ following from this analysis and in
Fig.~\mbox{\ref{figKpi}(b,d)} the corresponding result from the QCD
factorization analysis (with and without the use of branching
fractions). Finally Figs.~\mbox{\ref{figKpi}(c,e,f)} present the
constraints obtained by combining the above results with $\phi K_S, \eta' K_S$ 
data. In the case of the $SU(3)$ $K\pi$ analysis we repeat the combined
fit either using naive or QCD factorization for the $\eta',\phi K$ channels.

\section{Correlations}\label{Correlations}

\subsection{Adding together $\Delta F=2$ and $\Delta F=1$}

\begin{figure}[!htp]
\begin{center}
$\begin{array}{c@{\hspace{0.2in}}c}
\includegraphics[width=.5\textwidth]{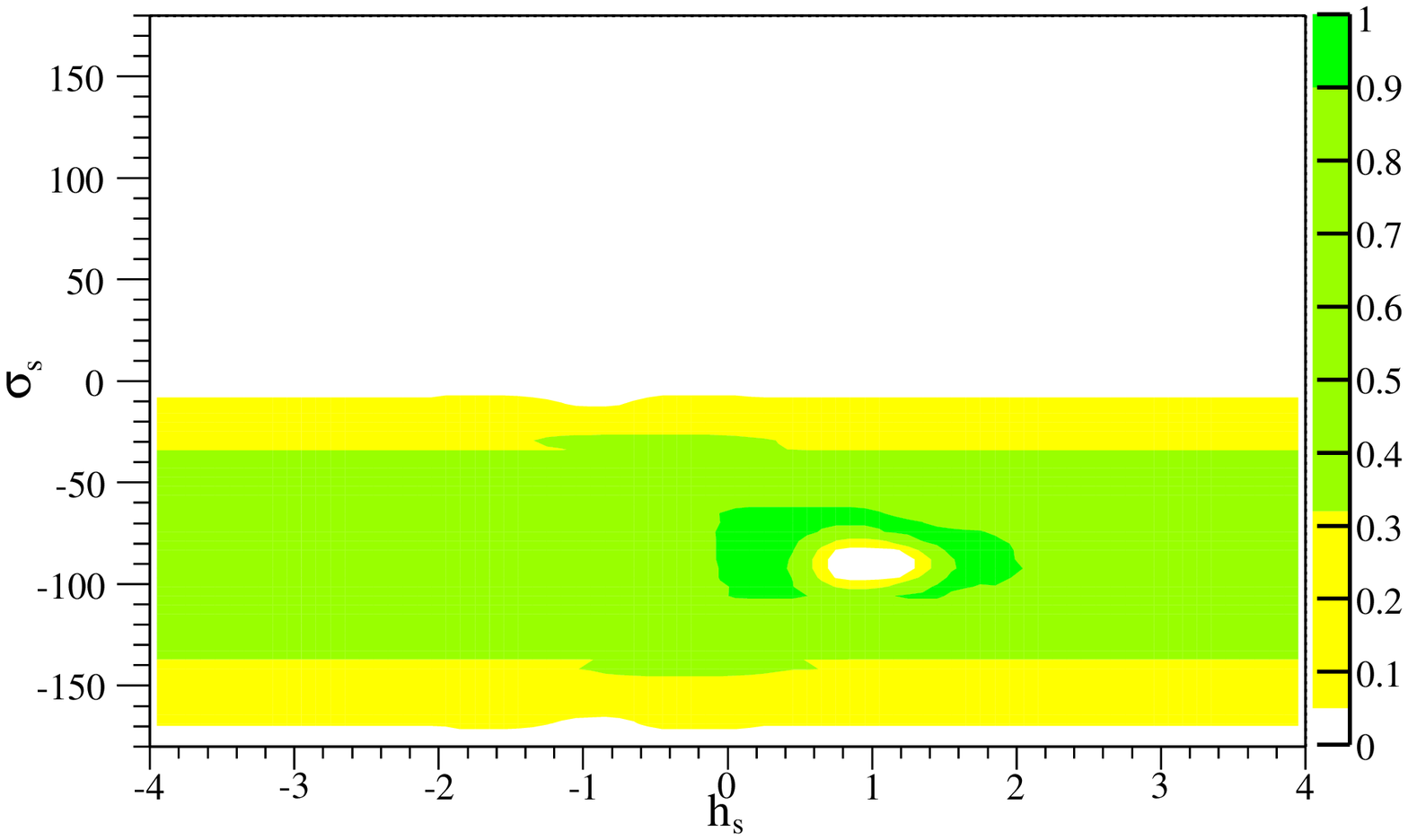} & 
\includegraphics[width=.5\textwidth]{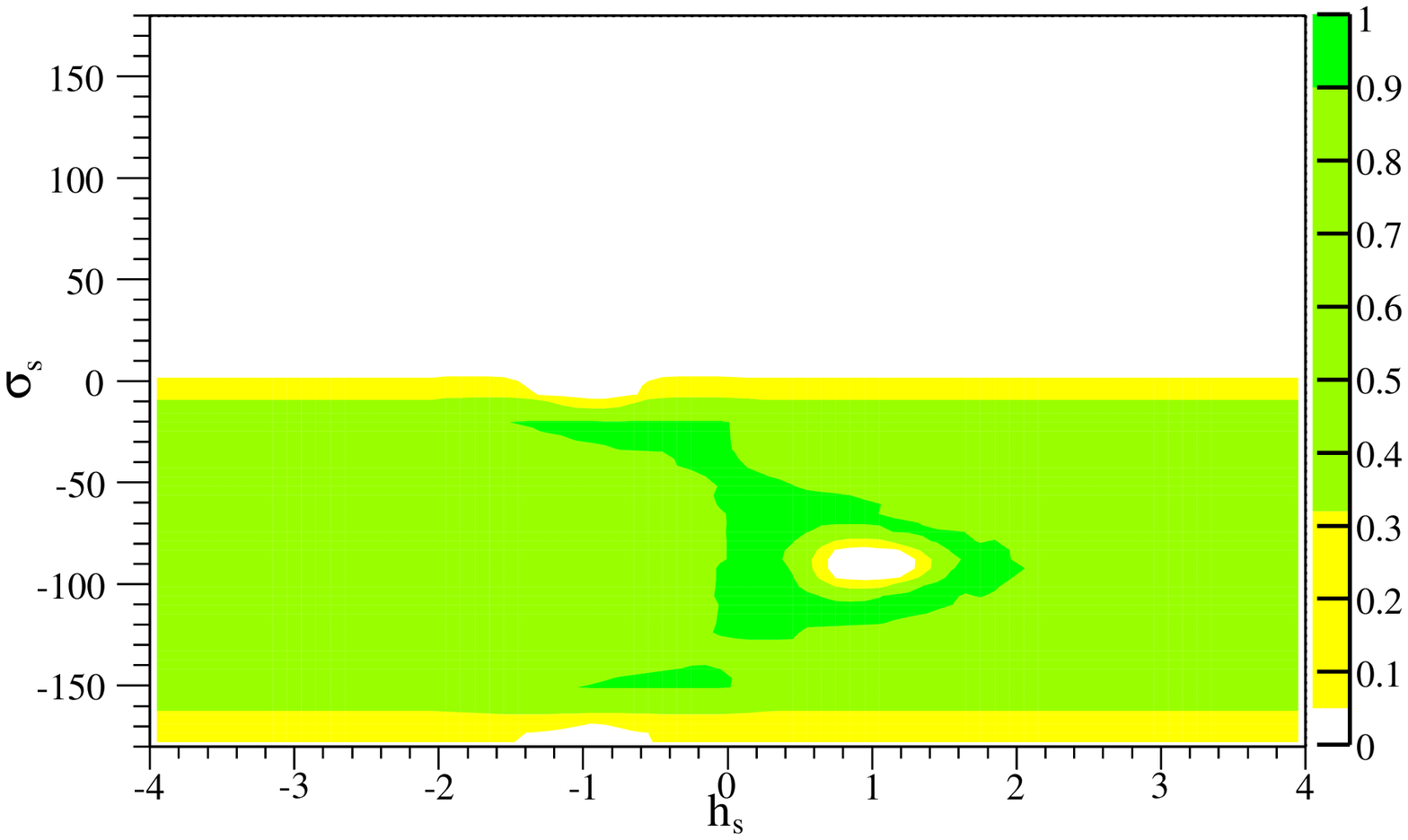} \\ 
\mbox{\bf (a)} & \mbox{\bf (b)} \\
\includegraphics[width=.5\textwidth]{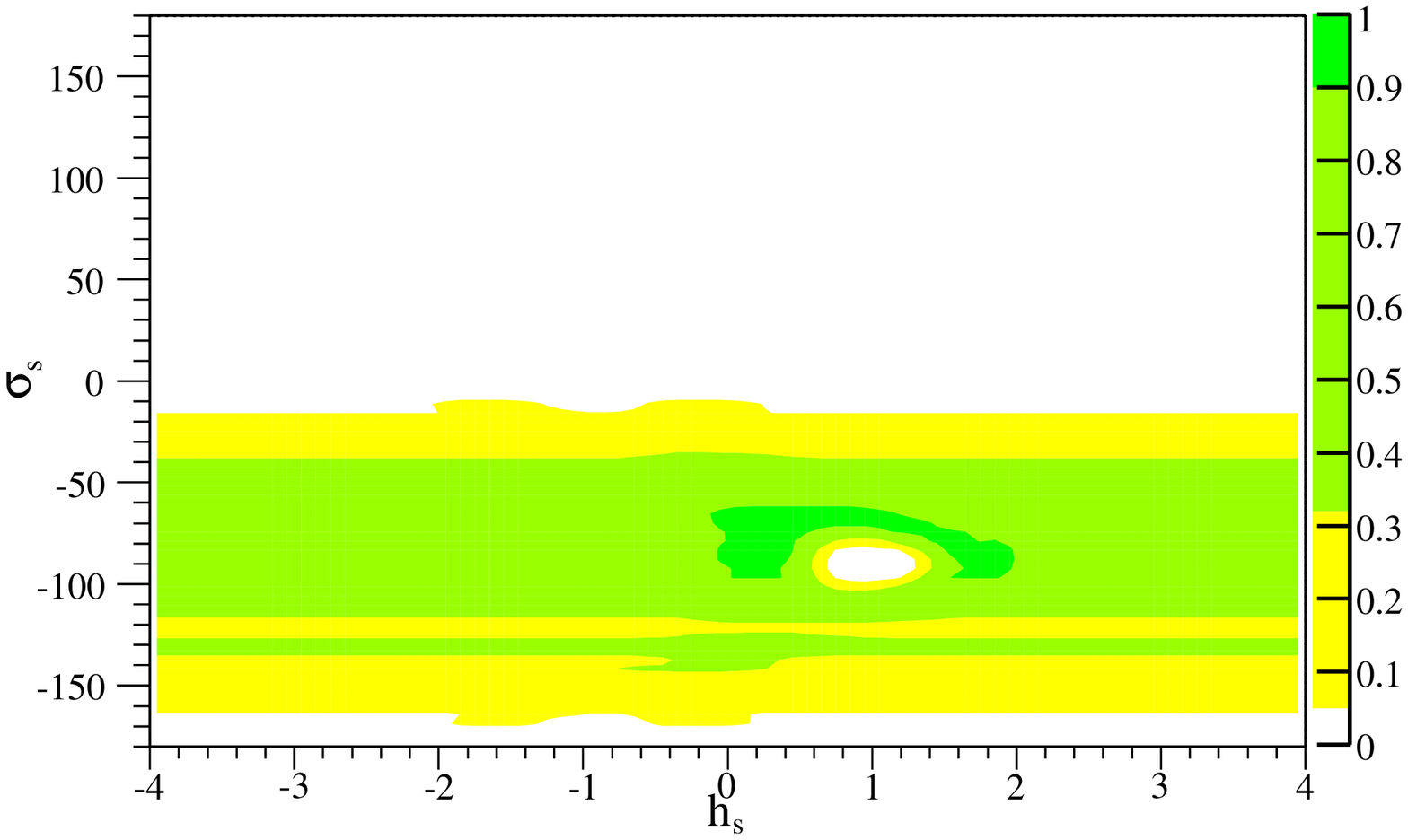} & 
\includegraphics[width=.5\textwidth]{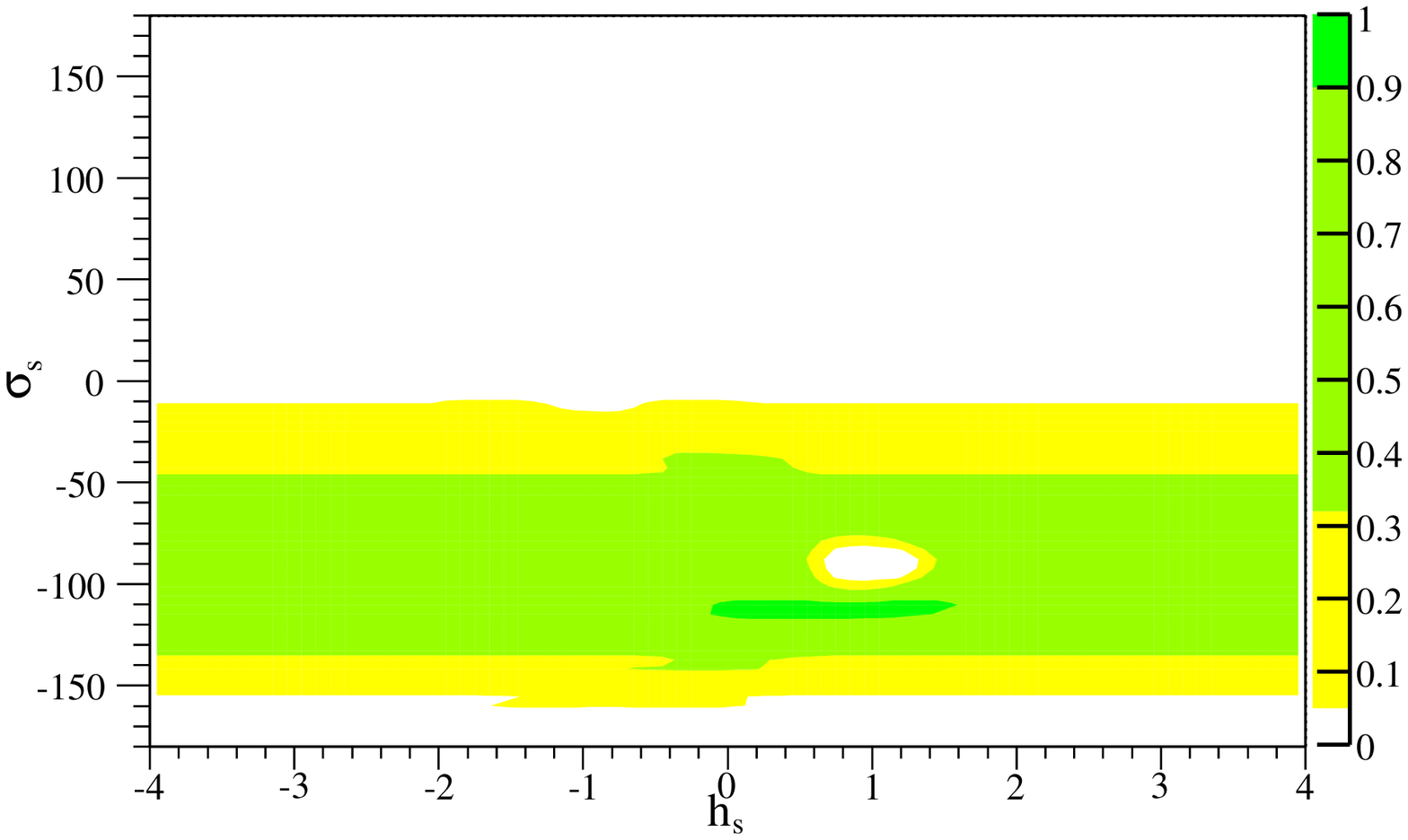} \\ 
\mbox{\bf (c)} & \mbox{\bf (d)} \\
\end{array}$
\end{center}
\caption{The allowed range for $h_s-\sigma_s$ combining the present data on
  $\Delta m_s$ with the bound on $\sigma_s$ coming from the $\Delta F=1$
  analysis.
  The first row is obtained considering $B\rightarrow \phi K_S, \eta' K_S$ only, while
  in the second row the $K \pi$ data is also added. Left column: naive factorization
  results. Right column: QCD factorization results. Naive
  factorization in $\phi,\eta' K_S$ is combined with the $SU(3)$
  analysis in $K \pi$.}
\label{fig:combms}
\end{figure}

\begin{figure}[!htp]
\begin{center}
$\begin{array}{c@{\hspace{0.2in}}c}
\includegraphics[width=.5\textwidth]{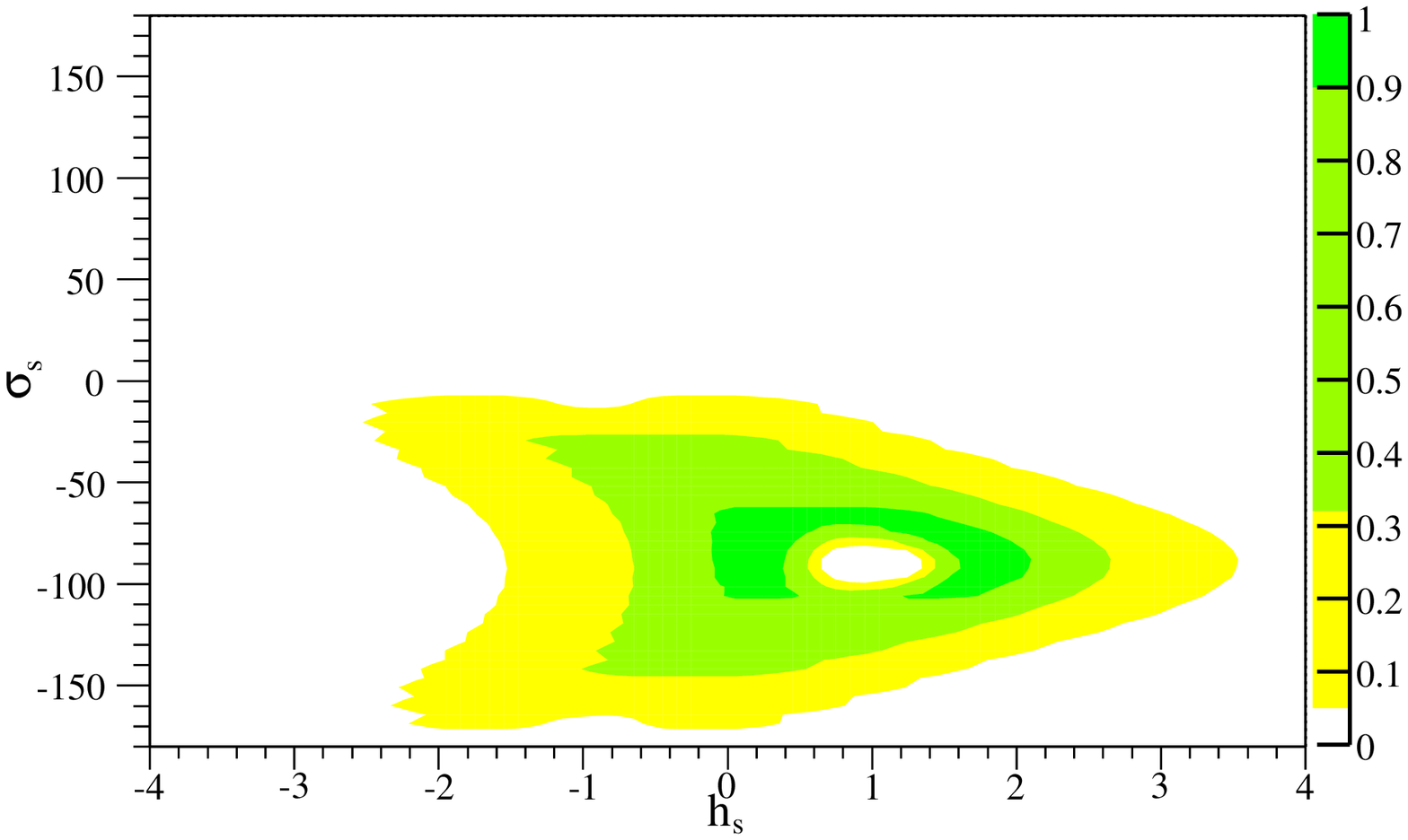} & 
\includegraphics[width=.5\textwidth]{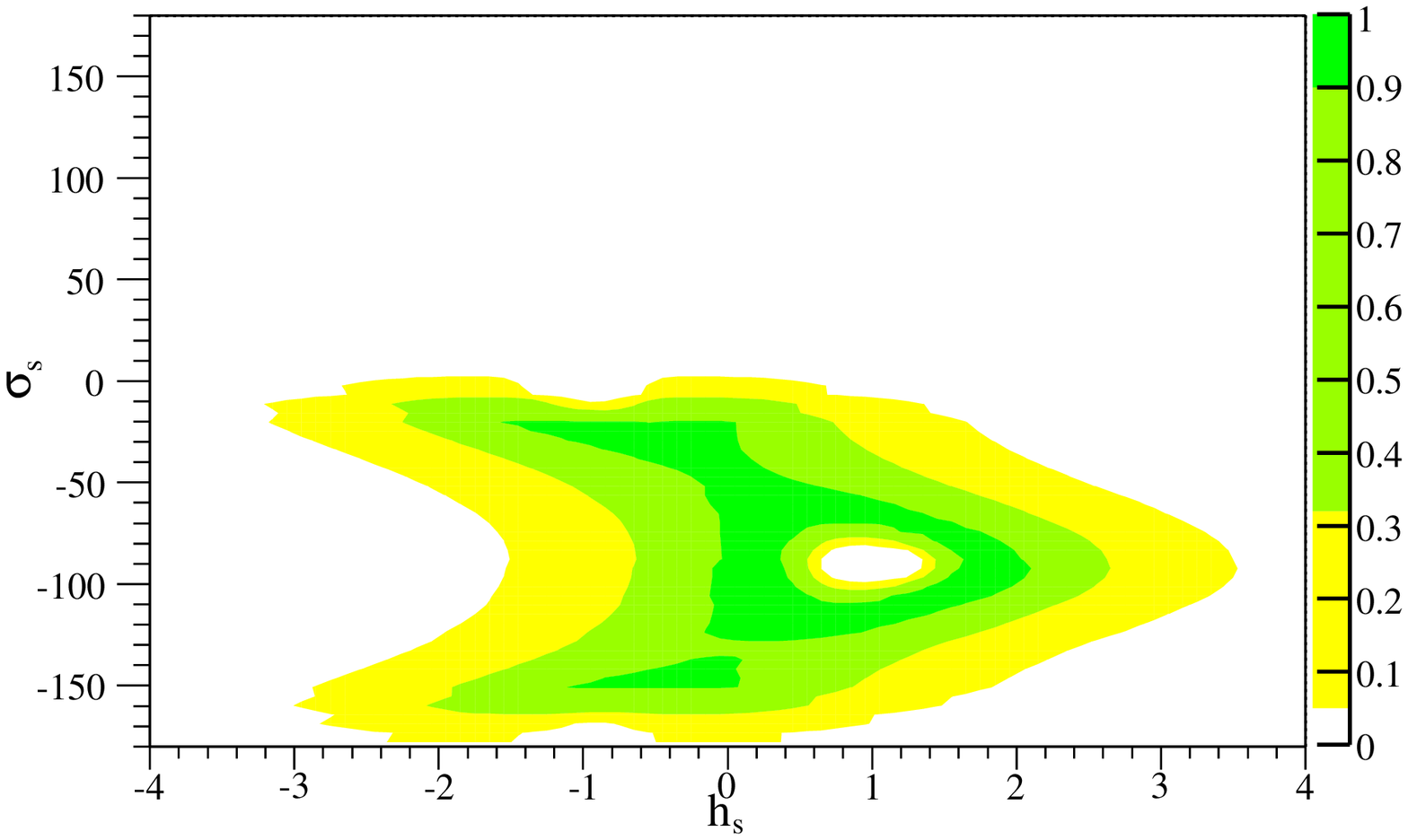} \\ 
\mbox{\bf (a)} & \mbox{\bf (b)} \\
\includegraphics[width=.5\textwidth]{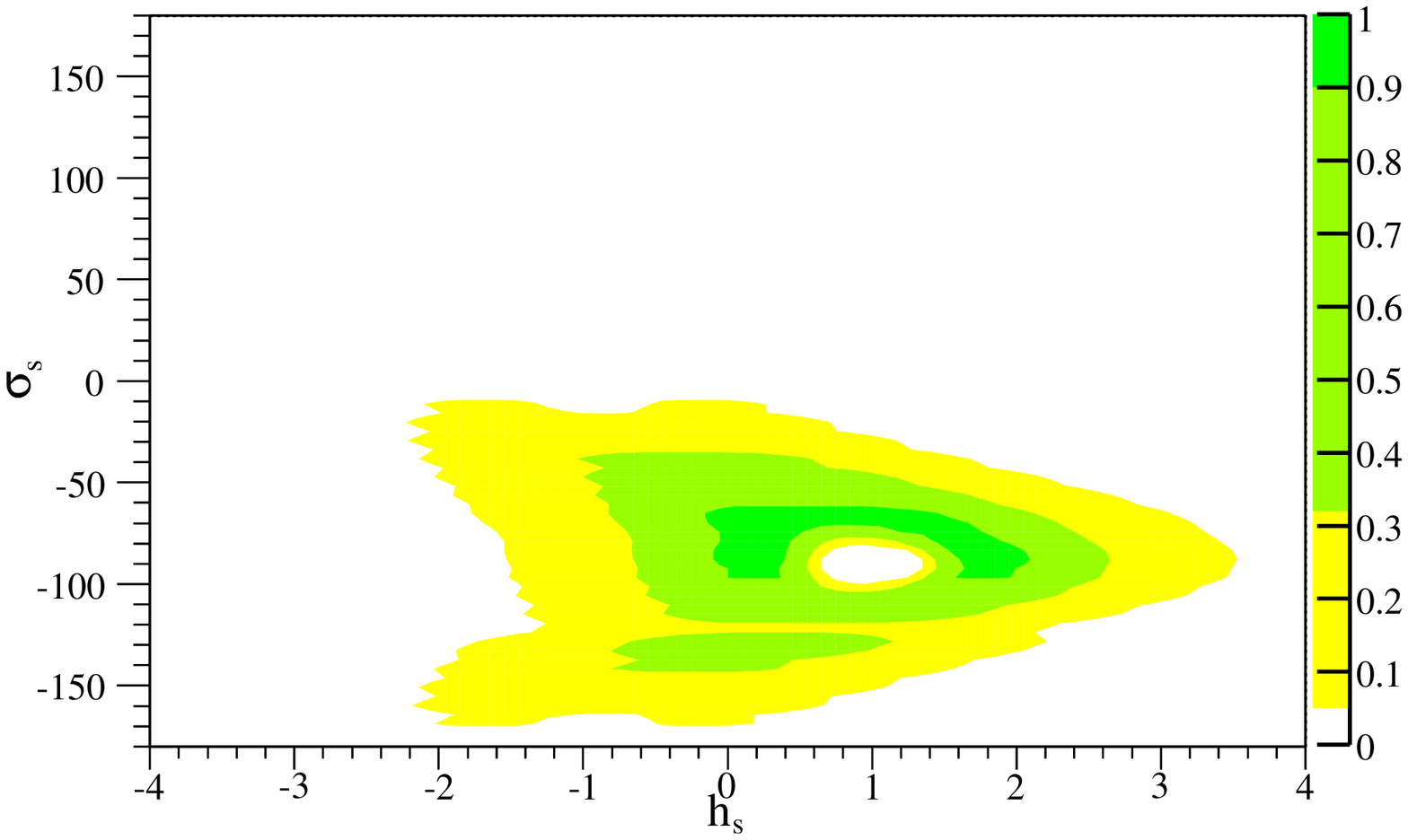} & 
\includegraphics[width=.5\textwidth]{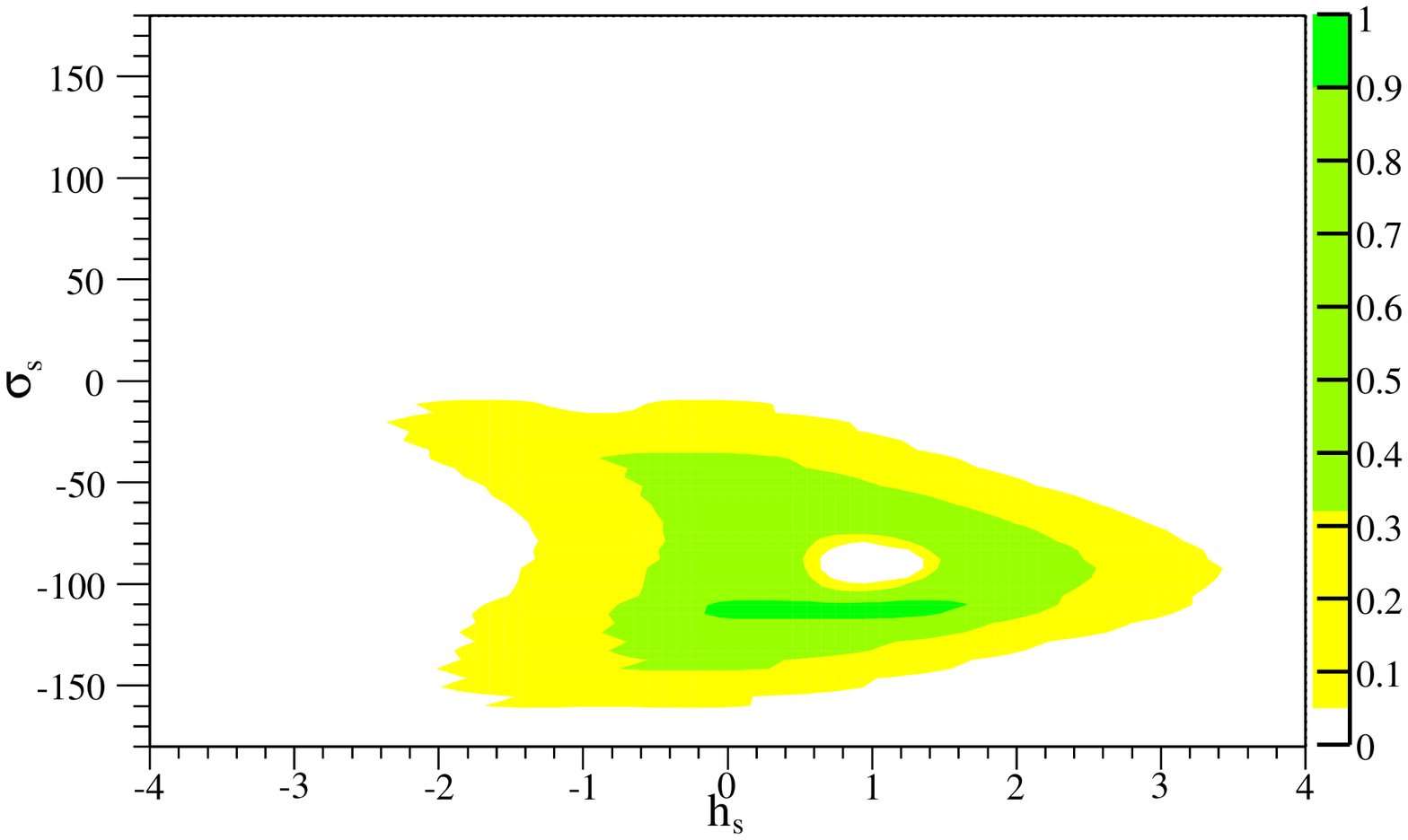} \\ 
\mbox{\bf (c)} & \mbox{\bf (d)} \\
\end{array}$
\end{center}
\caption{The allowed range for $h_s-\sigma_s$ combining a future measured
  $\Delta m_s=(18.3\pm0.3)ps^{-1}$ with the bound on $\sigma_s$ coming from $\Delta F=1$
  analysis.
  The first row is obtained considering $B\rightarrow \phi K_S, \eta' K_S$ only, while
  in the second row the $K \pi$ data is also added. Left column: naive factorization
  results. Right column: QCD factorization results. Naive
  factorization in $\phi,\eta' K_S$ is combined with the $SU(3)$
  analysis in $K \pi$.}
\label{fig:combmsfut}
\end{figure}

Since in our framework NP enters with the same phases in both $\Delta
F=2$ and $\Delta F=1$ processes, one might try to combine together the
constraints we obtained in Sections \ref{sec:deltaf2} and
\ref{sec:deltaf1}. We will do it here for the $b\rightarrow s$
transitions only, because it is where we have interesting
constraints both in $\Delta F=1$ and $\Delta F=2$ sectors.
The combined $\phi K_S$, $\eta' K_S$ analysis provided us with an
allowed range for $\sigma_s$. We can use  this constraint together with the
$\Delta m_s$ bound to get a restricted allowed region in the
$h_s-\sigma_s$ plane. This is illustrated in Fig.~\ref{fig:combms} for
the present bounds on ($h_s,\, \sigma_s$) (Fig.~\mbox{\ref{fighsss}(a)})
and in Fig.~\ref{fig:combmsfut} for future bounds coming from a
measured $\Delta m_s$ (Fig.~\mbox{\ref{fighsss}(b)}).  

Even if the combination does not constrain the NP ($h_s, \sigma_s$) parameters too much,
experimental improvements on $\Delta m_s$ can change the situation
dramatically.

\begin{figure}[tcbh]
\begin{center}
$\begin{array}{c@{\hspace{0.2in}}c}
\includegraphics[width=.5\textwidth]{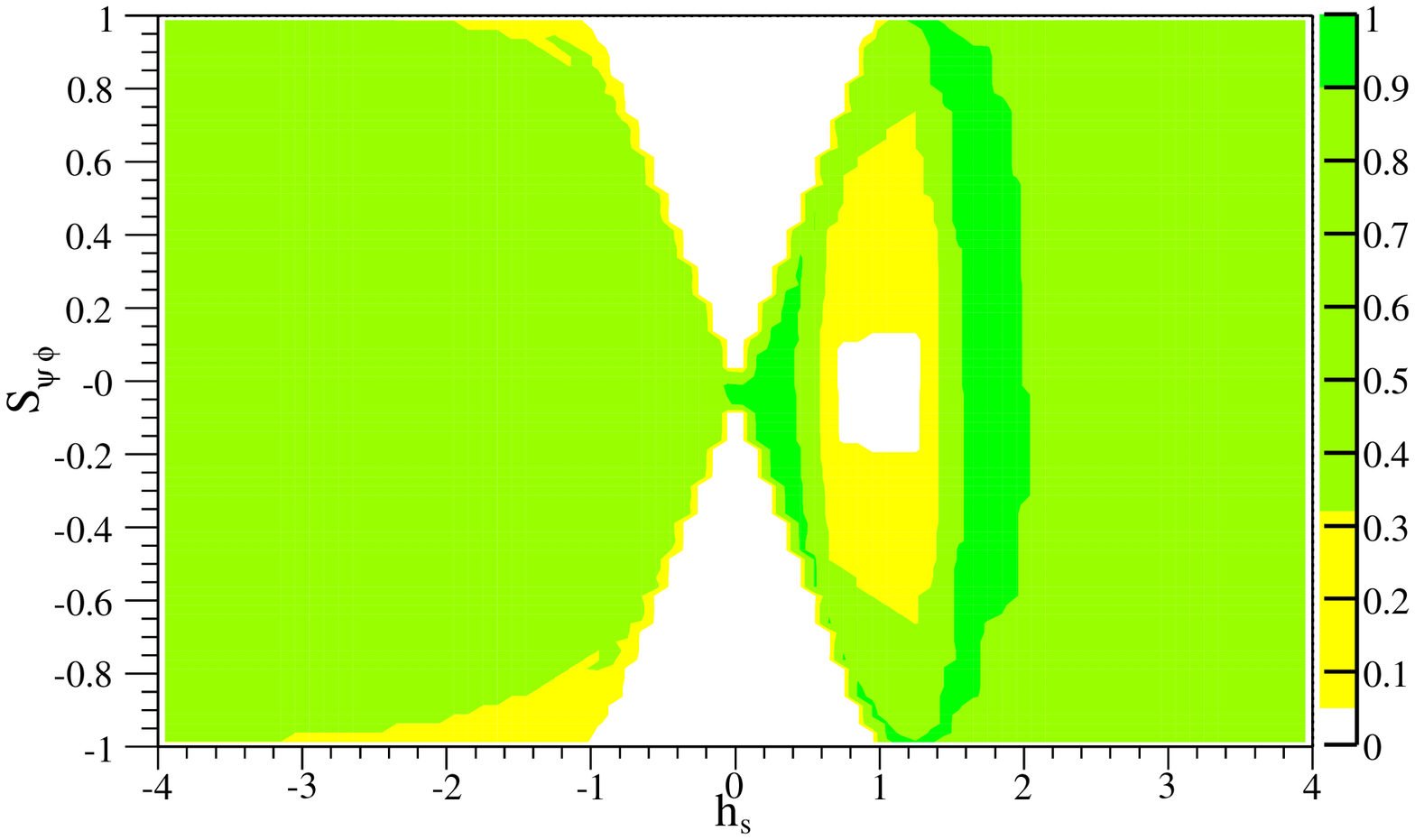} & 
\includegraphics[width=.5\textwidth]{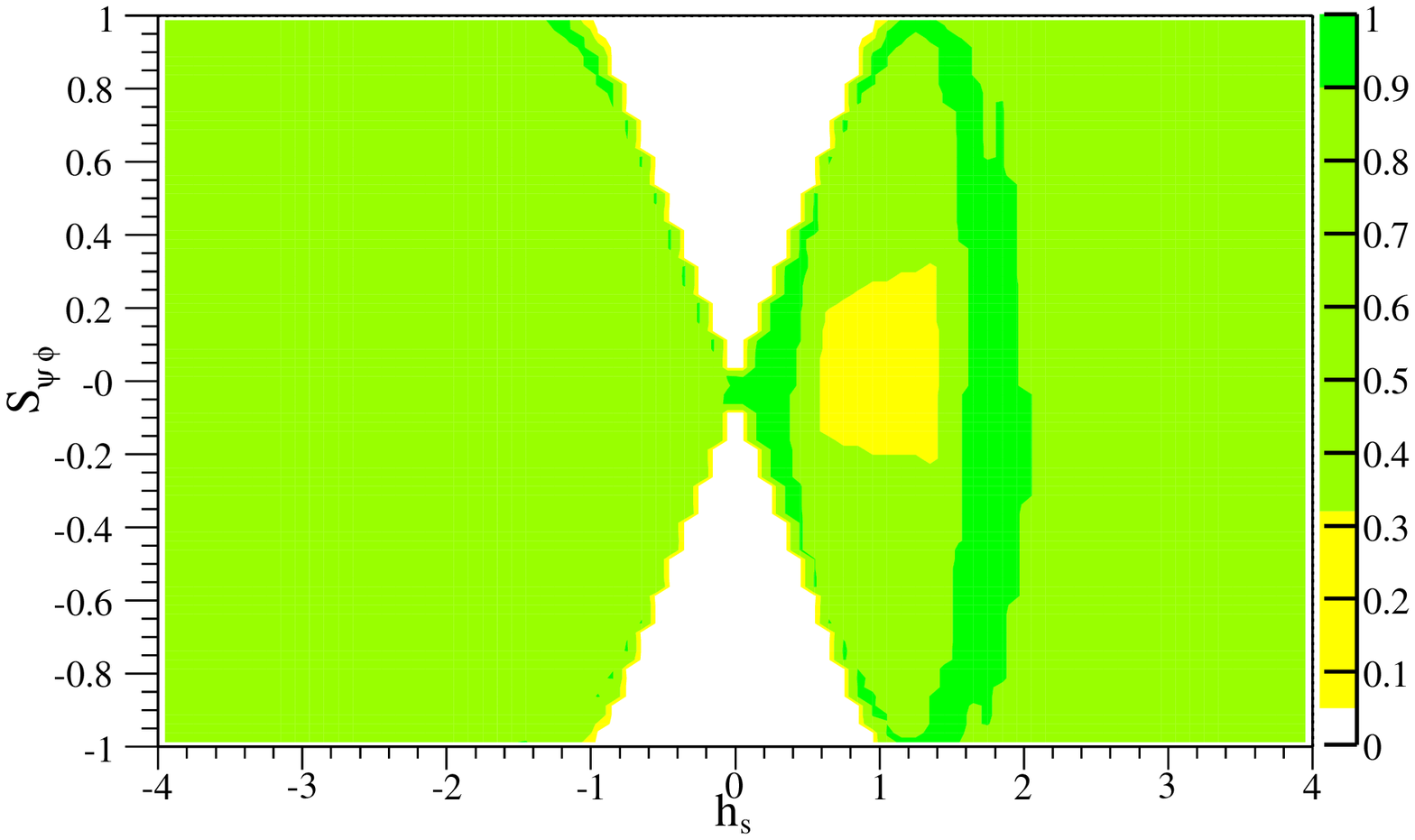} \\ 
\mbox{\bf (a)} & \mbox{\bf (b)} \\
\includegraphics[width=.5\textwidth]{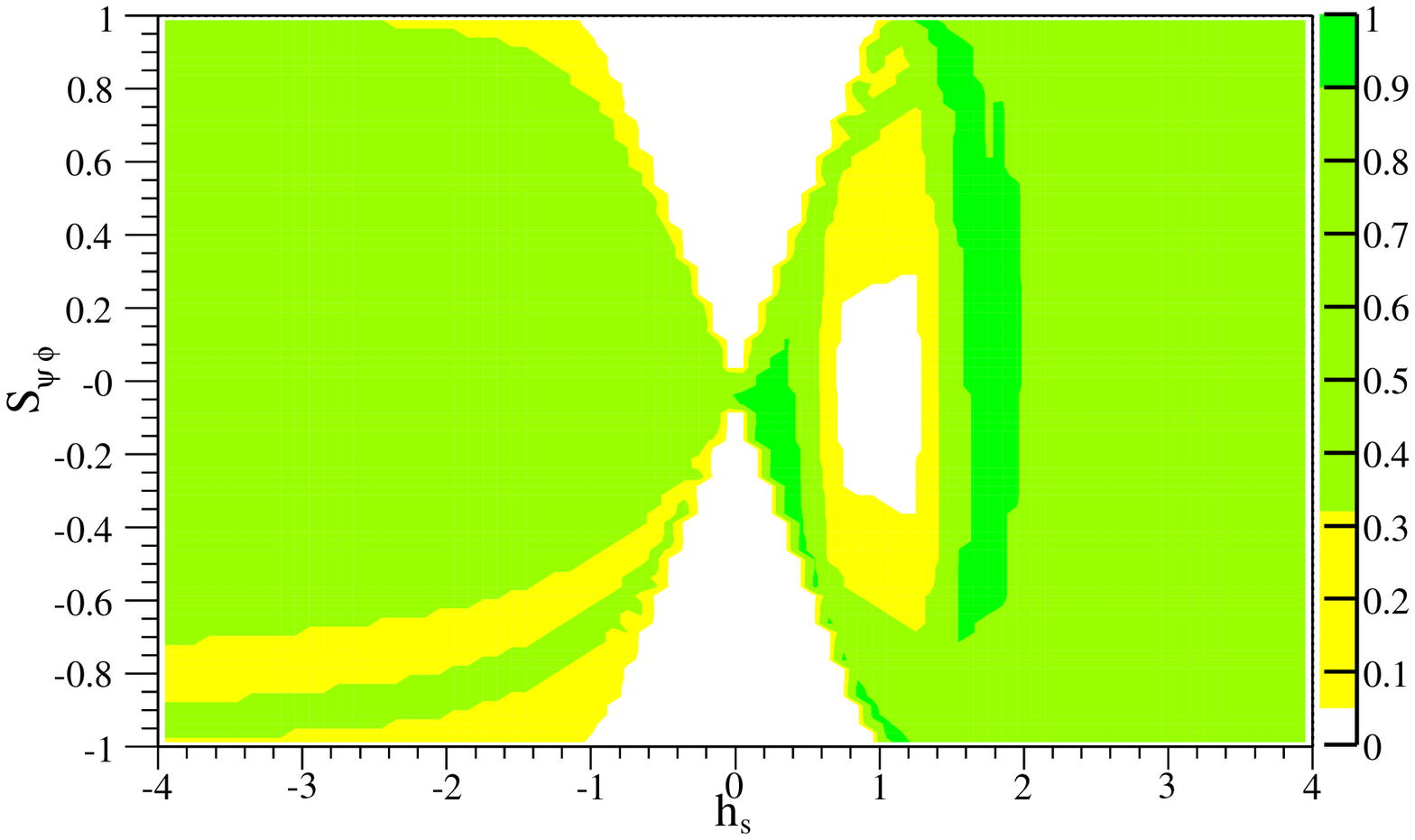} & 
\includegraphics[width=.5\textwidth]{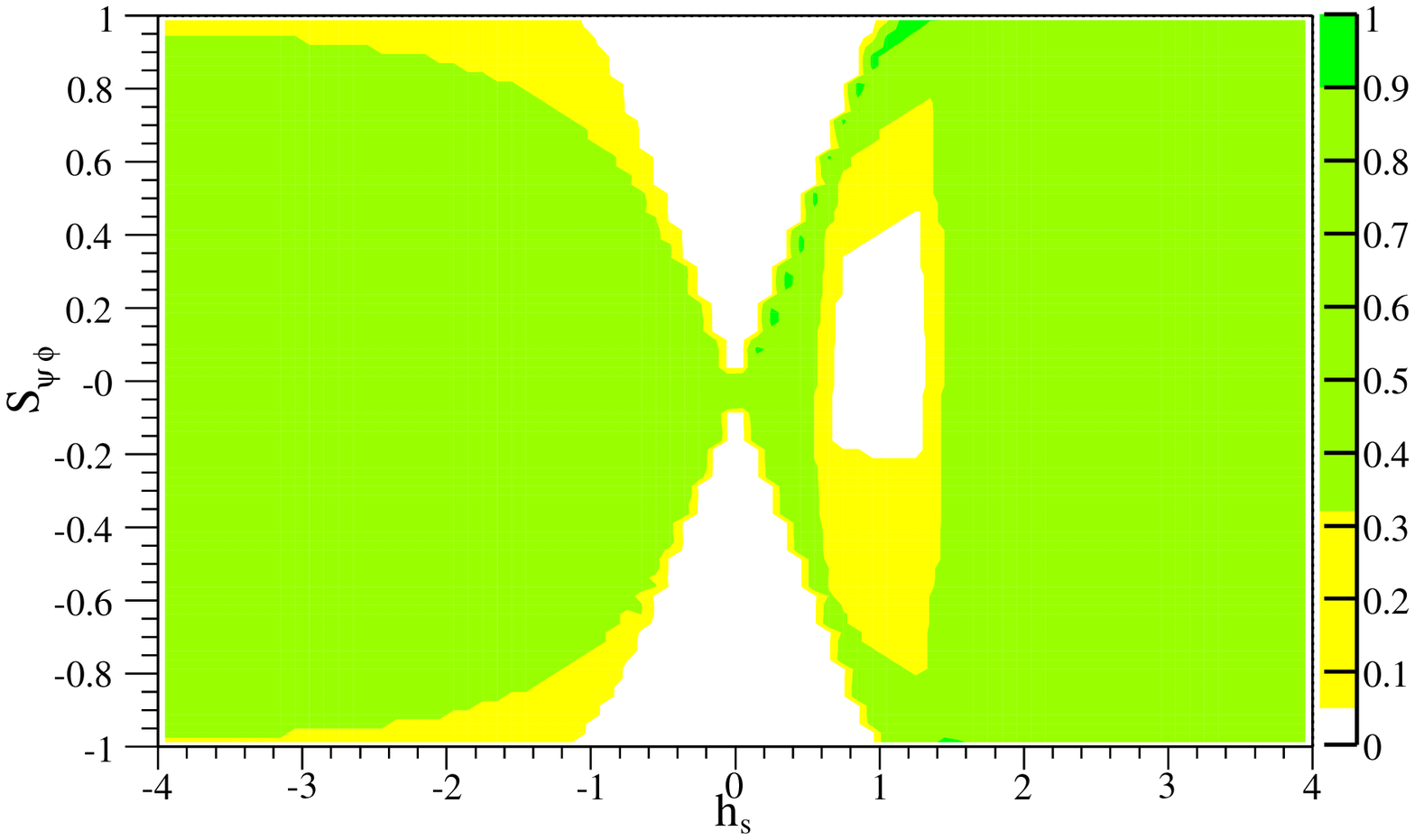} \\ 
\mbox{\bf (c)} & \mbox{\bf (d)} \\
\end{array}$
\end{center}
\caption{The allowed values of $S_{\psi\phi}$ as a function of $h_s$
  for the CL regions of Fig.~\mbox{\ref{fig:combms}(a-d)}.}
\label{fig:psiphi}
\end{figure}

\begin{figure}[tcbh]
\begin{center}
$\begin{array}{c@{\hspace{0.2in}}c}
\includegraphics[width=.5\textwidth]{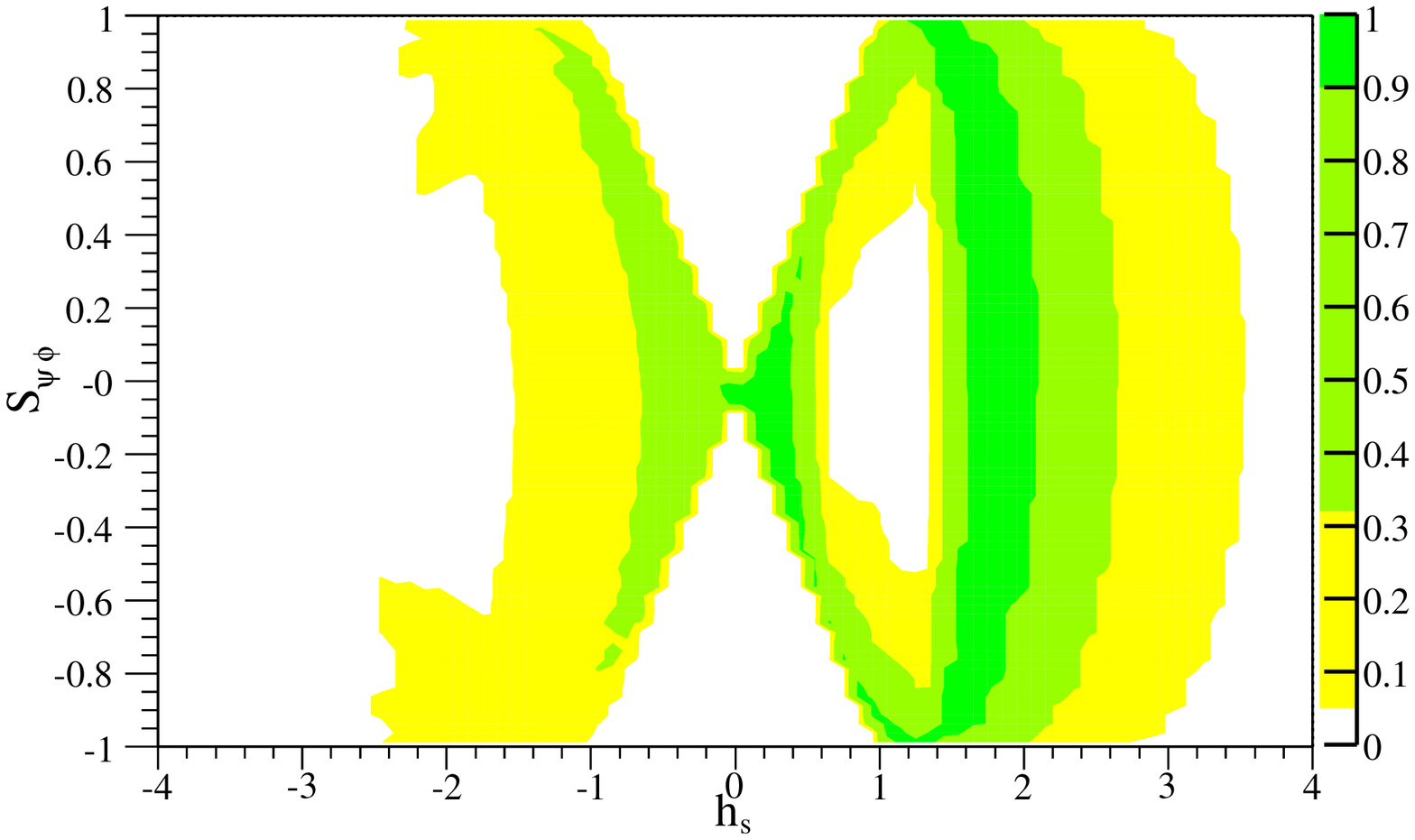} & 
\includegraphics[width=.5\textwidth]{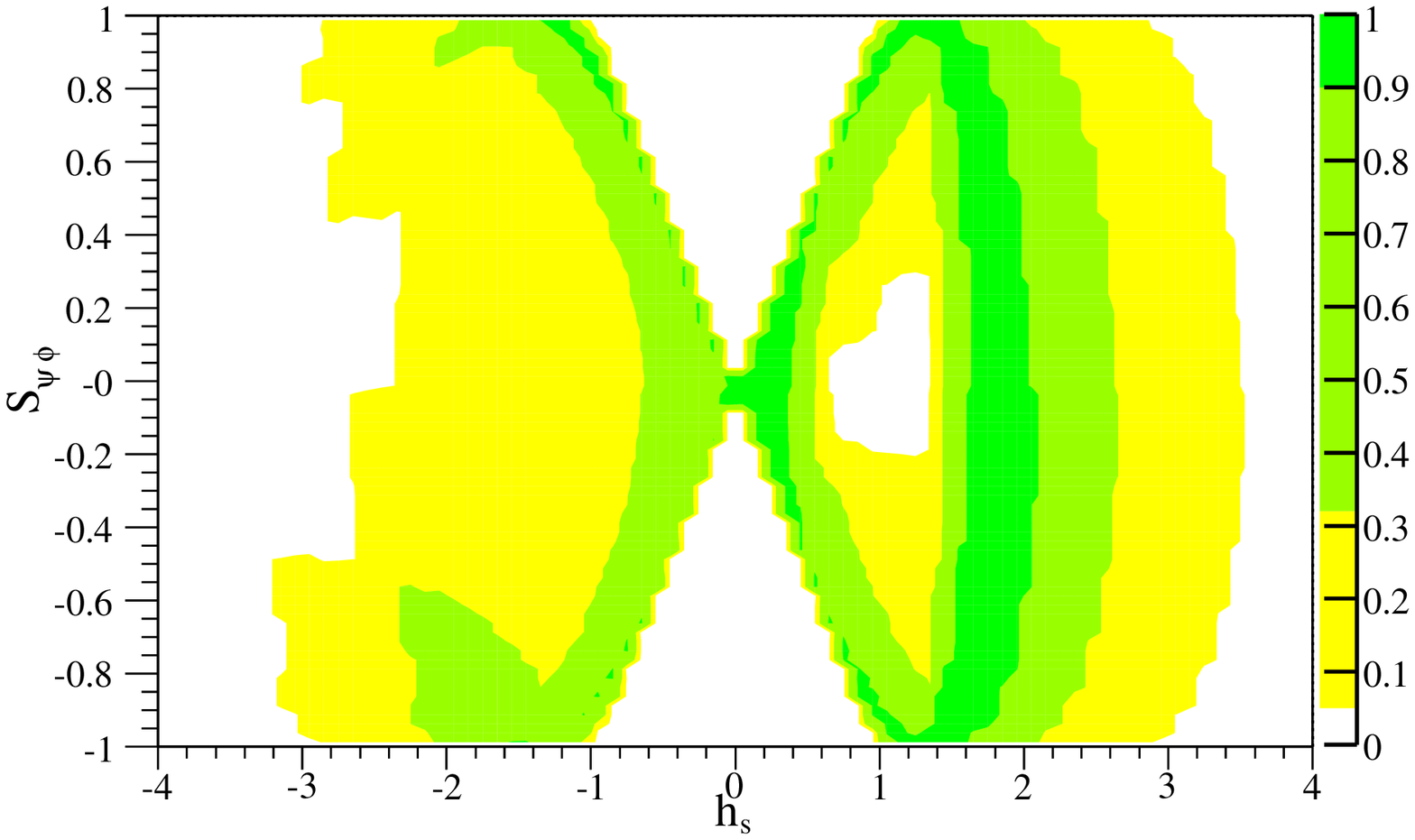} \\ 
\mbox{\bf (a)} & \mbox{\bf (b)} \\
\includegraphics[width=.5\textwidth]{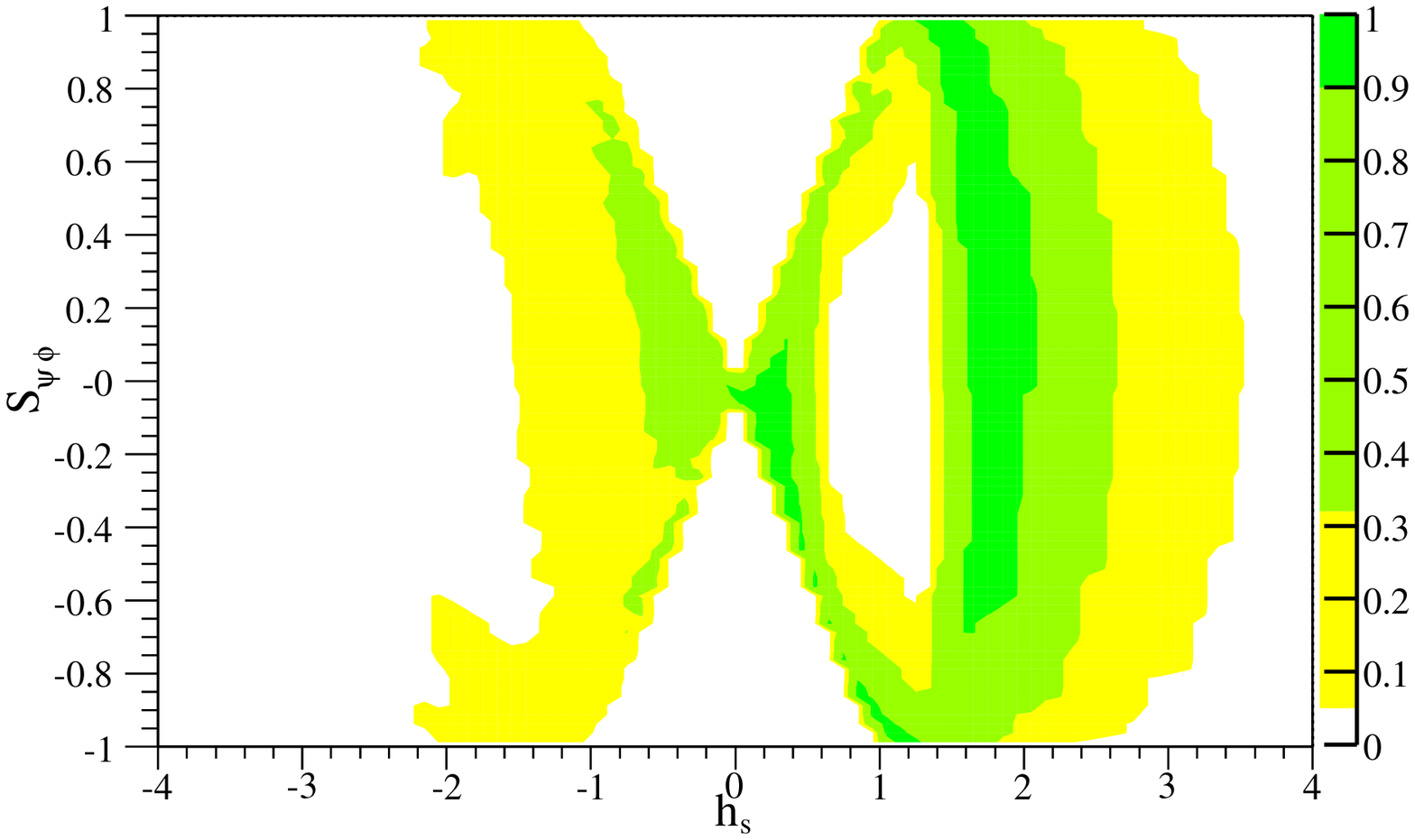} & 
\includegraphics[width=.5\textwidth]{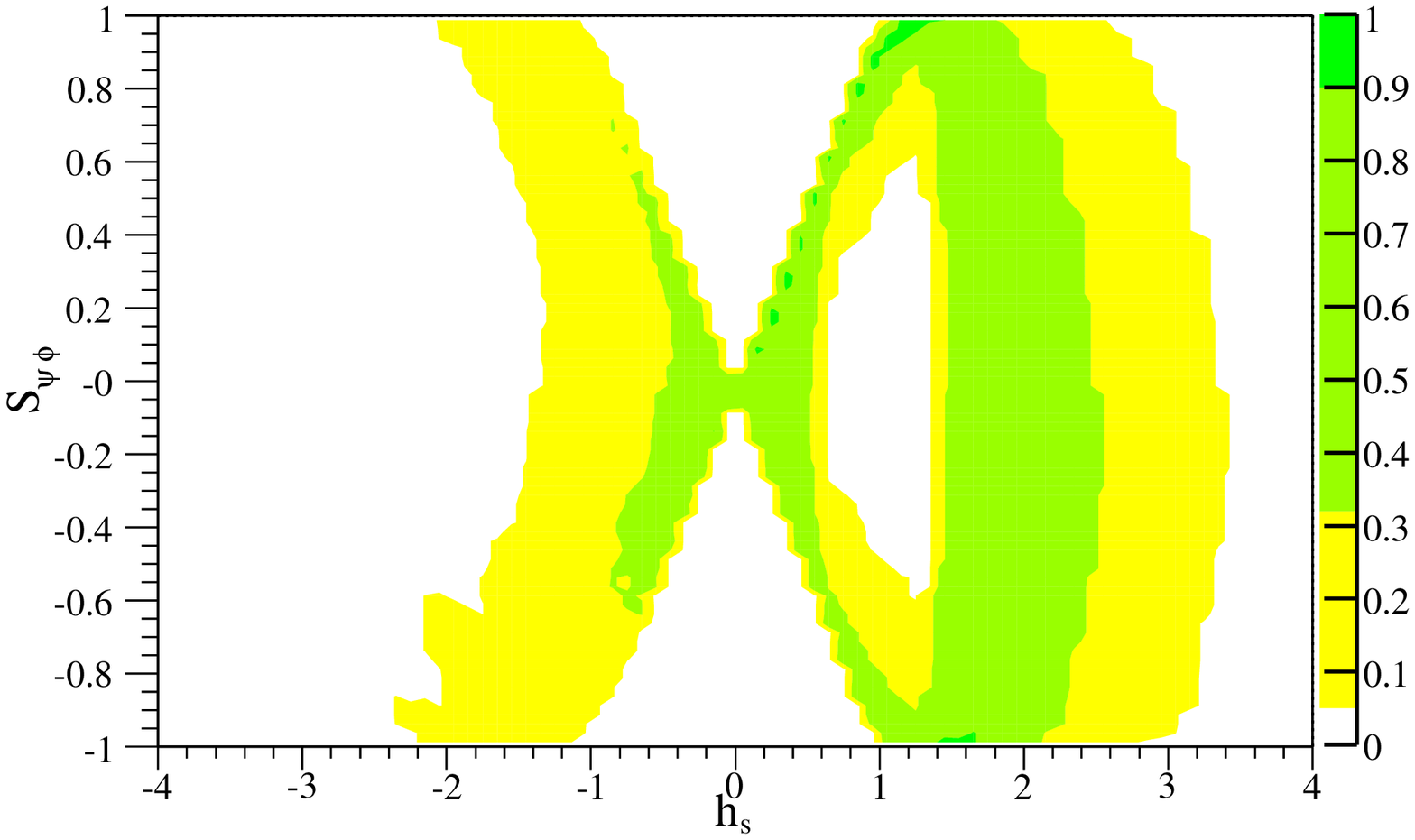} \\ 
\mbox{\bf (c)} & \mbox{\bf (d)} \\
\end{array}$
\end{center}
\caption{The allowed values $S_{\psi\phi}$ as a function of $h_s$
  for the CL regions of Fig.~\mbox{\ref{fig:combmsfut}(a-d)}.}
\label{fig:psiphifut}
\end{figure}

Given this more restricted region in $h_s-\sigma_s$ plane, we can look for correlations with
$S_{\psi \phi}$ that will be measured in the future. This is shown in
Fig.~\ref{fig:psiphi} for the CL regions of
Fig~\mbox{\ref{fig:combms}(a-d)} and in Fig.~\ref{fig:psiphifut} for Fig~\mbox{\ref{fig:combmsfut}(a-d)}, where we have represented
the allowed values of $S_{\psi\phi}$ as a function of $h_s$. 
The crucial point to
understand the plots of Fig.~\ref{fig:psiphi} is that
$\Delta F = 1$ analysis prefers the 
region {\em centered} (roughly) around $\sigma_s \approx 
-90^{\circ}$ which results in
constructive (destructive) interference (for $h_s < (>) 0$). Thus, 
$S_{ \psi \phi } \sim 0$ (see Eq. (\ref{par})) near the center
of the allowed range of $\sigma_s$. As we vary
$\sigma_s$ in preferred region and move away from center, both signs for 
$S_{ \psi \phi }$ are obtained. Thus, for the region preferred by the current
$\Delta F = 1$ data, we cannot make a prediction for 
$S_{ \psi \phi }$!
In particular and as mentioned earlier, $h_s \sim +1$
is excluded for $\sigma_s \approx
-90^{\circ}$, {\it i.e.}, near the center of the preferred region so that
$S_{ \psi \phi } \sim 0$ is not allowed for $h_s \sim +1$. This is
due to the large destructive interference reducing $\Delta m_s$
below limit.
Note that $\sigma_s \sim -45^{\circ}, -135^{\circ}$ 
are allowed  which corresponds to the 
maximally misaligned phase ($2 \sigma_s
\sim 5 \pi / 2, 7 \pi / 2$) for NP (relative to SM). This implies that
for these values of $\sigma_s$
and for $| h_s | \gtrsim 1$, the maximal ($\sim \pm 1)$ 
$S_{ \psi \phi }$ is obtained. In particular, this holds
even for $h_s \sim + 1$ (although, as mentioned above, $S_{ \psi \phi}
\sim 0$ is excluded for $h_s \sim +1$).
Finally, it is clear from Eq. (\ref{par}) that for
$| h_s | < 1$ ({\it i.e.}, small NP), again both signs for $S_{ \psi \phi }$
are allowed, 
but $S_{ \psi \phi }$ is restricted to be small (i.e., cannot be
maximal).


\subsection{Relations among different flavor transitions}

We now want to address the question whether it is possible to relate NP
appearing in {\em different} kind of flavor transitions, say, $b \rightarrow s$
and $b \rightarrow d$. Our framework
already assumes that given a {\em specific} flavor transition, the flavor
violating currents in the $\Delta F=2$ and $\Delta F=1$ processes have
the same origin. The $\Delta F=1$ processes depend also on the
flavor {\em diagonal} current and for this reason we distinguished among
$h_X$ and $h_X^1$. However in most of the models where NP in flavor
violating processes enters through a tree-level exchange or a
penguin-like process, the properties of flavor diagonal vertex does
not really depend whether the flavor violating vertex is a
$b\rightarrow s$ or $b \rightarrow d$ or $d \rightarrow s$ transition.
In particular, this is certainly true for models
with $Z$-alignment like the RS1 models.
With this additional input we are able to relate different flavor
transitions. In fact this implies that
\beq
  \label{eq:1}
  \sqrt{\frac{h_s}{h_d}}\frac{h_d^1}{h_s^1}=1 \qquad
  \sqrt{\frac{h_s}{h_K}}\frac{h_K^1}{h_s^1}=k \qquad \sqrt{\frac{h_d}{h_K}}\frac{h_K^1}{h_d^1}=k
\eeq
where the constants depend only on the flavor diagonal vertex and are
determined for a given model. For example, in the case of $Z$-alignment $k$
is the ratio between the neutrino Z-charge (since we have defined
$h_K^1$ for $K\rightarrow \pi \nu \bar \nu$) and the down quark Z
charge.
The origin of this relation is clear: the details of the flavor
violating vertex cancel in the ratio $\sqrt{h_X}/h_X^1$
({\it i.e.}, $\Delta F =2$ and $\Delta F =1$ for {\em same} transition), while the
details of the flavor diagonal vertex cancel in the ratio between two
{\em different} $h_X^1$'s under the assumptions above.
In principle testing these relations can provide a sensitive probe of the
subclass of NMFV models we analyzed here. However experimentally this is not
so simple since we need enough information in both $\Delta F=2$ and
$\Delta F=1$ for \emph{two} different transitions. Presently only $\Delta
F=1$ $b\rightarrow s$ is constrained by data while $h_K^1$ and $h_d^1$
are unconstrained. This means that at the moment we can use these
relations, within
our framework, only to say something about the expected size of NP  in
$\Delta F=1$  $b \to d$ and $d \to s$ processes, given our knowledge
in $b\to s$ transitions.

\section{Flavor structure \&  implications for SUSY models 
  and the RS1 framework}\label{RS1}
In this part we try to explain in more detail what we mean
by the NMFV framework: in particular, we give a more systematic 
analysis for the flavor structure of NMFV models. We also describe various flavor
models
and demonstrate that they belong to the NMFV class.

\subsection{Flavor structure of the NMFV}

Our main theoretical motivation to introduce the NMFV framework
is related to the following observation, regarding models which solve
the hierarchy problem and account for the flavor puzzle:
in many cases only the third
generation couples strongly to the NP sector in order to account for 
the heaviness of the top
quark. This usually implies that
the flavor violation is quasi-aligned with the SM flavor
sector. Consequently
the usual tension with FCNC measurements is largely avoided even though
the NP is at few TeV: the most stringent
constraint on such low flavor scale is from $K - \bar{K}$ mixing which is suppressed 
in this case since
NP couples weakly (or degenerately) to 1st and 2nd generation (see later for effects
of small (approximately degenerate)
couplings).
Below we shall try to make a more precise definition of the NMFV
framework by showing the operators induced.
The above definition implies that generically we expect the NP terms to conserve, to
leading order (see section VIB for sub-leading effects),
a U(2)$_Q\times$U(2)$_d\times$U(2)$_u\times$U(1)$_3$ subgroup of the SM
U(3)$_Q\times$U(3)$_d\times$U(3)$_u$ flavor group where U(1)$_3$
corresponds to an overall third generation charge.

Let us start with the generic form of the
helicity conserving, operators that are allowed
according
to our above description,\footnote{As mentioned in the introduction,
here we are concerned mostly with operators which
contributes to FCNC in the down type sector since this gives at
present and in the near future the most stringent constraints. Similar
analysis would apply for the up sector which is only mildly
constrained at present by the $D$-meson system since, with quasi-alignment and with 
$\Lambda_{
  \rm NMFV}\sim 2-3\,$TeV the current constraints are easily avoided.}
\beq
\left(\bar Q_3 Q_3\over \Lambda_{
  \rm NMFV}\right)^2\,,\ \ \ \ \ \left(\bar d_3 d_3\over \Lambda_{
  \rm NMFV}\right)^2 \,,\ \ \ \ \ \left(\bar Q_3 d_3\over \Lambda_{
  \rm NMFV}\right)^2\,,
\label{DelF2}
\eeq
where $d$ stands for a down type singlet quark and the Lorentz and
gauge structure is implicit so that each of the above operators
actually corresponds to several ones. Note that the operators are
given
in the special basis in which, by assumption, NP only couples to
the third generation to leading order.
The above operators are important for mediating $\Delta F=2$
processes. The scale $ \Lambda_{
\rm NMFV}\sim 2-3\,$TeV is equivalent to the one induced by a SM
electroweak loop. 
Note that the above operators are self hermitian and therefore
real in the special basis.

Although there are many operators (with more than one weak phase),
our analysis presented in section III 
is completely general and hence still useful and applicable
to this case.
Indeed, since such processes are governed by short distance physics, there
are no {\em relative} strong phases between
the matrix elements of these operators. Then it is possible to write 
the total amplitude collecting the above contributions in the form:
real parameter $\times$ strong phase\footnote{This common 
strong phase is of course irrelevant for the analysis.} $\times$ weak
phase. 
The latter is 
combination of weak phases (there are $5$ of them: see below) weighted by magnitude of
NP in the different operators
and (magnitude of) matrix element. Hence, our analysis (which assumes single weak phase)
effectively constrains the weak phase of this linear combination of
the above operators, whereas
other combinations are
unconstrained.

Let us now consider $\Delta F = 1$ transitions.
Assuming that the misalignment between the special basis and the mass
basis is at most a CKM-like rotation, it is clear that the above
operators
would give negligible contribution to $\Delta F=1$ process.
In general we expect that another large class of operators in which
flavor universality is broken by only two quarks would be present in
the theory (these are also self-Hermitian):
\beq
{\bar Q_3 Q_3 \bar Q_l Q_l \over \Lambda_{\rm NMFV}^2}\,,\ \ \ \ \
{\bar Q_3 Q_3 \bar d_l d_l \over \Lambda_{\rm NMFV}^2}\,,\ \ \ \ \
{\bar Q_3 Q_3 \bar u_l u_l \over \Lambda_{\rm NMFV}^2}\,,\ \ \ \ \
{\bar d_3 d_3 \bar Q_l Q_l \over \Lambda_{\rm NMFV}^2}\,,\ \ \ \ \
{\bar d_3 d_3 \bar d_l d_l \over \Lambda_{\rm NMFV}^2}\,,\ \ \ \ \
{\bar d_3 d_3 \bar u_l u_l \over \Lambda_{\rm NMFV}^2}\,,
\label{DelF1}
\eeq
where $l=1,2\,.$
In addition, generically, operators
which induce helicity flip are also expected,
\beq
{ \bar Q_3 d_3 \bar d_l Q_l\over  \Lambda_{
    \rm NMFV}^2}\,,\ \ \ \ \
&& {m_b\over  \Lambda_{
    \rm NMFV}^2}\, \bar Q_3\sigma^{\mu\nu} d_3 F_{\mu\nu}\,,\ \ \ \ \
  {m_b\over \Lambda_{
     \rm NMFV}^2}\, \bar d_3\sigma^{\mu\nu} Q_3 F_{\mu\nu}\,,\label{LR}
 \eeq
 where one should note that, apart from the first one, the dimension five operators are not SU(2) gauge invariant.
We allow ourself to add those since the exact form of the electroweak sector is yet to be
 determined and thus is left implicit (in principle we can make the
 dimension five operator with a Higgs field insertion).
 Furthermore based on the experimental constraints on helicity
 flipping processes
 such as $b\to s\gamma,l^+ l^-$ and the strange EDMs~\cite{hfag,PDG} we assume that
 the amplitude of these operators cannot be larger than the SM
 one. Hence, the chirality breaking scale of the coefficient of the operators in (\ref{LR}) is chosen to
 be $m_b\,.$

Given the most general flavor structure, presented in
Eqs. (\ref{DelF1},\ref{LR}), part of the predictive power is lost and
basically very little information can be extracted from the $\Delta
F=1$ transitions (apart from clear deviation from the SM predictions).
In particular in the generic case for
the down sector the above operators will induce five new CPV phases as follows.
There are a total of $18$ phases in the two Yukawa matrices. We can use
the U(2)$_Q\times$U(2)$_d\times$U(3)$_u\times$U(1)$_3$ transformations (under 
which NP operators are invariant)
to remove $13 - 1 = 12$ phases\footnote{Note that
since we are interested only in flavor violation in 
the down sector we can use the full $U(3)_u$ subgroup. Also,
one phase in these transformations corresponds to baryon-number
and hence does not result in reduction of total number of phases.} leaving the CKM phase
and $5$ new CPV phases.
Since we expect (process dependent) relative strong phases between the matrix elements
of the different operators (for $\Delta F = 1$ transitions), we cannot write the amplitude
in the form of a strong phase $\times$
weak phase (which is a combination of the $5$ weak phases) unlike in
the above case.
This implies in particular that there is effectively 
more than a single weak phase per
$\Delta F=1$ process. So, it seems
that our analysis (which assumes single weak phase
per transition) is not applicable even indirectly (for example, even with
a simple rescaling)\footnote{Even if matrix elements are known, 
since there are too many NP parameters ($5$ weak phases and many operators
and hence magnitudes of NP Wilson coefficients)
entering in process-dependent combinations, there is 
(currently) not enough $\Delta F = 1$ data to 
constrain all combinations in a model-independent way.}.
Also, since the new weak phases enter in a different combination in
$\Delta F = 1$ transition as compared to $\Delta F = 2$, we 
do not have a correlation between the effective weak phases entering
the two effects. 

Suppose however that, motivated by the absence of deviation from the
SM prediction in chirality
flipping processes, we assume that the operators in Eq. (\ref{LR})
are absent. Even in that case our theory would still contains four
new CPV phases\footnote{Since there are no dipole operators, we can perform
separate rotations on $Q_3$ and $d_3$ (unlike before) 
removing one more phase.} and again present data would not suffice to
yield a sensible constraint.

To obtain a manageable number of new CPV phases, we need to make further
assumptions. For example, if we assume that 
only a single type of operators ({\it e.g.} only the ones with
left handed or right handed flavor violation) are induced by the new
physics, then there are only two new CPV phases in the
model\footnote{We can now use the full $U(3)_d$ or $U(3)_Q$ group to remove $3$ more 
phases.}. 
Consequently we expect
our analysis above which assumes (to leading order)
a single weak phase per transition to yield an excellent constraint
on the above class of models (see below for more on this). 
The only difference being that since there are only $2$ new CPV phases
in this case
there is a relation between
the $3$ phases $\sigma_{K, d, s}$ used in our analysis 
(due to the fact that the $1-2$ flavor violation originates
via mixing with 3rd generation). It is also clear that
with the assumption
of only RH or LH operators,
we obtain correlation between $\Delta F = 2$ and $\Delta F = 1$ transitions.
The counting of phases is summarized in table
\ref{down}.

In this context we want to argue that models in which flavor violation
(or non-universality)
appears dominantly in the LH sector are motivated both theoretically
and experimentally as follows:
On the experimental side the only possible hint for deviation from the
SM is in the $b\to s$ penguin modes. It is interesting that the mean
values
of all the modes (but  $S_{f_0 K_S}$) is below the SM prediction which seems to 
favors LH models~\cite{Endo:2004dc}.
On the theoretical side, 
recall that the main motivations for considering such
a framework (in particular NP coupling dominantly 
to 3rd generation) was that the top quark is heavy and
hence it (and therefore its $SU(2)_L$ partner, the {\em left}-handed bottom) 
must couple strongly to the TeV scale
NP  related to EWSB: in particular, there is
no such motivation for $b_R$ to couple strongly to NP. 
It is quite plausible that the flavor violation
from the NP in {\em down} 
sector will reflect this fact and therefore be dominated by LH current.
Indeed in many flavor models that we describe below (but certainly not
in all of those) one typically finds strong tendency in this direction.
In that sense we might argue the the truly {\emph{ next to minimal
    flavor violation}} class of models is when flavor violation is
dominantly induced via left-handed operators.

\begin{center}
   \begin{table}[!hb]\begin{center}
\begin{tabular}{||c|c|c|c||}
\hline\hline
 &Generic& No left-right& Only left (or right)\\ \hline  
\# of new CPV phases&5&4&2\\
\hline\hline
\end{tabular}
\caption{\small Number of flavor-violating CPV phases for the three types of NMFV
  models starting from the generic case to the most restrictive one.
Only flavor violation for the down sector is considered.}
\label{down}\end{center}                               
 \end{table}\end{center}


We now explain why one last assumption is necessary is order to obtain constraints
from current data on $\Delta F = 1$ processes.
The point is that even with only LH structure for flavor violation, 
there are many operators due to the possibly different color structure
(more precisely there are six operators per transition, not shown for
simplicity in the discussion above). 
Although there is only a single weak phase,
appearing as an overall
factor in the total amplitude, as usual we need to know
the relative strong phases in the matrix elements of the different operators
in order to obtain constraints. This, in general, introduces lot of 
hadronic model-dependence. 
Actually, 
even if the matrix elements (including strong phases) are known
(say, using naive or QCD factorization), 
there are still too many NP parameters (there is only a single weak phase,
but six Wilson coefficients) and not enough data (as explained earlier). 
Thus, it is still 
difficult to constrain
generic NP (contributing to all operators with
independent strengths). This is inspite of the fact that it is clear that the correlation
in the weak phases in $\Delta F = 1$ and $\Delta F = 2$ transitions is 
still present.
Thus, in order to obtain constraints, we need further assumptions. 
Clearly if the NP only affects a smaller set of operators, then the
number of NP parameters and also hadronic 
model-dependence is reduced (since we need to know less number of
strong phases).
For example,
in our analysis of section IV, we assumed that NP appears only in
EWP that is aligned
with the SM $Z$ coupling.

Finally we want to comment about
how to relax the assumption regarding alignment of NP with the $Z$ couplings,
used in section IV when we analysed the $\Delta
F=1$ processes.
We claim that since the most stringent bound (at present and in
the near future) comes from  $B\to{\eta^{ \prime } K_S}$ 
for which the related strong phases are well constrained,
we can easily apply our results to other models belonging to
the NMFV class.
For example, if NP is present dominantly in QCD penguins (instead of $Z$ penguins),  
the constraint we obtained from a particular decay mode ({\it e.g.}
$\eta' K_S$) is still applicable by simply rescaling 
$h_s^1$ by ratio of matrix elements of QCD and $Z$ penguin operators
in that given mode.
This implies that, by using the channel yielding the most
stringent constraint, we can always
adapt our analysis to constrain other models
(provided that even in the latter models, the specific mode used  
yields the dominant constraint)\footnote{Of course, in general, the  
constraint obtained from the {\em combined} $b \rightarrow s$
data is no longer valid for such NP 
due to the ratio of matrix elements being process dependent.}.

\subsection{Relation with flavor models}

We now briefly show how SUSY non-abelian and alignment 
models
belong to this class and then discuss the connection with the RS1
framework in more details. For this purpose, we go beyond the strict
NMFV limit and consider NP coupling to 1st and 2nd generation also.
Consequently in the special interaction basis the Lagrangian mediated
by the NP degrees of freedom is given by
\begin{equation} 
{\cal L} = \sum_{i=1}^3 c_i \frac{ \left( \bar{Q}_i Q_i \right)^2 }
{ \Lambda_{\rm  NMFV }^2 }
\end{equation}
where 
up to a universal effect (which is irrelevant for flavor
violation), we can set $c_1 = 0$.

\subsubsection{SUSY models}

As is well-known, in SUSY models, there are
contributions to $\Delta F = 2$ processes
(which are too large in
generic SUSY models) from box diagrams with squarks and gluinos:
a combination of few $100$ GeV sparticle masses and loop factor
gives $\Lambda_{\rm  NMFV } \sim$ few TeV as mentioned earlier.
Here, the flavor
violation is due to squark and quark mass matrices being misaligned.

For our purpose it is useful to divide the models in which  the flavor
problem is solved in SUSY into two main classes.
The first contains models with non-abelian flavor
symmetries~\cite{nonabelian} and various hybrid models~\cite{hybrid}. The second contains alignment models.
In models with non-abelian flavor symmetries, the  
1st and 2nd generation squarks are 
approximately degenerate (typically the non-degeneracy is
$O\left( \lambda_c^4 \right)$ which is $\sim m_s^2 / m_b^2$),
whereas the 3rd generation squarks are split from
these two. This implies that $c_2 \sim \lambda_C^4$
above (or $c_1 \approx c_2 + O( \lambda_C^4 )$ in general).
This avoids too large contribution
to $K - \bar{K}$ mixing even with all
mixing angles at gluino vertices being CKM-like, in particular
$1-2$ mixing in LH sector, $\left( D_L \right)_{12}$ can be $\sim \lambda_C$. In fact,
non-degeneracy of such size leads to an additional
contribution to $\epsK$ comparable to the one induced via
3rd generation ({\it i.e.}, $c_3$).
Similar relations are obtained in the various hybrid models.
As we will show below,
these models are similar to RS in this respect.
It is clear that with this additional effect and assuming either LH or RH
flavor-violation (but not both) 
and no helicity-flipping
operators, there is $1$ additional weak phase\footnote{The $U(2)_Q$ or
$U(2)_d$ subgroup is only approximate in this case so that
we can remove 1 less phase.}. Thus, $\sigma_{ K, d,s }$ 
in our analysis are independent phases.

Whereas in SUSY alignment models~\cite{align},
the squarks are not degenerate so that $c_2 \sim O(1)$ (and different from
$c_3$: again, in general, all three $c$'s are $O(1)$ and different), but
$1-2$ mixing angle is of order $\left( D_L \right)_{12}
\sim \lambda_C^5$ or smaller so that again the contribution to
$\epsK$ from direct $1-2$ mixing can be (at most) comparable to
the one from mixing with 3rd generation.

Note that
flavor violation 
induced by RH operators ({\it i.e.}, Eq. (\ref{NMFV}) with $Q_i \rightarrow d_i$) 
is comparable to that from
LH operators in typical non-abelian models
and in some alignment models (unlike in RS below). 
We can show that in this case there are 
a total of $6$ CPV phases (assuming as usual no
helicity flipping operators) which can be thought of as $2$ per transition: one each
for LH and RH mixing.
These two weak phases per transition, in general, enter in different
combinations in $\Delta F = 2$ and $\Delta F = 1$ processes resulting in
a loss of correlation between the two effects. 

\subsubsection{RS1}

Next, we show how our definition
of the general class of models exactly applies
to the RS1 case.
We present a brief review of flavor physics
in RS1: for more details, the reader is referred to 
\cite{APS}. 
The model consists of  
a compact warped extra dimension where
$4D$ gravity is localized near one end (called the Planck brane) 
while the EWSB sector
is near the other end (called the TeV brane). 
The warp factor of this geometry leads to characteristic mass scales
(and UV cut-off) 
being position-dependent, in particular, being exponentially
different at the $2$ ends of the extra dimension
thus explaining the hierarchy between the scales of $4D$ gravity and EWSB sector.

The hierarchy of quark and lepton masses is explained
by the idea of split fermions: localization of the 
light fermions
near Planck brane implies small Yukawa coupling to
Higgs localized near the TeV brane, whereas top quark 
is localized near TeV brane to account for its large mass~\cite{Grossman:1999ra,
Gherghetta:2000qt}.

The novel aspect of split fermions in warped
extra dimension is that,
unlike in flat extra dimension,
FCNC due to exchange of gauge KK modes are small~\cite{Gherghetta:2000qt}. This 
is due to KK modes being 
localized near TeV brane so that the non-universal part of coupling
to light fermions (which are near the Planck brane) is suppressed. 
Thus, 
this suppression of FCNC is related 
to the lightness of these fermions and hence we 
refer to this feature as RS-GIM/approximate flavor symmetries.

In analogy with the SM, the RS-GIM/approximate symmetries are
violated by heavy top quark as follows (see references~\cite{APS} for more details).
$(t,b)_L$ is quasi-localized near the TeV brane to account for
the large top mass which results in
couplings of {\em left}-handed $b$ 
to KK modes being larger than expected on the basis of $m_b$~\cite{Burdman}.
To be precise, we can show that, in interaction
basis, the 
coupling of $b_L$ to gauge KK modes is of the same 
size as the SM gauge couplings, {\it i.e.}, the coupling
to KK gluon, $g^b_{ G^{ KK } }$ is $\sim g_s$ and that to the KK $Z$,
$g^b_{ Z^{ KK } } $ is $\sim g_Z$.
Whereas the coupling of $s_L$, $d_L$ 
(and, in general, all light fermions) to gauge KK modes is smaller than the
SM gauge coupling by factor of $\sim \sqrt{ \log \left(
M_{ Pl } / \hbox{TeV} \right)} \sim 5$. To repeat, all
these sizes of the couplings follow from considering the overlaps of the wave-functions.

This implies that, after
performing a unitary 
rotation to go to mass basis from interaction basis,
there is a flavor violating coupling of the KK gluon: 
$b_L - s_L (d_L) $ vertex
is $\sim g^b_{ G^{KK} } \left( D_L \right)_{ 23 (13) }$
(and similarly for the KK $Z$), where $D_L$ is the unitary
transformation for left-handed down quarks.

To estimate the sizes of these couplings, we need to know $D_L$.
We will assume structureless
(or anarchic) $5D$ Yukawa couplings, {\it i.e.},  
all the entries in the $5D$ Yukawa matrices are
of same order and hierarchies in the $4D$ Yukawas ({\it i.e.}, in
masses and mixing angles)
are explained by the overlap of the fermions' wave-functions in the
extra dimension. 
This results in $D_L \sim V_{ CKM }$.

We now briefly describe the features of the FCNC induced by these
flavor violating coupling to the gauge KK modes. First of all, it is clear that
NP is dominantly only in {\em left}-handed operators since the couplings
of only
$b_L$ to gauge KK mode violate RS-GIM/approximate flavor symmetries (due to the 
heaviness of top quark).

Let us begin with tree-level KK gluon exchange. 
The computation of this diagram can be estimated as:
\begin{eqnarray}
\frac{ M_{ 12 }^{\rm RS} }{ M_{ 12 }^{\rm SM} }
 & \sim & 16 \pi^2\,\frac{ \left(g_{ G^{ KK } }^b\right)^2  }{ g_2^4}\,
\frac{ m_W^2  }{ m_{\rm KK }^2 } 
\sim C \left( g^b_{ G^{ KK } } \right)^2
\left( \frac{ 3 \hbox{TeV} }{ m_{\rm KK } } \right)^2\,,
\label{F2}
\end{eqnarray}
where $C$ is an order one complex coefficient, 
mixing angles are of same size in both RS1 and SM contributions and
$M_{12}^{\rm SM,\,RS}$ is the SM (box diagram) 
and RS1 (KK gluon exchange) $\Delta F=2$
transition amplitudes respectively.
Using the above couplings,
it is 
easy to see that, for KK mass $\sim 3$ TeV, 
KK gluon exchange contribution to $\Delta F = 2$
processes is comparable to the short-distance part of the
SM box diagram. 
Note that the KK gluon exchange generates a $\Delta F = 2$ operator
with $V- A$, but color octet structure. However, using identities for $SU(3)$
Gell-Mann matrices and Fierz transformation, this operator can be converted to
that in the SM.
Then, it is clear that
the effect of the KK gluon exchange can be parameterized
as in Eq. (\ref{par}) with $h_{ K,d,s} \sim O(1)$ -- the crucial point
is that $h_{ K,d,s}$ are 
simply the ratios of WC's, {\it i.e.}, they are
given in terms of NP parameters only (since matrix elements are the same as in
SM).

Recall that the data after summer of 2004 constrains $h_d$ ($h_K$) 
to be smaller than $\sim 0.4$ ($0.6$).
This can be accommodated in RS1 with a very mild tuning as follows.
It is clear from the above discussion that
if the $(1, 3)$  
entry of $5D$ Yukawa is suppressed by $\sim 2$ relative
to other entries, then $\left( D_L \right)_{ 13 } \sim 1/2 \; V_{ td }$.
Since $h_d \propto 
\big[ \left( D_L \right)_{ 13 } \big]^2$, this gives $h_d \sim 1/4$ as desired
(and similarly for $h_K$).

Note that
there are two comparable contributions to $\Delta S = 2$ 
transition as follows.
The KK gluon $d_L - s_L$ vertex
has a contribution
$\sim g^b_{ G^{KK} } \left( D_L \right)_{13} \left( D_L \right)_{23}$ from
mixing with 3rd generation and a direct $1-2$ mixing contribution:
the latter involves large $1-2$ mixing angle $\left( D_L \right)_{ 12 }
\sim \lambda_c$ multiplied by a suppressed
coupling of KK gluon to $s_L$ (see references~\cite{APS} for more details).

Next, we consider
$\Delta F = 1$ transitions. It is clear that
the contribution from KK gluon (color octet) exchange
in $\Delta F = 1$ transitions is of 
the same type of the QCD penguins (QCDP) operators. 
However, using the above couplings (in particular, the small coupling
of KK gluon to light quarks) 
its contribution is suppressed by $\sim 1/5$ compared to the SM QCDP. 
Moreover, there is a dilution of this effect in RG scaling from the TeV scale
to $m_b$. Hence, KK gluon contribution in $\Delta F = 1$ QCDP is negligible.

The contribution from KK $Z$ exchange is smaller than
that of KK gluon by $\sim g_Z^2 / g_s^2$.
However, the 
KK $Z$ mixes with zero-mode of $Z$ due to
EWSB/Higgs vev. Moreover, the coupling of Higgs to KK $Z$ is enhanced 
by $\sim \sqrt{ \log \left(
M_{ Pl } / \hbox{TeV} \right)}$
relative to SM. This results in a flavor violating $Z$ vertex 
and, in turn, to a contribution of $Z$ exchange to $\Delta F = 1$ transition
which is comparable to SM
$Z$ penguin:
\begin{eqnarray}
\frac{ C^{ Z,{\rm RS} }_{ 7-10 } }{ C^{ Z, {\rm SM}}_{ 7 - 10 } } 
& \sim & \frac{ 16 \pi^2 }{ g_2^2 }  
\frac{ g^b_{Z^{\rm KK} }}{ g_Z } \sqrt{ \log \left( M_{ Pl} 
\hbox{TeV} \right) } 
\frac{ m_Z^2 }{ m_{\rm KK }^2 } 
\nonumber \\
 & \sim & \frac{ g^b_{Z^{\rm KK} }}{ g_Z } \left( \frac{ 3 \hbox{TeV} }{ m_{\rm KK } } \right)^2\,,
\end{eqnarray}
where the superscript 
$Z$ on $C_{ 7-10 }$ denotes 
$Z$ penguin part and, as for $\Delta F = 2$ case, the SM contribution is from
top 
quark loop
and mixing angles are of same size in both contributions. 
Thus,
$h_s^1 \sim 1$ in the notation of model-independent analysis
\footnote{We can show that 
$\Delta F = 2$ effect due to $Z$ exchange is smaller than due to KK gluon. 
Of course,
all the three contributions to $\Delta F = 2$ transition, namely KK gluon, KK $Z$
and physical $Z$ exchange generate the same operator with the
same weak phase and hence their
combined effect can be included in $h_{ K, d, s}$ as in Eq. (\ref{par}).}.

Two other features of NP also follow
from consideration of the above couplings.
We can see that the weak phase in NP contributions
to both $\Delta F = 2$ and $\Delta F = 1$ EWP come from phase of $\left( D_L 
\right)_{ 23 (13)}$ and hence we have the same phase ($\sigma_{ K, d, s }$:
up to the obvious factor of $2$) in Eqs.
(\ref{par}) and (\ref{NPEWP}) of the model independent analysis.

Secondly, it is clear that NP contributions to 
$\Delta F = 2$
and $\Delta F = 1$ are proportional to two and one power, respectively, of 
the coupling of $b_L$ to gauge KK mode.
Moreover, NP effect in $\Delta F =1$ depends, in addition, on
the coupling of KK $Z$ to Higgs. Thus, in general
$h_s \neq h_s^1$ in Eqs. (\ref{par}) ad (\ref{NPEWP}), 
although both are $\sim O(1)$ as explained above.

Thus, NP in RS1 has all the features we assumed in the model-independent analysis.

Before concluding this section, 
we point out that 
we have neglected mixing between zero and KK fermions induced by the Higgs vev
which results
in new flavor-violating effects. We can show that
the correlation between $\Delta B = 2$ and $\Delta B = 1$ is affected only at the
$\sim 10 \%$ level (even for maximal 
$5D$ Yukawa) since the new
effects are proportional to $\left( D_L \right)_{33} 
\left( D_L \right)_{3i}$ (just like
the
effects from the gauge KK modes). 
Whereas, for $s \rightarrow d$ transition correlation between 
$\Delta F = 2$ and $\Delta F = 1$ is spoiled for maximal $5D$ Yukawa. The reason is that
this transition involves both
$\left( D_L \right)_{31} \left( D_L \right)_{32}$ and 
$\left( D_L \right)_{12} \left( D_L \right)_{22}$ and the
combinations involved are {\em different} for the gauge KK and fermion mixing effects.
However, 
the KK fermion effect rapidly decreases as we reduce $5D$ Yukawa allowing us
to recover the correlation.



\section{Discussion and Conclusions}\label{Con}

In a few years, LHC will hopefully unravel the mystery of EWSB: 
unless nature is fine-tuned, the 
Planck-weak hierarchy
must be stabilized by
NP at $\sim$ TeV and the LHC will most likely discover this physics
beyond the SM. An interesting question is can
we indirectly see this NP before then in flavor physics? 
The reason to hope for such a possibility is that it is likely that this 
NP couples dominantly to the 3rd generation
due to heaviness of the top quark, whereas 
it couples weakly to 1st and 2nd generations,
giving rise to flavor violating effects. However, in
the scenario with minimal flavor violation (MFV), it is possible that at low energies there
are no new (in addition to SM Yukawa) 
spurions which break flavor symmetry, {\it i.e.}, the only
surviving imprint of
origins of flavor in NP at TeV 
comes from SM Yukawa. Such models are easily consistent
with data and on the flip side, it will be be
difficult to have any clue of flavor mechanism in low energy experiments
in this case.

In this paper, we have considered a more promising possibility
by extending the minimal scenario (we denote it as Next to MFV). Specifically,
we include new spurions which break flavor symmetries
in the form of 4-fermion operators involving only 3rd generation. These
effects have
CKM-like misalignment with up-Yukawa and are
suppressed by few TeV since they arise from the {\em same}
physics which stabilizes the weak scale.
Large contributions to $K$ mixing from such low scale physics is
avoided due to weak (or degenerate) coupling of 1st and 2nd generation to the NP.
We showed that this framework results in 
NP contributions (with new phases) in $\Delta F = 2$ 
processes being comparable to SM 
short distance effects.

The success of SM unitarity triangle fit, especially 
after the recent results on $B \rightarrow D K , \rho, \rho$. 
have led to the lore that such a scenario is ruled out.  
However, we showed that $\sim 30 \%$ NP effects (with arbitrary phase)
compared to SM
are still allowed so that more data is required
to rule out 
NMFV with a few TeV mass scale! Conversely, there is
an opportunity to discover it, especially in $B_s$ mixing
currently
at Run II (and LHC in a few years) or in more precise measurements
of $B \rightarrow D K , \rho \rho,\pi\pi$ (and other related modes) at
BABAR and BELLE.

We also considered NP effects in
$\Delta F = 1$  processes resulting from another class of 4-fermion
operators.  
In particular, we showed 
that NP comparable to SM $Z$ penguin
can explain the recent anomalies in $B \rightarrow \phi K_S,
\eta^{\prime} K_S$
and also be consistent with the data on $B\to K\pi$ transitions.
In a certain class of models, we showed that
there are correlations between NP effects
in $\Delta F = 2$ and $\Delta F = 1$ processes, 
in particular of the above effect in $B \rightarrow 
( \eta^{ \prime }, \phi ) K_S$
with $B_s$ mixing 
resulting
in predictions in latter if we choose parameters
to explain the anomalies in the former. 
It will be interesting to consider
processes such as $b \rightarrow s \gamma$ and $b \rightarrow s l^+ l^-$
in this framework.

We briefly showed how SUSY non-abelian and alignment and various
hybrid models fall 
in this class and in more detail how RS is in this class.
We hope that our work will
provide motivation to push further and continue vigorously the $B$ factory
program before the LHC and even during the LHC 
in order to complement the direct search for such NP.

\acknowledgments{We thank Alessandro Cerri,  Andreas H\"ocker,  Heiko
  Lacker, Zoltan Ligeti, Ann Nelson, Matthias Neubert,
  and Matt Strassler for useful discussions.
  This work was supported in part by the DOE under
contracts 
DE-FG02-90ER40542 (KA),
DE-FG02-90ER40542 and DE-AC02-05CH11231 (MP and GP), and
by the DOE under the cooperative
research agreement DOE-FC02-94ER40818 (DP).}



\begin{thebibliography}{99}

%
%
\bibitem{SUSY}
See review by H.~E.~Haber in S.~Eidelman {\it et al.} [Particle Data Group],
Phys.\ Lett.\ B {\bf 592}, 1 (2004)
and references therein.
%
%
\bibitem{TC}
S.~Weinberg,
Phys.\ Rev.\ D {\bf 13}, 974 (1976);
Phys.\ Rev.\ D {\bf 19}, 1277 (1979);
L.~Susskind,
Phys.\ Rev.\ D {\bf 20}, 2619 (1979).

%
%
\bibitem{TCreview}
For a review, see C.~T.~Hill and E.~H.~Simmons,
  Phys.\ Rept.\  {\bf 381}, 235 (2003)
  [Erratum-ibid.\  {\bf 390}, 553 (2004)]
  [arXiv:hep-ph/0203079].


%
%
\bibitem{CH}
D.~B.~Kaplan and H.~Georgi,
Phys.\ Lett.\ B {\bf 136}, 183 (1984)
 B {\bf 136}, 187 (1984); \\
H.~Georgi, D.~B.~Kaplan and P.~Galison,
Phys.\ Lett.\ B {\bf 143}, 152 (1984); \\
H.~Georgi and D.~B.~Kaplan,
Phys.\ Lett.\ B {\bf 145}, 216 (1984); \\
M.~J.~Dugan, H.~Georgi and D.~B.~Kaplan,
Nucl.\ Phys.\ B {\bf 254}, 299 (1985).

%
%
\bibitem{topcolor}
  C.~T.~Hill,
  Phys.\ Lett.\ B {\bf 266}, 419 (1991);
  C.~T.~Hill,
  Phys.\ Lett.\ B {\bf 345}, 483 (1995)
  [arXiv:hep-ph/9411426];
  K.~D.~Lane and E.~Eichten,
  Phys.\ Lett.\ B {\bf 352}, 382 (1995)
  [arXiv:hep-ph/9503433].

%
%
\bibitem{LH}
N.~Arkani-Hamed, A.~G.~Cohen, E.~Katz, A.~E.~Nelson, 
T.~Gregoire and J.~G.~Wacker,
JHEP {\bf 0208}, 021 (2002)
[arXiv:hep-ph/0206020];
N.~Arkani-Hamed, A.~G.~Cohen, E.~Katz and A.~E.~Nelson,
JHEP {\bf 0207}, 034 (2002)
[arXiv:hep-ph/0206021].
For a review and more references, see 
M.~Schmaltz and D.~Tucker-Smith,
arXiv:hep-ph/0502182.
For string models which realize the ADD scenario,
see 
D. Cremades, L.E.Ibanez and F.Marchesano,
Nucl. Phys. B643 (2002) 93, [hep-th/0205074];
C. Kokorelis,
Nucl. Phys. B677 (2004) 115, [hep-th/0207234].
%
%
\bibitem{ADD}
  N.~Arkani-Hamed, S.~Dimopoulos and G.~R.~Dvali,
  Phys.\ Lett.\ B {\bf 429}, 263 (1998)
  [arXiv:hep-ph/9803315];
I.~Antoniadis, N.~Arkani-Hamed, S.~Dimopoulos and G.~R.~Dvali,
Phys.\ Lett.\ B {\bf 436}, 257 (1998)
[arXiv:hep-ph/9804398].
%
\bibitem{RS}
  L.~Randall and R.~Sundrum,
  Phys.\ Rev.\ Lett.\  {\bf 83}, 3370 (1999)
  [arXiv:hep-ph/9905221].
%
\bibitem{GMSB}
 M.~Dine and A.~E.~Nelson,
 Phys.\ Rev.\ D {\bf 48}, 1277 (1993)
 [arXiv:hep-ph/9303230];
  M.~Dine, A.~E.~Nelson and Y.~Shirman,
  Phys.\ Rev.\ D {\bf 51}, 1362 (1995)
  [arXiv:hep-ph/9408384];
  M.~Dine, A.~E.~Nelson, Y.~Nir and Y.~Shirman,
  Phys.\ Rev.\ D {\bf 53}, 2658 (1996)
  [arXiv:hep-ph/9507378].
For a review and more references, see
G.~F.~Giudice and R.~Rattazzi,
Phys.\ Rept.\  {\bf 322}, 419 (1999)
[arXiv:hep-ph/9801271].
For AdS$_5$ duals of GMSB, see,
for example,
Z.~Chacko, Y.~Nomura and D.~Tucker-Smith,
arXiv:hep-ph/0504095.

%
%
\bibitem{AMSB}
L.~Randall and R.~Sundrum,
Nucl.\ Phys.\ B {\bf 557}, 79 (1999)
[arXiv:hep-th/9810155];
G.~F.~Giudice, M.~A.~Luty, H.~Murayama and R.~Rattazzi,
JHEP {\bf 9812}, 027 (1998)
[arXiv:hep-ph/9810442].


%
\bibitem{MFV}
 A.~Ali and D.~London,
  Eur.\ Phys.\ J.\ C {\bf 9}, 687 (1999)
  [arXiv:hep-ph/9903535];
  A.~J.~Buras, P.~Gambino, M.~Gorbahn, S.~Jager and L.~Silvestrini,
  Phys.\ Lett.\ B {\bf 500}, 161 (2001)
  [arXiv:hep-ph/0007085];
G.~D'Ambrosio, G.~F.~Giudice, G.~Isidori and A.~Strumia,
Nucl.\ Phys.\ B {\bf 645}, 155 (2002)
[arXiv:hep-ph/0207036].

\bibitem{MFVmore}
  C.~Bobeth, M.~Bona, A.~J.~Buras, T.~Ewerth, M.~Pierini, L.~Silvestrini and A.~Weiler,
  arXiv:hep-ph/0505110;
  A.~J.~Buras,
  Acta Phys.\ Polon.\ B {\bf 34}, 5615 (2003)
  [arXiv:hep-ph/0310208];
  A.~J.~Buras,
  Phys.\ Lett.\ B {\bf 566}, 115 (2003)
  [arXiv:hep-ph/0303060];
   A.~J.~Buras and R.~Fleischer,
  Phys.\ Rev.\ D {\bf 64}, 115010 (2001)
  [arXiv:hep-ph/0104238];
   A.~J.~Buras and R.~Buras,
  Phys.\ Lett.\ B {\bf 501}, 223 (2001)
  [arXiv:hep-ph/0008273];
 S.~Laplace, Z.~Ligeti, Y.~Nir and G.~Perez,
  Phys.\ Rev.\ D {\bf 65}, 094040 (2002)
  [arXiv:hep-ph/0202010];
  S.~Bergmann and G.~Perez,
  Phys.\ Rev.\ D {\bf 64}, 115009 (2001)
  [arXiv:hep-ph/0103299];
S.~Bergmann and G.~Perez,
  JHEP {\bf 0008}, 034 (2000)
  [arXiv:hep-ph/0007170].


                                %
%
%
\bibitem{IL}
  T.~Inami and C.~S.~Lim,
  Prog.\ Theor.\ Phys.\  {\bf 65}, 297 (1981)
  [Erratum-ibid.\  {\bf 65}, 1772 (1981)].

%
\bibitem{align}
Y.~Nir and N.~Seiberg,
Phys.\ Lett.\ B {\bf 309}, 337 (1993)
[arXiv:hep-ph/9304307];
M.~Leurer, Y.~Nir and N.~Seiberg,
Nucl.\ Phys.\ B {\bf 398}, 319 (1993)
[arXiv:hep-ph/9212278];
Nucl.\ Phys.\ B {\bf 420}, 468 (1994)
[arXiv:hep-ph/9310320].
For a recent study, see
Y.~Nir and G.~Raz,
Phys.\ Rev.\ D {\bf 66}, 035007 (2002)
[arXiv:hep-ph/0206064].

%
%
\bibitem{nonabelian}
M.~Dine, R.~G.~Leigh and A.~Kagan,
Phys.\ Rev.\ D {\bf 48}, 4269 (1993)
[arXiv:hep-ph/9304299];
A.~Pomarol and D.~Tommasini,
Nucl.\ Phys.\ B {\bf 466}, 3 (1996)
[arXiv:hep-ph/9507462];
R.~Barbieri, G.~R.~Dvali and L.~J.~Hall,
Phys.\ Lett.\ B {\bf 377}, 76 (1996)
[arXiv:hep-ph/9512388].

  
\bibitem{hybrid} See {\it e.g.}:
A.~G.~Cohen, D.~B.~Kaplan, F.~Lepeintre and A.~E.~Nelson,
  Phys.\ Rev.\ Lett.\  {\bf 78}, 2300 (1997)
  [arXiv:hep-ph/9610252];
  R.~Barbieri, L.~J.~Hall, G.~Marandella, Y.~Nomura, T.~Okui, S.~J.~Oliver and M.~Papucci,
  Nucl.\ Phys.\ B {\bf 663}, 141 (2003)
  [arXiv:hep-ph/0208153];bibitem{Barbieri:2002uk}
  R.~Barbieri, G.~Marandella and M.~Papucci,
  Phys.\ Rev.\ D {\bf 66}, 095003 (2002)
  [arXiv:hep-ph/0205280]; R.~Barbieri, L.~J.~Hall and Y.~Nomura,
  Phys.\ Rev.\ D {\bf 63}, 105007 (2001)
  [arXiv:hep-ph/0011311];
  A.~E.~Nelson and M.~J.~Strassler,
  JHEP {\bf 0009}, 030 (2000)
  [arXiv:hep-ph/0006251];  
  JHEP {\bf 0207}, 021 (2002)
  [arXiv:hep-ph/0104051];
  Phys.\ Rev.\ D {\bf 56}, 4226 (1997)
  [arXiv:hep-ph/9607362];
   M.~A.~Luty and J.~Terning,
  Phys.\ Rev.\ D {\bf 62}, 075006 (2000)
  [arXiv:hep-ph/9812290].


\bibitem{Frame}
C.~Dib, D.~London and Y.~Nir,
Int.\ J.\ Mod.\ Phys.\ A {\bf 6}, 1253 (1991);
Y.~Nir and D.~J.~Silverman,
Nucl.\ Phys.\ B {\bf 345}, 301 (1990).
  
\bibitem{Para}
Y.~Grossman, Y.~Nir and M.~P.~Worah,
Phys.\ Lett.\ B {\bf 407}, 307 (1997)
[arXiv:hep-ph/9704287];
J.~M.~Soares and L.~Wolfenstein,
Phys.\ Rev.\ D {\bf 47}, 1021 (1993);
A.~G.~Cohen, D.~B.~Kaplan, F.~Lepeintre and A.~E.~Nelson,
Phys.\ Rev.\ Lett.\  {\bf 78}, 2300 (1997)
[arXiv:hep-ph/9610252];
J.~P.~Silva and L.~Wolfenstein,
Phys.\ Rev.\ D {\bf 55}, 5331 (1997)
[arXiv:hep-ph/9610208];
N.~G.~Deshpande, B.~Dutta and S.~Oh,
Phys.\ Rev.\ Lett.\  {\bf 77}, 4499 (1996)
[arXiv:hep-ph/9608231];
S.~Bergmann and G.~Perez,
Phys.\ Rev.\ D {\bf 64}, 115009 (2001)
[arXiv:hep-ph/0103299].

\bibitem{ENP}
G.~Eyal, Y.~Nir and G.~Perez,
JHEP {\bf 0008}, 028 (2000)
[arXiv:hep-ph/0008009].

\bibitem{LLNP}
S.~Laplace, Z.~Ligeti, Y.~Nir and G.~Perez,
Phys.\ Rev.\ D {\bf 65}, 094040 (2002)
[arXiv:hep-ph/0202010].

\bibitem{epsKc}
See {\it e.g}: A.~J.~Buras,
Acta Phys.\ Polon.\ B {\bf 34}, 5615 (2003)
[arXiv:hep-ph/0310208];
A.~J.~Buras, P.~Gambino, M.~Gorbahn, S.~Jager and L.~Silvestrini,
Nucl.\ Phys.\ B {\bf 592}, 55 (2001)
[arXiv:hep-ph/0007313].



\bibitem{epsQCD}
A.~J.~Buras, M.~Jamin and P.~H.~Weisz,
Nucl.\ Phys.\ B {\bf 347}, 491 (1990).



\bibitem{charm} See {\it e.g.}:
S.~Herrlich and U.~Nierste,
Nucl.\ Phys.\ B {\bf 476}, 27 (1996)
[arXiv:hep-ph/9604330].

\bibitem{CKMfitter}
J.~Charles {\it et al.}  [CKMfitter Group Collaboration],
Eur.\ Phys.\ J.\ C {\bf 41}, 1 (2005)
  [arXiv:hep-ph/0406184] and periodic updates at 
http://www.slac.stanford.edu/xorg/ckmfitter/.



\bibitem{Endo:2004dc}
  A.~L.~Kagan,
  arXiv:hep-ph/0407076;
M.~Endo, S.~Mishima and M.~Yamaguchi,
arXiv:hep-ph/0409245.

\bibitem{LMP}
   D.~T.~Larson, H.~Murayama and G.~Perez,
  JHEP {\bf 0507}, 057 (2005)
  [arXiv:hep-ph/0411178].


\bibitem{Ligeti04}
Z.~Ligeti,
arXiv:hep-ph/0408267. 

\bibitem{Nirera}
Y.~Nir,
Nucl.\ Phys.\ Proc.\ Suppl.\  {\bf 117}, 111 (2003)
[arXiv:hep-ph/0208080].

\bibitem{APS}
K.~Agashe, G.~Perez and A.~Soni,
  Phys.\ Rev.\ D {\bf 71}, 016002 (2005)
  [arXiv:hep-ph/0408134];
Phys.\ Rev.\ Lett.\  {\bf 93}, 201804 (2004)
[arXiv:hep-ph/0406101].

\bibitem{BBNR}
  F.~J.~Botella, G.~C.~Branco, M.~Nebot and M.~N.~Rebelo,
  arXiv:hep-ph/0502133.

\bibitem{Lap}
  S.~Laplace,
  arXiv:hep-ph/0209188.


\bibitem{hfag} 
Heavy Flavor Averaging Group, http://www.slac.stanford.edu/xorg/hfag/~.

\bibitem{BBNS}
M.~Beneke, G.~Buchalla, M.~Neubert and C.~T.~Sachrajda,
Phys.\ Rev.\ Lett.\  {\bf 83}, 1914 (1999);
[arXiv:hep-ph/9905312].
M.~Beneke, G.~Buchalla, M.~Neubert and C.~T.~Sachrajda,
Nucl.\ Phys.\ B {\bf 606}, 245 (2001)



\bibitem{Beneke:2003zv}
M.~Beneke and M.~Neubert,
Nucl.\ Phys.\ B {\bf 675}, 333 (2003)
[arXiv:hep-ph/0308039].

\bibitem{BPRS}
C.~W.~Bauer, D.~Pirjol, I.~Z.~Rothstein and I.~W.~Stewart,
Phys.\ Rev.\ D {\bf 70}, 054015 (2004)

\bibitem{BBL} 
G.~Buchalla, A.~J.~Buras and M.~E.~Lautenbacher,
Rev.\ Mod.\ Phys.\  {\bf 68}, 1125 (1996)

\bibitem{Zp}
 L.~S.~Durkin and P.~Langacker,
  Phys.\ Lett.\ B {\bf 166}, 436 (1986);
 Y.~Nir and D.~J.~Silverman,
  Phys.\ Rev.\ D {\bf 42}, 1477 (1990);
P.~Langacker and M.~Plumacher,
  Phys.\ Rev.\ D {\bf 62}, 013006 (2000)
  [arXiv:hep-ph/0001204];
D.~Atwood and G.~Hiller,
  arXiv:hep-ph/0307251;
 A.~J.~Buras, R.~Fleischer, S.~Recksiegel and F.~Schwab,
  Nucl.\ Phys.\ B {\bf 697}, 133 (2004)
  [arXiv:hep-ph/0402112].


\bibitem{GNR}
  Y.~Grossman, Y.~Nir and R.~Rattazzi,
  Adv.\ Ser.\ Direct.\ High Energy Phys.\  {\bf 15}, 755 (1998)
  [arXiv:hep-ph/9701231].


\bibitem{Zprecent}
 V.~Barger, C.~W.~Chiang, P.~Langacker and H.~S.~Lee,
  Phys.\ Lett.\ B {\bf 580}, 186 (2004)
  [arXiv:hep-ph/0310073];
  D.~A.~Demir, G.~L.~Kane and T.~T.~Wang,
  Phys.\ Rev.\ D {\bf 72}, 015012 (2005)
  [arXiv:hep-ph/0503290];
 R.~Mohanta,
  Phys.\ Rev.\ D {\bf 71}, 114013 (2005)
  [arXiv:hep-ph/0503225];
V.~Barger, C.~W.~Chiang, P.~Langacker and H.~S.~Lee,
  Phys.\ Lett.\ B {\bf 598}, 218 (2004)
  [arXiv:hep-ph/0406126];
T.~Han, P.~Langacker and B.~McElrath,
  Phys.\ Rev.\ D {\bf 70}, 115006 (2004)
  [arXiv:hep-ph/0405244];
V.~Barger, C.~W.~Chiang, J.~Jiang and P.~Langacker,
  Phys.\ Lett.\ B {\bf 596}, 229 (2004)
  [arXiv:hep-ph/0405108];
 X.~G.~He and G.~Valencia,
  Phys.\ Rev.\ D {\bf 70}, 053003 (2004)
  [arXiv:hep-ph/0404229];

\bibitem{kpn}
  S.~Bergmann and G.~Perez,
  Phys.\ Rev.\ D {\bf 64}, 115009 (2001)
  [arXiv:hep-ph/0103299];
 A.~J.~Buras, F.~Schwab and S.~Uhlig,
  arXiv:hep-ph/0405132;
 G.~Isidori, F.~Mescia and C.~Smith,
  Nucl.\ Phys.\ B {\bf 718}, 319 (2005)
  [arXiv:hep-ph/0503107];
A.~J.~Buras, M.~Gorbahn, U.~Haisch and U.~Nierste,
arXiv:hep-ph/0508165.

\bibitem{Grossman:1996ke}
Y.~Grossman and M.~P.~Worah,
Phys.\ Lett.\ B {\bf 395}, 241 (1997)
[arXiv:hep-ph/9612269].

\bibitem{London:1997zk}
D.~London and A.~Soni,
Phys.\ Lett.\ B {\bf 407}, 61 (1997)
[arXiv:hep-ph/9704277].


\bibitem{BaBe}
K.~Abe {\it et al.}  [Belle Collaboration],
arXiv:hep-ex/0507037;
B.~Aubert {\it et al.}  [BABAR Collaboration],
arXiv:hep-ex/0507087;
K.~Abe {\it et al.}  [Belle Collaboration],
arXiv:hep-ex/0507037;
B.~Aubert {\it et al.}  [BABAR Collaboration],
Phys.\ Rev.\ D {\bf 71}, 091102 (2005)
[arXiv:hep-ex/0502019];
B.~Aubert {\it et al.}  [BABAR Collaboration],
Phys.\ Rev.\ Lett.\  {\bf 94}, 161803 (2005)
[arXiv:hep-ex/0408127].

\bibitem{LP05}22nd International Symposium On Lepton-Photon
  Interactions At High Energy (LP 2005),
  Jun 2005, Uppsala, Sweden, http://www.uu.se/lp2005/.

\bibitem{Grossman:2003qp}
Y.~Grossman, Z.~Ligeti, Y.~Nir and H.~Quinn,
Phys.\ Rev.\ D {\bf 68}, 015004 (2003)
[arXiv:hep-ph/0303171].

\bibitem{Gronau:2003kx}
M.~Gronau, Y.~Grossman and J.~L.~Rosner,
Phys.\ Lett.\ B {\bf 579}, 331 (2004)
[arXiv:hep-ph/0310020].

\bibitem{3body}
G.~Engelhard, Y.~Nir and G.~Raz,
  arXiv:hep-ph/0505194.


\bibitem{boundsplus}
M.~Gronau and J.~L.~Rosner,
  Phys.\ Rev.\ D {\bf 71}, 074019 (2005)
  [arXiv:hep-ph/0503131].
M.~Gronau, J.~L.~Rosner and J.~Zupan,
  Phys.\ Lett.\ B {\bf 596}, 107 (2004)
  [arXiv:hep-ph/0403287].


\bibitem{Beneke:2005pu}
  M.~Beneke,
  Phys.\ Lett.\ B {\bf 620}, 143 (2005)
  [arXiv:hep-ph/0505075].

\bibitem{Kagan}
A.~L.~Kagan,
arXiv:hep-ph/0407076;
Talk at the 30th SLAC summer Institute on Particle Physics, "Secret of
the B mesons", Stanford, USA (02).

\bibitem{Naive}
A.~Ali and C.~Greub,
Phys.\ Rev.\ D {\bf 57}, 2996 (1998)
[arXiv:hep-ph/9707251];
A.~Ali, G.~Kramer and C.~D.~Lu,
Phys.\ Rev.\ D {\bf 58}, 094009 (1998)
[arXiv:hep-ph/9804363];
A.~Ali, G.~Kramer and C.~D.~Lu,
Phys.\ Rev.\ D {\bf 59}, 014005 (1999)
[arXiv:hep-ph/9805403].

\bibitem{BFRS}
G.~Buchalla, G.~Hiller, Y.~Nir and G.~Raz,
arXiv:hep-ph/0503151.


\bibitem{xx}
Y.~Grossman, M.~Neubert and A.~L.~Kagan,
  JHEP {\bf 9910}, 029 (1999)
  [arXiv:hep-ph/9909297].

\bibitem{xx1}
R.~Fleischer and J.~Matias,
  Phys.\ Rev.\ D {\bf 61}, 074004 (2000)
  [arXiv:hep-ph/9906274].

\bibitem{xx2}
A.~Datta and D.~London,
  Phys.\ Lett.\ B {\bf 595}, 453 (2004)
  [arXiv:hep-ph/0404130];
A.~Datta, M.~Imbeault, D.~London, V.~Page, N.~Sinha and R.~Sinha,
  Phys.\ Rev.\ D {\bf 71}, 096002 (2005)
  [arXiv:hep-ph/0406192];
S.~Baek, P.~Hamel, D.~London, A.~Datta and D.~A.~Suprun,
  Phys.\ Rev.\ D {\bf 71}, 057502 (2005)
  [arXiv:hep-ph/0412086];
A.~Datta and P.~J.~O'Donnell,
Phys.\ Rev.\ D {\bf 72}, 113002 (2005)
[arXiv:hep-ph/0508314].


\bibitem{xx3}
S.~Mishima and T.~Yoshikawa,
  Phys.\ Rev.\ D {\bf 70}, 094024 (2004)
  [arXiv:hep-ph/0408090];
C.~S.~Kim, Y.~J.~Kwon, J.~Lee and T.~Yoshikawa,
  arXiv:hep-ph/0509015.

\bibitem{xx4}
M.~Gronau and J.~L.~Rosner,
  Phys.\ Rev.\ D {\bf 71}, 074019 (2005)
  [arXiv:hep-ph/0503131].



\bibitem{GHLR}
M.~Gronau, O.~F.~Hernandez, D.~London and J.~L.~Rosner,
Phys.\ Rev.\ D {\bf 50}, 4529 (1994)
[arXiv:hep-ph/9404283];
M.~Gronau, O.~F.~Hernandez, D.~London and J.~L.~Rosner,
Phys.\ Rev.\ D {\bf 52}, 6374 (1995)
[arXiv:hep-ph/9504327].

\bibitem{NeRo} 
M.~Neubert and J.~L.~Rosner,
Phys.\ Lett.\ B {\bf 441}, 403 (1998);
[arXiv:hep-ph/9808493];
M.~Neubert and J.~L.~Rosner,
Phys.\ Rev.\ Lett.\  {\bf 81}, 5076 (1998)
[arXiv:hep-ph/9809311];
M.~Neubert,
JHEP {\bf 9902}, 014 (1999)
[arXiv:hep-ph/9812396].

\bibitem{BuFl}
A.~J.~Buras and R.~Fleischer,
Eur.\ Phys.\ J.\ C {\bf 11}, 93 (1999)
[arXiv:hep-ph/9810260].

\bibitem{GPY}
M.~Gronau, D.~Pirjol and T.~M.~Yan,
Phys.\ Rev.\ D {\bf 60}, 034021 (1999)
[Erratum-ibid.\ D {\bf 69}, 119901 (2004)]
[arXiv:hep-ph/9810482].

\bibitem{Grossman:1999av}
Y.~Grossman, M.~Neubert and A.~L.~Kagan,
JHEP {\bf 9910}, 029 (1999)
[arXiv:hep-ph/9909297].

\bibitem{London}
M.~Imbeault, A.~L.~Lemerle, V.~Page and D.~London,
Phys.\ Rev.\ Lett.\  {\bf 92}, 081801 (2004)
[arXiv:hep-ph/0309061].


\bibitem{PDG}
  S.~Eidelman {\it et al.}  [Particle Data Group],
  Phys.\ Lett.\ B {\bf 592} (2004) 1.

\bibitem{Grossman:1999ra}
Y.~Grossman and M.~Neubert,
Phys.\ Lett.\ B {\bf 474}, 361 (2000)
[arXiv:hep-ph/9912408].

\bibitem{Gherghetta:2000qt}
T.~Gherghetta and A.~Pomarol,
Nucl.\ Phys.\ B {\bf 586}, 141 (2000)
[arXiv:hep-ph/0003129];
S.~J.~Huber and Q.~Shafi,
Phys.\ Lett.\ B {\bf 498}, 256 (2001)
[arXiv:hep-ph/0010195];
S.~J.~Huber,
  Nucl.\ Phys.\ B {\bf 666}, 269 (2003)
  [arXiv:hep-ph/0303183].

\bibitem{Burdman}
G.~Burdman,
Phys.\ Lett.\ B {\bf 590}, 86 (2004)
[arXiv:hep-ph/0310144].























  
  

\end{thebibliography}
\end{document}